\documentclass[12pt]{article}
\makeatletter \@addtoreset{equation}{section} \makeatother
\renewcommand{\theequation}{\thesection.\arabic{equation}}
\newcommand{\ra}{\rightarrow}
\newcommand{\ba}{\begin{array}}
\newcommand{\ea}{\end{array}}
\newcommand{\beq}{\begin{equation}}
\newcommand{\eeq}{\end{equation}}
\newcommand{\bea}{\begin{eqnarray}}
\newcommand{\eea}{\end{eqnarray}}

\def\bce{\begin{center}}
\def\ece{\end{center}}

\def\nonu{\nonumber}

\def\pa{\partial}

\def\be{\beta}

\def\ep{\epsilon}

\def\la{\lambda}

\def\eps6{{\displaystyle \mathop{\epsilon}^{6}}{}}
\def\g6{{\displaystyle \mathop{g}^{6}}{}}

\def\nab6{{\displaystyle \mathop{\nabla}^{6}}{}}

\def\0{{\sst{(0)}}}
\def\1{{\sst{(1)}}}
\def\2{{\sst{(2)}}}
\def\3{{\sst{(3)}}}
\def\4{{\sst{(4)}}}
\def\5{{\sst{(5)}}}
\def\6{{\sst{(6)}}}
\def\7{{\sst{(7)}}}
\def\8{{\sst{(8)}}}

\def\ba{\begin{array}}
\def\ea{\end{array}}
\def\beq{\begin{equation}}
\def\eeq{\end{equation}}
\def\be{\begin{equation}}
\def\ee{\end{equation}}

\def\la{\lambda}
\def\eps{\epsilon}

\def\ba{\begin{array}}
\def\ea{\end{array}}
\def\beq{\begin{equation}}
\def\eeq{\end{equation}}
\def\be{\begin{equation}}
\def\ee{\end{equation}}

\def\la{\lambda}
\def\eps{\epsilon}

\def\eps6{{\displaystyle \mathop{\epsilon}^{6}}{}}

\def\nab6{{\displaystyle \mathop{\nabla}^{6}}{}}

\newcommand{\bean}{\begin{eqnarray*}}
\newcommand{\eean}{\end{eqnarray*}}

\begin{document}
\thispagestyle{empty} \addtocounter{page}{-1}
   \begin{flushright}
\end{flushright}

\vspace*{1.3cm}
  
\centerline{ \Large \bf   
Three Point Functions in the Large ${\cal N}=4$ Holography } 
\vspace*{1.5cm}
\centerline{{\bf Changhyun Ahn } and {\bf Hyunsu Kim}
} 
\vspace*{1.0cm} 
\centerline{\it 
Department of Physics, Kyungpook National University, Taegu
702-701, Korea} 
\vspace*{0.8cm} 
\centerline{\tt ahn@knu.ac.kr, \qquad kimhyun@knu.ac.kr 
} 
\vskip2cm

\centerline{\bf Abstract}
\vspace*{0.5cm}

The $16$ higher spin currents with spins $(1, \frac{3}{2}, \frac{3}{2}, 2)$,
$(\frac{3}{2}, 2, 2, \frac{5}{2})$, $(\frac{3}{2}, 2, 2, \frac{5}{2})$
and $(2, \frac{5}{2}, \frac{5}{2}, 3)$ in an extension of 
large ${\cal N}=4$ `nonlinear' superconformal algebra in two dimensions
were obtained previously.
By analyzing the zero mode eigenvalue equations carefully, the 
three point functions of bosonic (higher spin) currents with two scalars  
for any finite $N$ (where $SU(N+2)$ is the group of coset) 
and $k$ (the level of spin-$1$ Kac Moody current) 
are obtained.
Furthermore, the $16$ higher spin currents with above spin contents
in an extension of large ${\cal N}=4$ `linear' superconformal algebra 
are obtained for generic $N$ and $k$ implicitly.
The corresponding three point functions are also determined.
Under the large $N$ 't Hooft limit, the two corresponding 
three point functions 
in the nonlinear and linear versions coincide with each other although they
are completely different for finite $N$ and $k$. 

\baselineskip=18pt
\newpage
\renewcommand{\theequation}
{\arabic{section}\mbox{.}\arabic{equation}}

\section{Introduction}

One of the consistency checks 
in the large ${\cal N}=4$ holography \cite{GG1305} can be 
described as the matching of correlation functions where one can see
more dynamical information.
One would like to match the two dimensional conformal field 
theory (CFT) answer with the predictions from the bulk Vasiliev 
(or its generalization) theory.
The simplest three-point functions  involve two scalar primaries with 
one higher spin current.
Then one needs to obtain the 
eigenvalue equations for the zero modes of the higher spin currents
acting on the coset scalar primaries. 
The main motivation of 
\cite{AK1411}
was to construct the higher spin currents for general 
$N$ and $k$ in order to see these eigenvalue equations explicitly.
Here the positive integer 
$N$ appears in the group $G=SU(N+2)$ of the 
${\cal N}=4$ coset theory in two dimensional CFT  
while the positive integer 
$k$ is the level of bosonic spin-$1$ (affine Kac-Moody) current. 
See also the relevant works in \cite{GG1011,GG1205} for the similar 
duality in the bosonic theory and there exists a review paper \cite{GG1207}
where one can find the relevant works in the context of 
higher spin AdS/CFT correspondence.

Before describing the eigenvalue equations,
let us recall what is
the large ${\cal N}=4$ coset theory in two dimensions.
More explicitly,
the ${\cal N}=4$ coset theory is described by the following 
`supersymmetric' coset
\bea
\mbox{Wolf} \times SU(2) \times U(1) = \frac{SU(N+2)}{SU(N)},
\nonu
\eea
where $N$ is odd.
The basic currents are given by 
the bosonic spin-$1$ current $V^a(z)$ and 
the fermionic spin-$\frac{1}{2}$
current $Q^b(z)$. 
The operator product expansion (OPE) 
between these currents does not have any singular term.
The indices run over 
$a, b, \cdots =1, 2, \cdots, \frac{(N+2)^2-1}{2}$, $1^{\ast}, 2^{\ast}, 
\cdots, (\frac{(N+2)^2-1}{2})^{\ast}$. The number 
$(N+2)^2-1$   is the dimension of 
$G=SU(N+2)$ group. 
For the extension of the ${\cal N}=4$ `nonlinear' superconformal algebra,
the relevant  coset is given by 
the Wolf space itself 
$\frac{SU(N+2)}{SU(N) \times SU(2) \times U(1)}$. 
For the extension of the ${\cal N}=4$ `linear' superconformal algebra,
the corresponding  coset is given by 
the Wolf space multiplied by $SU(2) \times U(1)$ 
which is equivalent to the above coset $\frac{SU(N+2)}{SU(N)}$.
In our previous work in \cite{AK1411}, 
the explicit $16$ lowest higher spin currents 
(which are multiple product of the above 
basic currents together with their derivatives)
were expressed in terms of the Wolf space coset fields explicitly.
These findings allow us to calculate 
the zero modes for the higher spin currents in terms of 
the generators of the $G=SU(N+2)$ because the zero modes of the spin-$1$
current satisfy the defining commutation  relations of the 
underlying finite dimensional Lie algebra $SU(N+2)$.
Furthermore, all the OPEs between the higher spin currents and the 
spin-$\frac{1}{2}$ current are determined explicitly by construction.    

The minimal representations of \cite{GG1305}
are given by two representations.
One of the minimal representation is
given by $(0;f)$ where  the nonnegative 
integer mode of the spin-$1$ current $V^a(z)$  in the $SU(N+2)$
acting on the state $|(0;f)>$ vanishes.
Under the decomposition of 
$SU(N+2)$ into the $SU(N) \times SU(2)$, the   
adjoint representation of $SU(N+2)$ breaks into as follows:
${\bf (N+2)^2-1} \rightarrow ({ \bf N^2-1}, {\bf 1}) 
\oplus ({\bf 1},{\bf 3} ) \oplus ({\bf 1},{\bf 1}) 
\oplus ({\bf N},{\bf 2} ) \oplus ({\bf \bar{N}},{\bf 2})$.
Among these representations, 
the fundamental representation for the $SU(N)$
is given by $({\bf N}, {\bf 2})$.
Therefore, the representation $(0;f)$ 
corresponds to these representations 
  $({\bf N}, {\bf 2})$.
Similarly, 
 the representation $(0;\bar{f})$ 
corresponds to these representations 
  $({\bf \bar{N}}, {\bf 2})$.
The corresponding states for the representation 
$(0;f)$ are given by 
$-\frac{1}{2}$ mode of the spin-$\frac{1}{2}$ current 
$Q^a(z)$ acting on the vacuum $|0>$ (corresponding to 
$(4.2)$ of \cite{GG1305}) where the index $a$
is restricted to the $2N$ coset index
\footnote{One can further classify two independent states
denoted by $|(0;f)>_{+}$ with $N$ coset indices and 
$|(0;f) >_{-}$ with other $N$ coset indices \cite{npb1989}
by emphasizing that $\pm$ refer to the doublet of $SU(2)$.}. 
As described in \cite{GG1305},
from the OPEs between 
the spin-$1$ currents $A^{+i}(z)$ 
and the spin-$\frac{1}{2}$ current $Q^a(w)$
with coset index, 
the states  $|(0;f)>$ are singlets with respect to  the spin-$1$ 
currents $A^{+i}(z)$. 
The eigenvalue for the zero mode in the (higher spin) currents (multiple product
of the above spin-$1$ and spin-$\frac{1}{2}$ currents)
acting on this state can be obtained 
from the highest pole of the OPE between    
the (higher spin) current and the spin-$\frac{1}{2}$ current
\footnote{
Furthermore, 
there exist the nontrivial states  for the 
negative half integer mode (as well as the $\frac{1}{2}$ mode) 
of the spin-$\frac{1}{2}$ current acting on this state 
$|(0;f)>$ because the action of those negative mode of spin-$\frac{1}{2}$ 
current
on the vacuum $|0>$ is nonzero.
Note that the action of  $\frac{1}{2}$ mode for the spin-$\frac{1}{2}$
current on the state $|(0;f)>$
can be written in terms of the anticommutator of these modes 
acting on the vacuum $|0>$ which is nonzero \cite{BBSS1,BBSS2,BS}.
The positive 
half integer modes of the spin-$\frac{1}{2}$ current ($\frac{3}{2}, 
\frac{5}{2}, \cdots$ modes)
acting on the state $|(0;f)>$ vanish.}.

Other of the minimal representation is given by
$(f;0)$ where 
the positive half integer mode of the spin-$\frac{1}{2}$ current $Q^a(z)$  
in the $SU(N+2)$
acting on the state $|(f;0)>$ vanishes.
The positive integer modes of the spin-$1$ current 
should annihilate this state.
They are  the 
singlets with respect to the $SU(N)$ in the $SU(N+2)$ representation 
based on the fundamental representation.
That is,
the fundamental representation $({\bf N+2})$ of $SU(N+2)$ transforms
as a singlet $ ({\bf 1},{\bf 2})_{-N}$ 
with respect to $SU(N)$ 
under the following branching 
$({\bf N+2})  \rightarrow 
({\bf N}, {\bf 1})_2 \oplus ({\bf 1},{\bf 2})_{-N}$
with respect to $SU(N) \times SU(2) \times U(1)$.
The indices $2$ and $-N$  
denote the $U(1)$ charge which will be described later
in (\ref{uf00f})
\footnote{In this case also, the states are further classified as
$|(f;0)>_{+}$ and $|(f;0)>_{-}$ with explicit $SU(2)$ indices.
For the antifundamental representation of $SU(N+2)$, one has
the following branching rule $({\bf \overline{N+2}})  \rightarrow 
({\bf \bar{N}}, {\bf 1})_{-2} \oplus ({\bf 1},{\bf 2})_{N}$
with respect to $SU(N) \times SU(2) \times U(1)$.}.  
On the other hand, $({\bf N}, {\bf 1})_2$
refers to the fundamental representation with respect to 
$SU(N)$ and describes the light state $|(f;f)>$. 
For the state  $|(f;0)>$, the $SU(N+2)$ generator
$T_{a^{\ast}}$ corresponds to the zero mode of 
the spin-$1$ current $V^a(z)$ because the 
the zero mode of the spin-$1$ current satisfies the commutation relation
of the underlying finite dimensional Lie algebra $SU(N+2)$. 
Then the nontrivial contributions to the zero mode (of (higher spin) 
currents) eigenvalue equation
associated with the state  $|(f;0)>$
come from the multiple product of the spin-$1$ current $V^a(z)$ 
in the (higher spin) currents \footnote{When the spin-$\frac{1}{2}$
current is present in the (higher spin) current $\pa^i Q J^{(s-i-\frac{1}{2})}$
where $J^{(s-i-\frac{1}{2})}$ stands for composite field between the spin-$1$ 
currents with spin 
$(s-i-\frac{1}{2})$, 
then the 
zero mode contains $\sum_{p=-i+\frac{1}{2}}^{-\frac{1}{2}} J_{-p}^{(s-i-\frac{1}{2})} 
(\pa^i Q)_p |(f;0) >$. 
The group indices are ignored.
From the mode expansion $Q^a(z) =\sum_{m =-\infty}^{\infty} 
\frac{Q^a_m}{z^{m+\frac{1}{2}}}$, the 
$p$ mode of $(\pa^i Q)$ can be written as 
$(p+\frac{1}{2})(p+\frac{3}{2}) \cdots (p+\frac{(2i-1)}{2}) Q_p$
up to an overall factor.
Therefore, one realizes that for each $p$ value in the summation,
the coefficient in $Q_p$ vanishes and there is no contribution in the 
eigenvalue equation. There are also  the terms containing 
$(\pa^i Q)_p J_{-p}^{(s-i-\frac{1}{2})} | (f;0)>$ with negative 
$p$ which do not produce any contribution.  }. 
After substituting the $SU(N+2)$ generator $T_{a^{\ast}}$ into the 
zero mode of spin-$1$ current $V_0^a$ in the multiple product 
of the (higher spin) currents, one obtains the $(N+2) \times (N+2)$
matrix acting on the state  $|(f;0)>$.
Then the last $2 \times 2$ subdiagonal matrix is associated with 
the above $SU(2) \times U(1)$ group. The eigenvalue 
can be read off from the last each diagonal matrix element
in this $2 \times 2$ matrix.   
Furthermore, the first  $N \times N$ subdiagonal matrix
provides the corresponding eigenvalues (for the higher spin currents) 
for the light state $|(f;f)>$ as mentioned before.

It is known under the large level limit 
that the perturbative Vasiliev  theory is 
a subsector of the tensionless string theory \cite{GG1406}.
The corresponding CFT 
is based on the small ${\cal N}=4$ linear superconformal 
algebra.  For finite level, 
the large ${\cal N}=4$ linear superconformal algebra 
plays an important role.
Then it is natural to consider the extension 
of large ${\cal N}=4$ linear superconformal algebra.
The coset realization for the large ${\cal N}=4$
linear superconformal algebra 
has been done by Saulina \cite{Saulina}.
See also \cite{ST}.
Compared to the spin-$\frac{3}{2}$ currents 
in the large ${\cal N}=4$ nonlinear superconformal 
algebra,  
the extra terms (which are cubic in the spin-$\frac{1}{2}$ currents) 
in the spin-$\frac{3}{2}$ currents 
of large ${\cal N}=4$ linear superconformal algebra
occur.
Moreover, the contractions between  the 
basic currents living in the coset $\frac{SU(N+2)}{SU(N)}$
contain  the extra coset indices
corresponding to the above $SU(2) \times U(1)$.
One can repeat the procedures described in \cite{AK1411}
and would like to construct the higher spin currents 
in an extension of the large ${\cal N}=4$ linear superconformal algebra.
In \cite{Ahn1311}, the explicit coset realization for $N=3$
in the large ${\cal N}=4$ linear superconformal algebra
has been found. 

In section $2$,
the large ${\cal N}=4$ nonlinear superconformal algebra and its 
extension are reviewed \cite{AK1411,Ahn1311,Ahn1408} 
and the $SU(N)$ subgroup appears in the 
first $N\times N$ matrix inside of $(N+2) \times (N+2)$ matrix.

In section $3$,
based on the section $2$, the eigenvalue equations of spin-$2$ 
stress energy tensor for the above 
two minimal states. 
Next, the eigenvalue equations of higher spin currents with 
spins-$1, 2$ and $3$ 
for the above 
two minimal states are presented.
The corresponding three-point functions are described.

In section $4$,
by summarizing the previous work by Saulina on the large ${\cal N}=4$
linear superconformal algebra, its extension is described. 

In section $5$, based on the section $4$, 
 the eigenvalue equations of spin-$2$ 
stress energy tensor (which is different from the one in section $2$) 
for the above 
two minimal states. 
Next, the eigenvalue equations of higher spin currents with 
spins-$1, 2$ and $3$ 
for the above 
two minimal states are given.
The corresponding three-point functions are described.

In section $6$,
the summary of this paper is given and the future directions 
are described briefly.

In Appendices $A$-$F$, some details in sections $2,3,4,5$ 
are presented. 

The Thielemans package \cite{Thielemans} is used in this paper.


\section{ The large $\mathcal N = 4$ nonlinear superconformal algebra and its
extension in 
the Wolf space coset: review}

In order to compare the previous results given in \cite{GG1305,GP1403}
with our findings explicitly, 
one should associate the subgroup of the group in the coset theory
with the first $N \times N$ matrix inside of $(N+2) \times (N+2)$ 
matrix.   
Then the corresponding coset indices occur 
in the remaining row and column elements 
except the last $2 \times 2$ 
matrix elements.

\subsection{ The $\mathcal N =1$ Kac-Moody current algebra in component 
approach }

The generators of $G=SU(N+2)$ in the Wolf space coset 
are given in Appendix $A$.
They satisfy the commutation relation
$\left[ T_a, T_b \right] = f_{a b}^{\;\;\;\; c} T_c $ where 
the indices run over $a, b, \cdots = 1, 2, \cdots,  \frac{(N+2)^2-1}{2},
1^{\ast}, 2^{\ast}, 
\cdots, (\frac{(N+2)^2-1}{2})^{\ast}$. 
The normalization is as follows:
$
g_{ab} = \frac{1}{2 c_G} f_{ac}^{\,\,\,\,\,\, d} f_{bd}^{\,\,\,\,\,\, c}
$
where $c_G$ is the dual Coxeter number of the group $G$.
The operator product expansions between the spin-$1$ and the spin-$\frac{1}{2}$
currents  are described as  \cite{KT1985}
\footnote{ In the work of \cite{GP1403}, the 
different normalization for the  spin-$\frac{1}{2}$ currents is used.
The right hand side of their OPE between $
\psi^{i,\alpha}(z)$ and $ \bar{\psi}^{j,\beta}(w)$
has the first order pole with weight $1$.  
Then our $\frac{1}{\sqrt{k+N+2}} Q^a(z)$ corresponds to their 
spin-$\frac{1}{2}$ currents. }
\bea
V^a(z) \, V^b(w) & = & \frac{1}{(z-w)^2} \, k \, g^{ab}
-\frac{1}{(z-w)} \, f^{ab}_{\,\,\,\,\,\,c} \, V^c(w) 
+\cdots,
\nonu \\
Q^a(z) \, Q^b(w) & = & -\frac{1}{(z-w)} \, (k+N+2) \, g^{ab} + \cdots,
\nonu \\
V^a(z) \, Q^b(w) & = & + \cdots.
\label{opevq}
\eea
Here $k$ is the level and a positive integer.
Note that there is no singular term in the OPE between 
the spin-$1$ current $V^a(z)$ and the spin-$\frac{1}{2}$ current 
$Q^b(w)$.

\subsection{ The large $\mathcal N = 4$ nonlinear superconformal algebra  in 
the Wolf space coset}

The Wolf space coset is given by 
\bea
\mbox{Wolf}= \frac{G}{H} = 
\frac{SU(N+2)}{SU(N) \times SU(2) \times U(1)}.
\label{coset}
\eea
The group indices are divided into 
\bea 
G \quad \mbox{indices} &:& a, b, c, \cdots = 1, 2, \cdots,
\frac{(N+2)^2-1}{2},  1^{\ast}, 2^{\ast}, \cdots,
(\frac{(N+2)^2-1}{2})^{\ast},
\nonu \\
\frac{G}{H} \quad \mbox{indices} &:& \bar{a},\bar{b},\bar{c},\cdots
=1, 2, \cdots, 2N, \cdots, 1^{\ast}, 2^{\ast}, \cdots, 2N^{\ast}.
\label{abnotation}
\eea
For given $(N+2) \times (N+2)$ matrix, 
one can associate the above $4N$ coset indices
as follows \cite{GG1305}:
\bea
\left(\begin{array}{rrrrr|rr}
 &&&&&{\ast} & {\ast}\\
 &&&&&{\ast} & {\ast}\\
 &&&&&\vdots & \vdots \\
 &&&&&{\ast} & {\ast}\\
 &&&&&{\ast} & {\ast}\\ \hline
 {\ast} & {\ast} & \cdots & {\ast} & {\ast}&& \\ 
 {\ast} & {\ast} & \cdots & {\ast} & {\ast}&&  \\ 
\end{array}\right)_{(N+2) \times (N+2)}.
\label{4nmatrix}
\eea
%
Because the different embedding is used in this paper, 
the different matrix representations of $h^i_{\bar{a} \bar{b}}$ and 
$d^0_{\bar{a} \bar{b}}$ will be given. 
However the large ${\cal N}=4$ nonlinear superconformal 
algebra remains unchanged although the $11$ currents look differently 
as one expects. 
The relevant structure corresponding to $SU(N+2)$ appears in Appendix $A$
where the adjoint index $a$ is written in terms of complex basis.

Then the 
$11$ currents of large $\mathcal N = 4$ nonlinear superconformal 
algebra in terms of 
 $\mathcal N =1$ Kac-Moody currents $V^a(z)$ and $Q^{\bar{b}}(z)$ 
together with the three almost complex structures
 $h^i_{\bar{a} \bar{b}} (i=1,2,3)$ are obtained. 
Furthermore, 
the three almost complex structures $(h^1,h^2,h^3)$ are 
 antisymmetric rank-two tensors and 
satisfy the algebra of imaginary quaternions \cite{Saulina}
\bea
h^{i}_{\bar{a} \bar{c} } \, h^{j \bar{c} }_{\,\,\,\,\,\, \bar{b} }
&=& 
\ep^{ijk} \, h^{k}_{\bar{a} \bar{b} } -  \delta^{ ij } \, g_{\bar{a} \bar{b}}. 
\label{hidentity}
\eea
By collecting the $11$ currents in \cite{AK1411} together,  the explicit 
$11$ currents of large $\mathcal N = 4$ nonlinear superconformal algebra 
with (\ref{abnotation}) are given by
\footnote{One has the following relations between the 
spin-$\frac{3}{2}$ currents with double index notation
and those with a single index notation
\label{doubleindex}
\bea
G_{11} (z) & = & \frac{1}{\sqrt{2}} (G^1- i G^2) (z) , \qquad
G_{12} (z) = -\frac{1}{\sqrt{2}} (G^3- i G^0) (z),
\nonu \\
G_{22} (z) &  = & \frac{1}{\sqrt{2}} (G^1+ i G^2)(z), \qquad
G_{21} (z) = -\frac{1}{\sqrt{2}} (G^3+ i G^0) (z).
\nonu
\eea
} 
\bea
G^{0}(z) &  = &   \frac{i}{(k+N+2)}  \, Q_{\bar{a}} \, V^{\bar{a}}(z),
\qquad
G^{i}(z)  =  \frac{i}{(k+N+2)} 
\, h^{i}_{\bar{a} \bar{b}} \, Q^{\bar{a}} \, V^{\bar{b}}(z),
\nonu \\
A^{+i}(z) &  = & 
-\frac{1}{4N} \, f^{\bar{a} \bar{b}}_{\,\,\,\,\,\, c} \, h^i_{\bar{a} \bar{b}} \, V^c(z), 
\qquad
A^{-i}(z)  =  
-\frac{1}{4(k+N+2)} \, h^i_{\bar{a} \bar{b}} \, Q^{\bar{a}} \, Q^{\bar{b}}(z),
\nonu \\
T(z)  & = & 
\frac{1}{2(k+N+2)^2} \left[ (k+N+2) \, V_{\bar{a}} \, V^{\bar{a}} 
+k \, Q_{\bar{a}} \, \pa \, Q^{\bar{a}} 
+f_{\bar{a} \bar{b} c} \, Q^{\bar{a}} \, Q^{\bar{b}} \, V^c  \right] (z)
\nonu \\
&&- \frac{1}{(k+N+2)} \sum_{i=1}^3 \left( A^{+i}+A^{-i}  \right)^2 (z),
\label{11currents}
\eea
where $i=1,2,3$. The $G^{\mu}(z)$ currents are four supersymmetry currents,
$A^{\pm i}(z)$ are six spin-$1$ generators of $SU(2)_{k} \times SU(2)_{N}$ and 
$T(z)$ is the spin-$2$ stress energy tensor. 
We will use these explicit results written in terms of the 
Wolf space coset fields in this paper all the times.

Finally the three almost complex structures (satisfying (\ref{hidentity})) 
using $4N \times 4N $ matrices
are given by
\bea
h^1_{\bar{a} \bar{b}} = 
\left(
\begin{array}{cccc}
0 & 0  & 0 & -i \\
0 & 0 & -i & 0 \\
0 & i & 0 & 0 \\
i & 0 & 0 & 0 \\
\end{array}
\right), \quad
h^2_{\bar{a} \bar{b}} = 
\left(
\begin{array}{cccc}
0 & 0  & 0 & 1 \\
0 & 0 & -1 & 0 \\
0 & 1 & 0 & 0 \\
-1 & 0 & 0 & 0 \\
\end{array}
\right), \qquad
h^{3}_{\bar{a} \bar{b}}
\equiv h^{1}_{\bar{a} \bar{c}}  \, h^{2 \bar{c} }_{\,\,\,\,\,\, \bar{b}}
\label{himatrix},
\eea
where each entry in (\ref{himatrix}) is  $N \times N$ matrix.
One sees that these complex structures occur in the above $11$ currents.
The defining OPE 
equations for the $11$ currents are given in Appendix $B$
for convenience. Note that the corresponding relations in different 
embedding appear in Appendix $B$ of \cite{AK1411}.

\subsection{ The higher spin currents in  the Wolf space coset (\ref{coset})}

The $16$ lowest higher spin currents 
have the following four ${\cal N}=2$ multiplets with spin contents
\bea
\left(1, \frac{3}{2}, \frac{3}{2}, 2 \right)
& : & (T^{(1)}, T_{+}^{(\frac{3}{2})}, T_{-}^{(\frac{3}{2})}, T^{(2)}), 
\nonu \\
 \left(\frac{3}{2}, 2, 2, \frac{5}{2} \right) & : & 
(U^{(\frac{3}{2})}, U_{+}^{(2)}, U_{-}^{(2)}, U^{(\frac{5}{2})} ), \nonu \\
\left(\frac{3}{2}, 2, 2, \frac{5}{2} \right) & : & 
(V^{(\frac{3}{2})}, V^{(2)}_{+}, V^{(2)}_{-}, V^{(\frac{5}{2})}),  \nonu \\
\left(2, \frac{5}{2}, \frac{5}{2}, 3 \right) & : &
 (W^{(2)}, W_{+}^{(\frac{5}{2})}, W_{-}^{(\frac{5}{2})}, W^{(3)}).
\label{lowesthigher}
\eea
The corresponding $16$ higher spin currents 
in different basis will appear in section $4$.
The higher spin-$1$ current which will be important in the linear 
version also
is
\footnote{
There is also normalization with overall sign where one has the following 
OPE
\bea
T^{(1)} (z) T^{(1)} (w)= \frac{1}{(z-w)^2} \left[ \frac{2 N k }{N+k+2}\right]+\cdots.
\nonu
\eea
\label{foot}
One can  introduce the $U(1)$ current $U(z)$ described in 
\cite{GP1403} from this higher spin-$1$ current.
See also $(4.20)$ and $(4.38)$ of \cite{GP1403}.
Then one can easily see that the corresponding their $U(z)$ is given by
\bea
U (z) &=&
(k+N+2)\left[-\frac{1}{2(k+N+2)} \, d^0_{\bar{a} \bar{b}} \,
f^{\bar{a} \bar{b}}_{\,\,\,\,\,\, c}  V^c  (z) \right]
- \frac{(N+2)(k+N+2)}{k} \left[ \frac{k}{2(k+N+2)^2} \, d^0_{\bar{a} \bar{b}} \, Q^{\bar{a}} \, Q^{\bar{b}} (z) \right]
\nonu \\
&=&
-\frac{1}{2} \, d^0_{\bar{a} \bar{b}} \,
f^{\bar{a} \bar{b}}_{\,\,\,\,\,\, c}  V^c  (z)
- \frac{(N+2)}{2(k+N+2)} \, 
d^0_{\bar{a} \bar{b}} \, Q^{\bar{a}} \, Q^{\bar{b}} (z).
\nonu
\eea
The OPEs between $U(z)$ and  the 
coset components of $\mathcal N = 1$ Kac-Moody currents 
$Q^{a}(w)$ and $V^{b}(w)$ satisfy the following 
OPEs
\bea
U(z) \, 
\left(
\begin{array}{c}
Q^{\bar{A}} \nonu \\
 Q^{\bar{A}^{\ast}} 
\end{array}
\right)(w) & = &  \mp \frac{1}{(z-w)} (N+2) 
\left(
\begin{array}{c}
Q^{\bar{A}} \nonu \\
 Q^{\bar{A}^{\ast}} 
\end{array}
\right)(w)+\cdots, \nonu \\
U(z) \, 
\left(
\begin{array}{c}
V^{\bar{A}} \nonu \\
 V^{\bar{A}^{\ast}}
\end{array}\right)
(w) & = &  \mp \frac{1}{(z-w)} (N+2) 
\left(
\begin{array}{c}
V^{\bar{A}} \nonu \\
 V^{\bar{A}^{\ast}}
\end{array}\right)
(w)+\cdots.
\nonu 
\eea
Note that the OPEs between 
$U(z)$ and the $11$ currents in (\ref{11currents})
are regular.
One can read off the corresponding $U(1)$ charges in the right hand side.
Then the $T^{(1)}$ charges for the coset fields  
are given by 
$ \pm \frac{k}{(k+N+2)}$ and $\mp \frac{(N+2)}{(k+N+2)}$ respectively.
}
\bea
T^{(1)} (z) &=&
-\frac{1}{2(k+N+2)} \, d^0_{\bar{a} \bar{b}} \,
f^{\bar{a} \bar{b}}_{\,\,\,\,\,\, c}  V^c  (z)
+ \frac{k}{2(k+N+2)^2} \, d^0_{\bar{a} \bar{b}} \, Q^{\bar{a}} \, Q^{\bar{b}} (z),
\label{finalspinone}
\eea
where 
the rank-two tensor 
$d^0_{\bar{a} \bar{b}}$ is antisymmetric and satisfies following properties
\bea
d^{0}_{\bar{a} \bar{c}} \, d^{0 \bar{c}}_{\,\,\,\,\,\, \bar{b}} 
=
g_{\bar{a} \bar{b}}, \quad \quad 
d^{0}_{\bar{a} \bar{b}} \, f^{\bar{b}}_{\,\,\,\, \bar{c} d} =
d^{0}_{\bar{c} \bar{b}} \, f^{\bar{b}}_{\,\,\,\, \bar{a} d}. 
\label{d0property}
\eea
The tensorial structure in (\ref{finalspinone})
is the same as the one in \cite{AK1411}.
Furthermore,
the $4N \times 4N $ matrix representation of $d^0_{\bar{a} \bar{b}}$ 
satisfying (\ref{d0property}) is 
\bea
d^0_{\bar{a} \bar{b}}  = 
\left(
\begin{array}{cccc}
0 & 0 & -1 & 0 \\
0 & 0 & 0 & -1 \\
1 & 0 & 0 & 0  \\
0 & 1 & 0 & 0 \\
\end{array}
\right).
\label{dzero}
\eea
Each entry is a $N \times N$ matrix as before.
Again the corresponding $d$ tensor in \cite{AK1411}
was appeared in Appendix $B$ of \cite{AK1411}.

Let us define the four higher spin-$\frac{3}{2}$ currents  $G'^{\mu} (z)$
from the first order pole of the following OPE
\bea
G^{\mu} (z) \, T^{(1)} (w) 
&=&
\frac{1}{(z-w)} \, G'^{\mu} (w) + \cdots.
\label{Gprime}
\eea
Then  the first order pole in (\ref{Gprime})
provides
\bea
G'^{\mu}(z) &=&
\frac{i}{(k+N+2)} \, 
d^{\mu}_{\bar{a} \bar{b}} \, Q^{\bar{a}} \, V^{\bar{b}} (z),
\label{gprimemu}
\eea
where  $d^{\mu}_{\bar{a} \bar{b}} \equiv d^{0 \bar{c} }_{ \bar{a} } \, 
h^{\mu }_{\bar{c} \bar{b}}$ and $h^0_{\bar{a} \bar{b}} \equiv g_{\bar{a} \bar{b}}$
with (\ref{dzero}).
These four independent higher spin-$\frac{3}{2}$ currents 
also appear in the linear version in section $4$.

Then the remaining $15$ currents of (\ref{lowesthigher}) can be written
 in terms of  $\mathcal N =1$ Kac-Moody currents $V^a(z), Q^{\bar{b}}(z)$, 
the three almost complex structures
 $h^i_{\bar{a} \bar{b}} $, antisymmetric rank-two tensor 
 $d^{0}_{\bar{a} \bar{b}}$ and symmetric rank-two tensors 
 $d^{i}_{\bar{a} \bar{b}}( \equiv d^{0 \bar{c} }_{ \bar{a} } \, 
h^{i }_{\bar{c} \bar{b}})$ as in \cite{AK1411}.

\section{ Three-point functions in an extension of 
large ${\cal N}=4$ nonlinear superconformal algebra}

This section describes the three-point functions 
with scalars for the currents of  
spins $s=1,2$ and the higher spin currents of spins $s=1,2,3$ 
explained in previous section.  
The large $N$ limit is defined by \cite{GG1305}
\bea
N,k \rightarrow \infty, \quad \lambda 
\equiv \frac{N+1}{N+k+2} \quad \mbox{fixed}.
\label{limit}
\eea

As described in the introduction, 
there are two simplest states  $|(f;0)>$ and $|(0;f)>$ in 
\cite{GG1305}.
These two representations (and their conjugate representations) 
play an important role 
in $\mathcal N = 4$ holographic duality which is conjectured 
in the large $N$ 't Hooft limit (\ref{limit}).

\subsection{ Eigenvalue equations for spin-$2$ current in an 
$\mathcal N = 4$ nonlinear superconformal algebra}


Let us focus on the eigenvalue equations for the stress energy tensor 
(\ref{11currents}) acting on the above two states.

\subsubsection{ Eigenvalue equation for spin-$2$ current acting on the 
state $|(f;0)>$}

As described in the introduction,
the terms containing the fermionic spin-$\frac{1}{2}$ currents
$Q^{a }(z)$ do not contribute to the eigenvalue equation 
when we calculate the zero mode eigenvalues for the 
bosonic spin-$s$ current 
$J^{(s)}(z)$  acting on the state  $|(f;0)>$.
The zero mode of the spin-$1$ current satisfies the commutation relation 
of the underlying finite dimensional Lie algebra $SU(N+2)$.
For the state $|(f;0)>$, the generator $T_{a^{\ast}}$ corresponds to 
the zero mode $V_0^a$ as follows (See also \cite{GH1101}):
\bea
 V^a_0 |(f;0)>  =  T_{a^{\ast}} |(f;0)>. 
\label{generator}
\eea
Then the eigenvalues are encoded in the  last 
$2 \times 2$ diagonal matrix.
Note that the nonvanishing components of the 
metric are given by $g_{aa^\ast}=1$ in Appendix (\ref{metricappendix}).  

For example, 
we can calculate the conformal dimension of $|(f;0)>$ when $N=3$.
The explicit form for the stress energy tensor is given by
(\ref{11currents}). 
The only $Q^a(z)$ independent terms are given by 
the first term and the $A^{+i} A^{+i}(z)$ dependent term because
the other terms contain $Q^a(z)$ explicitly.
Then the eigenvalue equation for the zero mode of
the spin-$2$ current acting on the state $|(f;0)>$ leads to
\bea
T_0 |(f;0)>  &\sim& 
\left[ \frac{1}{2(k+5)}   \, V_{\bar{a}} \, V^{\bar{a}}  
- \frac{1}{(k+5)} \sum_{i=1}^3  A^{+i} A^{+i} \right]_0   |(f;0)>
\nonu \\
&=&
\left[ 
\frac{1}{2(k+5)} \left( \sum_{a=1}^{6 } V^{a} V^{a^\ast} +
\sum_{a=1}^{6 } V^{a^\ast} V^{a} \right) \right]_0 |(f;0)>
+ \frac{1}{(k+5)} \left[ -\sum_{i=1}^3  A^{+i} A^{+i} \right]_0   |(f;0)>
\nonu \\
&=&
\left[ 
\frac{1}{2(k+5)} \left(  \sum_{a=1}^{6 } T_{a^\ast} T_{a} 
+\sum_{a=1}^{6 } T_{a} T_{a^\ast} \right) \right] |(f;0)>
+ \frac{1}{(k+5)} l^{+} (l^{+}+1)  |(f;0)>
\nonu \\
&=&
\frac{1}{2(k+5)} \left(
\begin{array}{ccc|cc}
 2 & 0 & 0 & 0 & 0 \\
 0 & 2 & 0 & 0 & 0 \\
 0 & 0 & 2 & 0 & 0 \\
\hline 
0 & 0 & 0 & 3 & 0 \\
 0 & 0 & 0 & 0 & 3 \\
\end{array}
\right) |(f;0)> + \frac{1}{(k+5)} \frac{3}{4}  |(f;0)>
= \frac{9}{4 (k+5)}  |(f;0)>,
\label{hfzero}
\eea
where $\sim$ in the first line of (\ref{hfzero}) 
means that we ignore the terms including $Q^a(z)$.
In the second line, the summation over the coset indices 
$\bar{a} = 1, 2, \cdots, 6, 1^{\ast}, 2^{\ast}, \cdots, 6^{\ast}$
is taken explicitly.
In the third line, 
the corresponding $SU(5)$ generators using the condition (\ref{generator})
are replaced and  moreover the eigenvalue equation for the zero mode of the 
quadratic 
spin-$1$ currents is used where $l^{+}$ is the spin of the affine $SU(2)$ 
algebra.   
In the fourth line, the $SU(5)$ matrix product is done 
where the first six  terms contribute to the first diagonal elements, $2$
and the last six terms contribute to the last diagonal elements, $3$.
Because the lower $2 \times 2$ diagonal matrix corresponding to 
the zero mode of quadratic spin-$1$ currents has 
two eigenvalues $\frac{3}{4}$, the spin $l^{+}$ can be identified with 
$\frac{1}{2}$
\footnote{
\label{footfoot}
The highest weight states of  the large $\mathcal N = 4$  (non)linear 
superconformal 
algebra can be characterized by
the conformal dimension $h$ and  two (iso)spins $l^{\pm}$ of 
$SU(2)\times SU(2)$ \cite{npb1989}
\bea
\left[ -\sum_{i=1}^3  A^{+i} A^{+i} \right]_0   |\mbox{hws}> 
=l^{+}(l^{+}+1) \, |\mbox{hws}>, 
\qquad
\left[ -\sum_{i=1}^3  A^{-i} A^{-i} \right]_0   |\mbox{hws}> 
=l^{-}(l^{-}+1) \, |\mbox{hws}>. 
\label{quadcasimir}
\eea
For example, in $G=SU(5)$, the expressions (\ref{11currents})
imply that 
  \bea
 \left[ -\sum_{i=1}^3  A^{+i} A^{+i} \right]_0 |(f;\star)> =
\left( \begin{array}{ccc|cc}
 0 & 0 & 0 & 0 & 0 \\
 0 & 0 & 0 & 0 & 0 \\
 0 & 0 & 0 & 0 & 0 \\
\hline
 0 & 0 & 0 & \frac{3}{4} & 0 \\
 0 & 0 & 0 & 0 & \frac{3}{4} \\
\end{array} \right)  |(f;\star)>, \quad
 \left[ -\sum_{i=1}^3  A^{-i} A^{-i} \right](z) \, 
Q^{\bar{A}^{\ast}}(w)|_{\frac{1}{(z-w)^2}}
 =\frac{3}{4} Q^{\bar{A}^{\ast}}(w),
 \nonu
\eea
where the representation 
$\star=0$ (trivial representation) or $f$ (fundamental representation) of
$SU(5)$.
Note that although all of these expressions are written in terms of 
the fields without boldface (in the nonlinear version), 
it is also true that 
all of these can be replaced by the fields with boldface 
(in the linear version).  
We can see $l^{+}(f;0)=\frac{1}{2}$ (from the two eigenvalues $\frac{3}{4}$),  
$l^{+}(f;f)=0$ (from the three 
eigenvalues $0$) and $l^{-}(0;f)=\frac{1}{2}$ (from 
the coefficient of the second order pole $\frac{3}{4}$).
Then the state $|(f;0)>$  has $l^{+}=\frac{1}{2}$, $l^{-}=0$, the state
$|(0;f)>$ has $l^{+}=0$, $l^{-}=\frac{1}{2}$
and the state $|(f;f)>$  has $l^{\pm}=0$. 
The eigenvalues for $l^{-}$ will be explained in next subsection.
Note the $(-1)$ sign in the left hand side of 
(\ref{quadcasimir}) comes from the anti-hermitian property
 \cite{npb1989,STVplb}.}.   

Now we want to obtain the general $N$ dependence for the above eigenvalue 
equation.
From the similar calculations for $N=5, 7, 9$
where all the (higher spin) currents are known explicitly, 
we can find $N$-dependence of the corresponding  $(N+2) \times (N+2)$ matrix.
In the generalization of third line of (\ref{hfzero}),
the first $2N$  terms contribute to the first $N$ diagonal elements, $2$
and the last $2N$ terms contribute to the last two diagonal elements, $N$.
That is, first $N$ diagonal elements are given by $2$ and 
the remaining last two diagonal elements are 
given by $N$. 
The eigenvalues $\frac{3}{4}$ obtained in the footnote
\ref{footfoot}
hold for any generic $N$. The total contribution is given by 
$(\frac{N}{2} +\frac{3}{4})$ multiplied by 
an obvious overall factor $\frac{1}{(k+N+2)}$. 
Therefore, the eigenvalue equation (for generic $N$) 
for the zero mode of spin-$2$ current 
which provides the conformal dimension for the 
state $|(f;0)>$ is given by 
\bea
T_0 |(f;0)>  &=& \left[ \frac{(2N+3)}{4(k+N+2)}  \right] |(f;0)>,
\label{t0f0}
\eea
where  the eigenvalue is the same value as 
$h(f;0)$ given in \cite{GG1305}.
One can also check that 
this leads to the following 
reduced eigenvalue 
equation  $ T_0 |(f;0)>  =  \frac{\la}{2}  |(f;0)> $
under the large $N$ 't Hooft limit (\ref{limit}). 

\subsubsection{ Eigenvalue equation for spin-$2$ current acting on the 
state $|(0;f)>$}

Let us move on the next simplest representation.
When we calculate the eigenvalue equations for the state  $|(0;f)>$,  
because the state is given by  \cite{GG1305,npb1989}
\bea
|(0;f)>=\frac{1}{\sqrt{k+N+2}} \, Q_{-\frac{1}{2}}^{\bar{A}^{\ast}}|0>,
\qquad \bar{A}^{\ast} = 1^{\ast}, 2^{\ast}, \cdots, 2N^{\ast}, 
\label{state0f}
\eea
the OPEs between the (higher spin) currents $J^{(s)}(z)$ 
and $Q^{\bar{A}^{\ast}}(w)$ are needed.
We need only the coefficient of highest order pole $\frac{1}{(z-w)^s}$
in the OPEs.
The lower singular terms do not contribute to the zero mode eigenvalue
equations.
Let us denote the highest order pole as follows \cite{AKP,MZ1211}:
\bea 
J^{(s)}(z) \,
Q^{\bar{A}^{\ast}}(w) |_\frac{1}{(z-w)^s}=  j(s) \, Q^{\bar{A}^{\ast}}(w),
\label{opejq}
\eea
where $j(s)$ stands for the corresponding coefficient of highest order pole.
One can write down $
J_0^{(s)} |(0;f)>$ as $ \frac{1}{\sqrt{k+N+2}} 
[ J_0^{(s)},  Q_{-\frac{1}{2}}^{\bar{A}^{\ast}}] |0>$ and this commutator acting on the 
vacuum can be written in terms of 
$ \frac{1}{\sqrt{k+N+2}} j(s) \, Q_{-\frac{1}{2}}^{\bar{A}^{\ast}} |0>$.
Then one obtains the following eigenvalue equation for the zero mode 
of the spin-$s$ current together with (\ref{state0f}) and (\ref{opejq})
\bea 
J_0^{(s)} |(0;f)> = j(s) |(0;f)>,
\label{j0f}
\eea
where the explicit relation between the 
current and its mode is given by
$J^{(s)}(z)=\sum_{n=-\infty}^\infty \frac{J_n^{(s)}}{z^{n+s}}$.
Therefore, in order to determine the above 
eigenvalue $j(s)$, one should calculate the explicit OPEs between 
the corresponding (higher spin) currents and the spin-$\frac{1}{2}$
current and read off the highest order pole.

Let us consider the eigenvalue equation for the spin-$2$
current acting on the above state.
Since the OPE between 
the spin-$1$ current $A^{+i}(z)$ and the spin-$\frac{1}{2}$ current 
$Q^{\bar{A^\ast}}(w)$ is regular, 
the terms containing $A^{+i}(z)$ in $T(z)$ of (\ref{11currents}) 
do not contribute to the highest order 
pole.
Furthermore the terms containing the spin-$1$ current 
$V^a(z)$ do not  contribute to the highest order pole.
Therefore, the relevant terms in $T(z)$
are given by purely the spin-$\frac{1}{2}$ current 
dependent terms (by ignoring the spin-$1$ current dependent terms
completely).
Then the conformal dimension of $|(0;f)>$ is
\bea
T_0 |(0;f)>  &\sim&
\left[ \frac{k}{2(k+N+2)^2} Q_{\bar{a}} \pa Q^{\bar{a}} 
-\frac{1}{(k+N+2)} \sum_{i=1}^3 A^{-i} A^{-i} \right]_0  |(0;f)> 
\nonu \\
&=& \left[ \frac{k}{2(k+N+2)^2}  Q_{\bar{a}} \pa Q^{\bar{a}} \right]_0 |(0;f)> 
+\frac{1}{(k+N+2)}  
\left[ -\sum_{i=1}^3 A^{-i} A^{-i} \right]_0  |(0;f)> 
\nonu \\
&=&  \frac{k}{2(k+N+2)}   |(0;f)> 
+\frac{1}{(k+N+2)}  l^{-}( l^{-}+1)  |(0;f)> 
\nonu \\ 
&=&  \left[ \frac{(2k+3)}{4(N+k+2)} \right] |(0;f)>.
\label{hzerof}
\eea
In the first line of (\ref{hzerof}), 
the spin-$1$ current dependent terms are 
ignored.
In the second line, the zero modes for each term are taken.
In the third line, 
we have used the fact that 
the eigenvalue equation 
$\left[ Q_{\bar{a}} \pa Q^{\bar{a}} \right]_0 |(0;f)>=(k+N+2)|(0;f)>$
(see (\ref{j0f})) can be obtained 
because  the highest order pole gives the corresponding eigenvalue
$ Q_{\bar{a}} \pa Q^{\bar{a}}(z) \, Q^{\bar{A}^{\ast}}(w) |_{\frac{1}{(z-w)^2}}
=(k+N+2) Q^{\bar{A}^{\ast}}(w)$ (see (\ref{opejq})) which can be checked from the 
defining relation in (\ref{opevq}).
Furthermore, the characteristic eigenvalue equation 
for the affine $SU(2)$ algebra described in the footnote \ref{footfoot} 
is used.
The total contribution is given by $(\frac{k}{2} + \frac{3}{4})$ 
multiplied by the overall factor $\frac{1}{(N+k+2)}$.
The above eigenvalue  is exactly the same as the  $h(0;f)$ described 
in \cite{GG1305}.
Under the large $N$ 't Hooft limit (\ref{limit}), the eigenvalue 
equation implies that $T_0 |(0;f)> = \frac{1}{2} (1-\la) |(0;f)>$.

There exists $N \leftrightarrow k$ and $0 \leftrightarrow f$ 
symmetry 
between the above two eigenvalue equations (\ref{t0f0}) and 
(\ref{hzerof}). In the classical symmetry (in the large 
$N$ 't Hooft limit), this is equivalent to the exchange 
of $ \la \leftrightarrow (1-\la)$ and $0\leftrightarrow  f$. 
The $U$-charge corresponding to 
twice of $\hat{u}$ charge in \cite{GG1305} 
can be determined as follows
:
\bea
U_0 |(f;0)> = -N   |(f;0)>, \qquad  U_0 |(0;f)>=( N+2 )  |(0;f)>,
\label{uf00f}
\eea
where we use the same normalization as in \cite{GP1403}.
The second relation can be seen from the explicit OPE result in footnote
\ref{foot} where the eigenvalue $(N+2)$ appears in the OPE 
between $U(z)$ and $Q^{\bar{A}\ast}(w)$.
The detailed description for these eigenvalue equations 
will be given in next subsection
\footnote{
Explicitly for $N=3$, one has
$
U(z)=\frac{1}{4}
(3 \sqrt{10}+5 i \sqrt{6})  V^{12}(z)+\frac{1}{4} (3
   \sqrt{10}-5 i \sqrt{6})  V^{12^*}(z)+
   \frac{5}{(5+k)} \sum_{a=1}^{6} Q^a Q^{a^*}(z)$.
As done in (\ref{hfzero}), 
one can construct
the following eigenvalue equation for  $G=SU(5)$,
\bea
U_0 |(f;\star)> = \left(
\begin{array}{ccc|cc}
 2 & 0 & 0 & 0 & 0 \\
 0 & 2 & 0 & 0 & 0 \\
 0 & 0 & 2 & 0 & 0 \\
\hline 
0 & 0 & 0 & -3 & 0 \\
 0 & 0 & 0 & 0 & -3 \\
\end{array}  \right) |(f;\star)>,
\nonu
\eea
where the representation $\star$ is trivial  representation $0$
or fundamental representation $f$ in $SU(5)$.
The eigenvalues $-3$ living in the lower $2 \times 2$ diagonal matrix 
can be generalized to $-N$ according to the next subsection.
By reading off the the first three eigenvalues 
$2$ which is valid for generic $N$, the  $U$-charge for the state 
$|(f;f)>$ is given by $2$ \cite{GG1305}. 
Furthermore  one can determine  the 
conformal dimension for the `light' state  $ |(f;f)>$ by using (\ref{hfzero})
and its generalization.
  Since $l^{+}(f;f)=0$ (from the analysis in the footnote \ref{footfoot}), 
the eigenvalue equation  is 
 $
 T_0 |(f;f)>  = \frac{1}{2(k+N+2)} \times 2  |(f;f)>
 = \frac{1}{(k+N+2)}   |(f;f)>$ observed in \cite{GG1305}. 
Under the large $N$ 't Hooft limit,
one has 
 $T_0 |(f;f)>  = \frac{\la}{(N+1)} |(f;f)>$ and this vanishes.
}.

\subsection{ Eigenvalue equation for higher spin currents of spins 
$1,2$ and $3$ }

Now one can consider the eigenvalue equations for the higher spin currents
by following the descriptions in previous subsection.

\subsubsection{ Eigenvalue equation for higher 
spin-$1$ current acting on the 
states $|(f;0)>$ and $|(0;f)>$}

From the explicit expression in (\ref{finalspinone}),
one starts with the first term which does not contain the 
spin-$\frac{1}{2}$ current, applies to the condition (\ref{generator})
and read off the last two diagonal matrix elements for $N=3$.
It turns out that
\bea
T^{(1)}_0 |(f;0)> &\sim&
\left[ -\frac{1}{2(k+5)} \, d^0_{\bar{a} \bar{b}} \,
f^{\bar{a} \bar{b}}_{\,\,\,\,\,\, c}  V^c   \right]_0 |(f;0)> 
\nonu \\
&=&
\left[ -\frac{1}{2(k+5)} \left(
-\frac{3 \sqrt{10}+5 i \sqrt{6}}{2 } V^{12}-\frac{3 \sqrt{10}-5 i
   \sqrt{6}}{2} V^{12^\ast} \right) \right]_0 |(f;0)>
   \nonu \\
&=&
\left[ -\frac{1}{2(k+5)} \left(
-\frac{3 \sqrt{10}+5 i \sqrt{6}}{2 } T_{12^\ast}-\frac{3 \sqrt{10}-5 i
   \sqrt{6}}{2} T_{12} \right) \right] |(f;0)>  
\label{newspinonef0}
\\
   &=&
    -\frac{1}{2(k+5)}
\left(
\begin{array}{ccc|cc}
 -4 & 0 & 0 & 0 & 0 \\
 0 & -4 & 0 & 0 & 0 \\
 0 & 0 & -4 & 0 & 0 \\
\hline 
0 & 0 & 0 & 6 & 0 \\
 0 & 0 & 0 & 0 & 6 \\
\end{array} 
\right)  |(f;0)>  =  -\frac{3}{(k+5)} |(f;0)>.
\nonu
\eea
In the first line of (\ref{newspinonef0}), one ignores the spin-$\frac{1}{2}$
dependent part. 
In the second line, one can substitute the $d$ tensor 
and $f$ structure constant for $N=3$ from the section $2$.
From the third line to the fourth line, the explicit generators 
are substituted from Appendix $A$.
Miraculously, the final $5 \times 5$ matrix is simple diagonal matrix.  
Finally, the two eigenvalues $6$ from the last $2 \times 2$ diagonal matrix 
are taken \footnote{The matrix $T_{12}$ is given by
the nonzero $11, 22, 33$ diagonal elements  
$ \frac{1}{60} \left(3 \sqrt{10}+5 i \sqrt{6}\right)$,
$44$ diagonal element
$
\frac{1}{20} \left(\sqrt{10}-5 i \sqrt{6}\right)$,
and $55$ diagonal element  $-\frac{\sqrt{10}}{5}$ in Appendix $A$.}.
How does one obtain the general $N$ behavior for the above eigenvalue 
equation?
One can find $N$-dependence of above $5 \times 5$ matrix by similar calculation 
for next several values for $N$:$N=5,7,9$. The $(N+2) \times (N+2)$ 
matrix in $G=SU(N+2)$ is given by  
$\mbox{diag}(-4, \cdots, -4, 2N, 2N)$ with the obvious overall factor
$-\frac{1}{2(N+k+2)}$. The $N$ dependence appears in the last 
$2 \times 2$ diagonal matrix.
Therefore the eigenvalue equation for generic $N$ can be summarized by
\footnote{
One can read off the eigenvalue equation for the `light' state  
$|(f;f)> $ by taking the eigenvalues $-4$ living in the first 
$N \times N$  diagonal matrix as follows:
\bea
T^{(1)}_0 |(f;f)> &=&  -\frac{1}{2(k+N+2)} \times (-4)|(f;f)>
= \left[\frac{2}{(k+N+2)} \right] |(f;f)>.
\nonu
\eea
}
\bea
T^{(1)}_0 |(f;0)> &=&  - \left[ \frac{N}{(k+N+2)} \right] |(f;0)>.
\label{t1f0}
\eea

The zero mode eigenvalue equation for the state $|(0;f)>$
can be obtained from the explicit OPE between the second term in 
(\ref{finalspinone}) and the spin-$\frac{1}{2}$ current $Q^{\bar{A}^\ast}(w)$ 
and one obtains
\bea
T^{(1)}_0 |(0;f)> &=&  - \left[ \frac{k}{(k+N+2)} \right] |(0;f)>.
\label{t10f}
\eea
It is obvious that 
there exists a $ N \leftrightarrow k$ and $0 \leftrightarrow f$ 
symmetry between the two eigenvalue equations in (\ref{t1f0}) and (\ref{t10f})
\footnote{
One can find the above two eigenvalue equations  by using $U$-charge
introduced in (\ref{uf00f}) indirectly. 
This is because 
the first term (the second term)  of $T^{(1)}(z)$ in  
(\ref{finalspinone})
contributes to the zero mode eigenvalue equation for the state 
$|(f;0)>$ ($|(0;f)>$) respectively. See the footnote \ref{foot}.
By taking  the numerical factors correctly, we find
that one of them is given by $
T^{(1)}_0 |(f;0)>  =   
\frac{1}{(k+N+2)} U_0 |(f;0)> =  -\frac{N}{(k+N+2)} |(f;0)>$
where the spin-$\frac{1}{2}$ dependent term is ignored and the other 
is given by
$T^{(1)}_0 |(0;f)>  =    -\frac{k}{(k+N+2)(N+2)} U_0 |(0;f)> = 
-\frac{k}{(k+N+2)} |(0;f)>$ where 
the spin-$1$ dependent term is ignored.}.
Under the large $N$ 't Hooft limit (\ref{limit}), one obtains the 
following eigenvalue equations 
\bea
T^{(1)}_0 |(f;0)> & = & -  \lambda |(f;0)>,
\nonu \\ 
T^{(1)}_0 |(0;f)> & = &   -(1-\lambda) |(0;f)>.
\label{spinoneexpexp}
\eea
Compared to the previous $U$ charge in (\ref{uf00f}), 
the above higher spin-$1$ current preserves the $N \leftrightarrow k$
symmetry.
If one replaces the fundamental representation
$f$ with the antifundamental representation $\bar{f}$ in (\ref{t1f0})
and (\ref{t10f}),
then the extra minus signs appear in the right hand side respectively.
Note that 
the corresponding equations are 
given by $V^a_0 |(\bar{f};0)> = - (T_{a^{*}})^T |(\bar{f};0)> $ 
and $|(0;\bar{f})>=\frac{1}{\sqrt{k+N+2}} \, Q_{-\frac{1}{2}}^{\bar{A}}|0>$
associated with (\ref{generator}) and (\ref{state0f}).

\subsubsection{ Eigenvalue equation for higher 
spin-$2$ currents acting on the 
states $|(f;0)>$ and $|(0;f)>$}

In order to understand the eigenvalue equations for the 
higher spin-$2$ currents, 
one should classify the $|(f;0)>$ state into the following two types
of column vectors
\bea
|(f;0)>_{+}= (0, \cdots, 0, 1, 0)^T, \qquad
|(f;0)>_{-}= (0, \cdots, 0, 0, 1)^T.
\label{eigenvector}
\eea
They are ${\bf 2}$ under the $SU(2)$ and 
transform as singlets under the $SU(N)$ characterized by the first $N$ zeros
in (\ref{eigenvector}).
They have nontrivial $U(1)$ charges (\ref{uf00f}).
 
On the other hand, 
the $|(0;f)>$ states are  expressed by the following forms
\bea
|(0;f)>_{+} & : & \frac{1}{\sqrt{k+N+2}} Q_{-\frac{1}{2}}^{1^{\ast}}|0>,
\qquad 
\cdots, 
\qquad \frac{1}{\sqrt{k+N+2}} Q_{-\frac{1}{2}}^{N^{\ast}}|0>,
\nonu \\
|(0;f)>_{-} & : & \frac{1}{\sqrt{k+N+2}} Q_{-\frac{1}{2}}^{(N+1)^{\ast}}|0>,
\qquad
\cdots, 
\qquad
\frac{1}{\sqrt{k+N+2}} Q_{-\frac{1}{2}}^{(2N)^{\ast}}|0>.
\label{plusminus}
\eea
They are fundamental representation ${\bf N}$ under the $SU(N)$ 
respectively and transform as a doublet 
under the $SU(2)$
\footnote{
Let us comment on the 
nontrivial action of the spin-$1$ currents into the above two states.
The result is as follows: 
\bea
A^{+ \pm}_0 |(f;0)>_{\mp}&=&
-i |(f;0)>_{\pm}, 
\nonu \\
A^{- \pm}_0 |(0;f)>_{\mp}&=&
i |(0;f)>_{\pm}. 
\nonu
\eea
One has nontrivial eigenvalues as one applies to one more zero modes 
at each expression.
The corresponding three point functions can be described.}.

It turns out that the 
eigenvalue equations for the higher spin-$2$ (primary) 
current $T^{(2)}(z)$ acting on (\ref{plusminus}) and (\ref{eigenvector}) 
are summarized by
\bea
T^{(2)}_0 |(f;0)>_{+} & = &  
\left[ \frac{N(2Nk+2N-k)}{2(N+k+2)(2Nk+N+k)} \right] |(f;0)>_{+},
\nonu \\
T^{(2)}_0 |(f;0)>_{-} & = & - \left[ 
\frac{Nk(2N+3)}{2(N+k+2)(2Nk+N+k)} \right] |(f;0)>_{-},
\nonu \\
T^{(2)}_0 |(0;f)>_{+} & = &  \left[
\frac{k(2kN+2k-N)}{2(N+k+2)(2Nk+N+k)}\right] |(0;f)>_{+},
\nonu \\
T^{(2)}_0 |(0;f)>_{-} & = & - \left[ 
\frac{Nk(2k+3)}{2(N+k+2)(2Nk+N+k)} \right] |(0;f)>_{-}.
\label{t2eigen}
\eea
These are the first examples where 
all the eigenvalues appear differently
\footnote{More precisely, 
the previous eigenvalue equations (\ref{t0f0}), (\ref{hzerof}), 
(\ref{t1f0}) and (\ref{t10f})
can be rewritten as $T_0 |(f;0)>_{\pm}  =   
\frac{(2N+3)}{4(N+k+2)} |(f;0)>_{\pm}$, $T_0 |(0;f)>_{\pm}  =   
\frac{(2k+3)}{4(N+k+2)} |(0;f)>_{\pm}$, $T^{(1)}_0 |(f;0)>_{\pm}  =   
-\frac{N}{(N+k+2)} |(f;0)>_{\pm}$, and $ T^{(1)}_0 |(0;f)>_{\pm}  =   
-\frac{k}{(N+k+2)} |(0;f)>_{\pm}$ respectively.}. 
There exists $N \leftrightarrow k$ and $0 \leftrightarrow f$ 
symmetry in (\ref{t2eigen}) such that
\bea
\left[ T^{(2)}_0 |(f;0)>_{\pm} \right]_{N \leftrightarrow k, 0 \leftrightarrow f}
=T^{(2)}_0 |(0;f)>_{\pm}.
\label{t2f0}
\eea
The states  $|(f;0)>_{\pm}$ are changed into 
$ |(0;f)>_{\pm} $ and vice versa.
Furthermore, if one replaces the fundamental representation
$f$ with the antifundamental representation $\bar{f}$ in (\ref{t2eigen}),
then the right hand sides remain unchanged.

The higher spin-$2$ primary current $\tilde{W}^{(2)}(z)$ 
was defined in \cite{AK1411} as follows:
\bea
\tilde{W}^{(2)}(z) \equiv \left[ W^{(2)} - \frac{2kN}{(k+N+2kN)} T \right](z).
\label{pripri}
\eea
Then the corresponding eigenvalue equations 
from (\ref{eigenvector}), (\ref{plusminus}) and (\ref{pripri})
can be obtained as follows:
\bea
\tilde{W}^{(2)}_0 |(f;0)>_{+} & = &
- \left[ \frac{Nk(2N+3)}{2(N+k+2)(2Nk+N+k)} \right] |(f;0)>_{+},
\nonu \\
\tilde{W}^{(2)}_0 |(f;0)>_{-} & = &
\left[ \frac{N(2Nk+2N-k)}{2(N+k+2)(2Nk+N+k)} \right] |(f;0)>_{-},
\nonu \\
\tilde{W}^{(2)}_0 |(0;f)>_{+} & = &
\left[ \frac{k(2Nk+2k-N)}{2(N+k+2)(2Nk+N+k)}\right] |(0;f)>_{+},
\nonu \\
\tilde{W}^{(2)}_0 |(0;f)>_{-} & = &
-\left[ 
\frac{Nk(2k+3)}{2(N+k+2)(2Nk+N+k)}\right] |(0;f)>_{-}.
\label{w2eigen}
\eea
There exists a $N \leftrightarrow k$ symmetry in (\ref{w2eigen}) such that
\bea
\left[ \tilde{W}^{(2)}_0 |(f;0)>_{\pm} \right]_{N \leftrightarrow k, 0 \leftrightarrow
f, + \leftrightarrow -}
=\tilde{W}^{(2)}_0 |(0;f)>_{\mp}.
\label{w2f0}
\eea
Note that the states  $|(f;0)>_{\pm}$ are changed into 
$ |(0;f)>_{\mp} $ and vice versa.
Furthermore, if one replaces the fundamental representation
$f$ with the antifundamental representation $\bar{f}$ in (\ref{w2eigen}),
then the right hand sides remain unchanged. 

One can easily see that 
there exist the precise relations between the above eigenvalue equations
as follows:
\bea
\left[ T^{(2)}_0 |(f;0)>_{\pm} \right]_{+ \leftrightarrow -} & = & \tilde{W}^{(2)}_0 |(f;0)>_{\mp}, 
\nonu \\
T^{(2)}_0 |(0;f)>_{\pm}  & = & \tilde{W}^{(2)}_0 |(0;f)>_{\pm}.
\label{somerel}
\eea
Furthermore, under the large $N$ 't Hooft limit (\ref{limit}), 
the above eigenvalue equations (\ref{t2eigen}) and 
(\ref{w2eigen}) (or (\ref{somerel}))
become
\bea
T^{(2)}_0 |(f;0)>_{\pm}  & = & \pm \frac{1}{2} \lambda |(f;0)>_{\pm},
\nonu \\
T^{(2)}_0 |(0;f)>_{\pm} & = & \pm  \frac{1}{2} (1-\lambda)  |(0;f)>_{\pm},
\nonu \\
\tilde{W}^{(2)}_0 |(f;0)>_{\pm} & = &
\mp \frac{1}{2} \lambda |(f;0)>_{\pm},
\nonu \\
\tilde{W}^{(2)}_0 |(0;f)>_{\pm} & = & \pm \frac{1}{2} (1-\lambda) |(0;f)>_{\pm}.
\label{express}
\eea
Up to the overall signs, 
these relations (\ref{express}) behave similarly as 
the ones in the spin-$2$ current described in the subsection $3.1$.


For the other spin-$2$ currents one obtains the following nonzero results
\footnote{More precisely, 
one has 
\bea
\left[ U^{(2)}_{+} \right]_0 \frac{1}{\sqrt{N+k+2}} Q^{a^*}_{-\frac{1}{2}} |0>
&=& \left[ \frac{k}{(N+k+2)}\right] 
\frac{1}{\sqrt{N+k+2}} Q^{(a+N)^*}_{-\frac{1}{2}} |0>, 
\nonu \\
\left[ V^{(2)}_{-} \right]_0 \frac{1}{\sqrt{N+k+2}} Q^{(a+N)^*}_{-\frac{1}{2}} |0>
&=&- \left[ \frac{k}{(N+k+2)} \right] 
\frac{1}{\sqrt{N+k+2}} Q^{a^*}_{-\frac{1}{2}} |0>, 
\nonu
\eea
where $a=1, 2, \cdots, N$.}
\bea
\left[ U^{(2)}_{+} \right]_0 |(0;f)>_{+}
&=& \left[ \frac{k}{(N+k+2)} \right] |(0;f)>_{-} \rightarrow 
 (1-\lambda) |(0;f)>_{-}, 
\nonu \\
\left[ U^{(2)}_{-} \right]_0 |(f;0)>_{+}
&=& - \left[ \frac{N}{(N+k+2)} \right] |(f;0)>_{-} \rightarrow
 -\lambda |(f;0)>_{-},
\nonu \\
\left[ V^{(2)}_{+} \right]_0 |(f;0)>_{-}
&=& \left[ \frac{N}{(N+k+2)} \right] |(f;0)>_{+} \rightarrow
 \lambda |(f;0)>_{+},
\nonu \\
\left[ V^{(2)}_{-} \right]_0 |(0;f)>_{-}
&=&- \left[ \frac{k}{(N+k+2)} \right] |(0;f)>_{+} \rightarrow
- (1-\lambda)  |(0;f)>_{+}. 
\label{nontrivial}
\eea
In (\ref{nontrivial}), 
one sees that as one multiplies the zero modes 
further, there exist nonzero eigenvalue equations which will be discussed 
in next subsection (the footnote \ref{otherthree})
\footnote{
One can easily check the eigenvalue equations for the higher spin-$2$ 
current $W^{(2)}$ which is a quasiprimary field
as follows:
\bea
W^{(2)}_0 |(f;0)>_{+} & = &  0,
\qquad
W^{(2)}_0 |(f;0)>_{-}  =   \left[ \frac{N}{(N+k+2)} \right] |(f;0)>_{-},
\nonu \\
W^{(2)}_0 |(0;f)>_{+} & = &  \left[ \frac{k}{(N+k+2)} \right] |(0;f)>_{+},
\qquad
W^{(2)}_0 |(0;f)>_{-}  =   0.
\nonu
\eea
Furthermore, its large $N$ 't Hooft limit (\ref{limit})
can be summarized by
\bea
W^{(2)}_0 |(f;0)>_{+} & = & 0, \qquad
W^{(2)}_0 |(f;0)>_{-}  =   \lambda |(f;0)>_{-}, 
\nonu \\
W^{(2)}_0 |(0;f)>_{+} & = & (1-\lambda) |(0;f)>_{+}, 
\qquad
W^{(2)}_0 |(0;f)>_{-}  =   0.
\nonu
\eea
}. 

\subsubsection{ Eigenvalue equation for the higher 
spin-$3$ current acting on the 
states $|(f;0)>$ and $|(0;f)>$}
 
It turns out that the eigenvalue equations  of the zero mode 
of the higher spin-$3$ current  $W^{(3)}(z)$
are described as
\footnote{
For $N=3$, the corresponding three diagonal matrix elements are 
given by 
$-\frac{4 (k-3) (23 k+31)}{3 (k+5)^2 (23 k+19)}$,
the $44$ element,
$\frac{(k+7) (46 k+11)}{(k+5)^2 (23 k+19)}$ and 
the $55$ element
$\frac{(46 k^2+403 k+55)}{(k+5)^2 (23 k+19)}$.
The eigenvalue equation for the zero mode of 
higher spin-$3$ current acting on the  `light' state
can be summarized as follows and its large $N$ 't Hooft limit 
is also given
\bea
W^{(3)}_0 |(f;f)> & = & -\left[ \frac{4(k-N)(5N+16+(6N+5)k)}
{3(N+k+2)^2 ( 5N+4 +(6N+5)k) }\right]  |(f;f)>
\rightarrow 
\frac{4 \lambda (2 \lambda-1)}{3 N} |(f;f)> 
\rightarrow 0.
\nonu
\eea}
\bea
W^{(3)}_0 |(f;0)>_{+} & = & 
\left[ \frac{N(2N+k+1)(12Nk+10k+4N-1)}{3(N+k+2)^2 ( 6Nk+5N+5k+4 ) }
\right] |(f;0)>_{+},
\nonu \\
W^{(3)}_0 |(f;0)>_{-} & = & 
\left[ 
\frac{N ( 24N^2 k+ 12N k^2+ 8N^2+10 k^2+48N k-6N+43 k+1 )}{3(N+k+2)^2 ( 6Nk+5N+5k+4 ) } \right] |(f;0)>_{-},
\nonu \\
W^{(3)}_0 |(0;f)>_{+} & = & 
-\left[ 
\frac{k ( 24k^2 N+ 12k N^2  +8k^2+10N^2+48 k N-6k +43N+1 )}{3(N+k+2)^2 ( 6Nk+5N+5k+4 ) } \right] |(0;f)>_{+},
\nonu \\
W^{(3)}_0 |(0;f)>_{-} & = & -
\left[ \frac{k  (2k+N+1)(12Nk+10N+4k-1) }
{3(N+k+2)^2 ( 6Nk+5N+5k+4 ) } \right] |(0;f)>_{-}.
\label{fourw}
\eea
There is a $N \leftrightarrow k$ symmetry in (\ref{fourw}) such that
\bea
\left[ W^{(3)}_0 |(f;0)>_{\pm}  \right]_{N \leftrightarrow k, 0 \leftrightarrow f, + 
\leftrightarrow -}
=-W^{(3)}_0 |(0;f)>_{\mp},
\label{spin3Nksymm}
\eea
which looks similar to (\ref{w2f0}) up to the sign.
Furthermore, if one replaces the fundamental representation
$f$ with the antifundamental representation $\bar{f}$ in (\ref{fourw}),
then the extra minus sign appears in the right hand side respectively. 

Under the large $N$ 't Hooft limit (\ref{limit}), one has 
\bea
W^{(3)}_0 |(f;0)> & = & \frac{2}{3} \lambda (1+\lambda)|(f;0)>,
\nonu \\
W^{(3)}_0 |(0;f)> & = & -\frac{2}{3} (1-\lambda) (2-\lambda)|(0;f)>.
\label{nonw3large}
\eea
The following relation holds
$\left[  W^{(3)}_0 |(f;0)> \right]_{\lambda \rightarrow (1- \lambda), 
0 \leftrightarrow f}
= - W^{(3)}_0 |(0;f)>$.
 This is the expected symmetry because of the relation (\ref{spin3Nksymm}).
 The $N \leftrightarrow k$ symmetry corresponds to 
 $\lambda \leftrightarrow (1- \lambda)$ symmetry in the large $N$ limit.

Let us describe the three point functions
\footnote{
We assume the following normalizations
\bea
_{\pm}<(\bar{f};0)|(f;0)>_{\pm}= _{\pm}<(0;\bar{f})|(0;f)>_{\pm}, 
\qquad
_{\pm}<(\bar{f};0)|(f;0)>_{\mp}= _{\pm}<(0;\bar{f})|(0;f)>_{\mp}=0.
\nonu
\eea
}.
From the diagonal modular invariant with 
pairing up identical representations on the left (holomorphic)
and the right (antiholomorphic) 
sectors \cite{CY1106}, 
one of the primaries is given by 
$(f;0) \otimes (f;0)$
which is denoted by 
${\cal O}_{+}$ and the other 
is given by $(0;f) \otimes (0;f)$
which is denoted by ${\cal O}_{-}$.
Then the three point functions with these two scalars  
are obtained and their ratios can be written as 
\bea
\frac{<\overline{{\cal O}}_{+ } 
{\cal O}_{+ }T^{(1)}>}{< \overline{{\cal O}}_{- } 
{\cal O}_{- } T^{(1)}>}
 &= & 
 \left[\frac{\lambda}{1-\lambda} \right],
\nonu \\
\frac{<\overline{{\cal O}}_{+ } 
{\cal O}_{+ }T^{(2)}>}{< \overline{{\cal O}}_{- } 
{\cal O}_{- } T^{(2)}>}
 &= & 
 \left[\frac{\lambda}{1-\lambda} \right],
\nonu \\
 \frac{< \overline{{\cal O}}_{+ } 
{\cal O}_{+ } \tilde{W}^{(2)}>}{< \overline{{\cal O}}_{- } 
{\cal O}_{- }
\tilde{W}^{(2)}>}
& = &  - \left[ \frac{\lambda}{1-\lambda} \right],
\nonu \\
\frac{<\overline{{\cal O}}_{+ } 
{\cal O}_{+ } W^{(3)}> }{< \overline{{\cal O}}_{- } 
{\cal O}_{- }  W^{(3)}> }
&=&- \left[ \frac{\lambda(1+\lambda)}{(1-\lambda)(2-\lambda)}
\right].
\label{threethree}
\eea
Compared to the bosonic higher spin AdS/CFT  duality 
in the context of  $W_N$ minimal model,
the behavior of (\ref{threethree})
looks similar in the sense that 
the factor $\frac{\la}{(1-\la)}$ which is present in the 
ratios of three point function of higher spin-$2$ currents 
appears in the right hand side of the  
ratio of the three point functions of the higher spin-$3$ current.
Furthermore,
the factor  $\frac{(1+\la)}{(2-\la)}$ contribute to the 
final ratio of the three point functions of higher spin-$3$ current.
 Then one expects that 
the ratio of the three point functions for the higher spin-$4$ current 
can be 
described as   $\left[ \frac{\lambda(1+\lambda)(2+\la)}{(1-\lambda)
(2-\lambda)(3-\la)}
\right]$.
Even for the higher spin-$s$ current one can expect that 
the ratio of the three point functions for the higher spin current 
of spin-$s$ can be 
written as
$ \prod_{n=1}^{s-1} \frac{(n-1 +\la) }{(n-\la )}$ up to sign.
Note that the bosonic case has same formula except
the numerator of this expression contains 
$n$ rather than $(n-1)$ \cite{AK1308}.  
It would be interesting to study 
this general spin behavior in details
\footnote{
\label{otherthree}
One describes the following three point functions
from (\ref{nontrivial}) by specifying the primaries further
with $\pm$ indices showing the $SU(2)$ doublet 
\bea
<\overline{{\cal O}}_{-,- } 
{\cal O}_{-,+ } U^{(2)}_{+} > &=&  (1-\lambda) 
<\overline{{\cal O}}_{-,- } 
{\cal O}_{-,- } >,
\nonu \\
<\overline{{\cal O}}_{+,- } 
{\cal O}_{+,+ } U^{(2)}_{-} >& =&  -\lambda
<\overline{{\cal O}}_{+,- } 
{\cal O}_{+,- } >,
\nonu \\
<\overline{{\cal O}}_{+,+ } 
{\cal O}_{+,- } V^{(2)}_{+} >& =&  \lambda
<\overline{{\cal O}}_{+,+ } 
{\cal O}_{+,+ } >,
\nonu \\
<\overline{{\cal O}}_{-,+ } 
{\cal O}_{-,- } V^{(2)}_{-} >& =&  - (1-\lambda) 
<\overline{{\cal O}}_{-,+ } 
{\cal O}_{-,+ } >.
\nonu
\eea
}.

As in the footnote \ref{footfoot}, one can calculate 
the sum of the square of the triplet (corresponding the higher spin-$2$
currents) in each $SU(2)$ group.  
For the similar calculations in the nonlinear version
where the equation $(4.23)$ of \cite{BCG1404} is used,
one obtains
\bea
\left[ \sum_{i=1}^3 \tilde{V}_{1}^{(1) +i} \tilde{V}_{1}^{(1) +i}  \right]_0 |(f;0)>
& = & - \left[ \frac{12N (5N+4k+2)}{(N+k+2)^2} \right]|(f;0)>,
\nonu \\
\left[ \sum_{i=1}^3 \tilde{V}_{1}^{(1) +i} \tilde{V}_{1}^{(1) +i}  \right]_0 |(0;f)>
& = & - \left[ \frac{24 k}{(N+k+2)^2}\right] |(0;f)>,
\nonu \\
\left[ \sum_{i=1}^3 \tilde{V}_{1}^{(1) -i} \tilde{V}_{1}^{(1) -i}  \right]_0 |(f;0)>
& = & - \left[ \frac{24 N }{(N+k+2)^2} \right]|(f;0)>,
\nonu \\
\left[ \sum_{i=1}^3 \tilde{V}_{1}^{(1) -i} \tilde{V}_{1}^{(1) -i}  \right]_0 |(0;f)>
& = & - \left[ \frac{12k (5k+4N+2)}{(N+k+2)^2} \right] |(0;f)>.
\label{spin2rel} 
\eea
There exist the symmetries in $N \leftrightarrow k$ and 
$0 \leftrightarrow f$.
As the large $N$ 't Hooft limits in (\ref{spin2rel}) are taken,
they become 
$-12 \lambda (4+ \lambda)$, $-\frac{24 \lambda (1- \lambda)}{N}$, 
$ -\frac{24 \lambda^2 }{N}$ and 
$-12 (1-\lambda) (5- \lambda)$ respectively.

\subsection{ 
Various eigenvalue equations for other states }

Using the following identification of the spin-$\frac{1}{2}$ current 
in \cite{GG1305},
\bea
\psi^{1,1}(z) & \equiv & \frac{1}{\sqrt{k+N+2}}Q^{1^\ast}(z),  
\qquad
\cdots,
\qquad 
\psi^{N,1}(z) \equiv \frac{1}{\sqrt{k+N+2}} Q^{N^\ast}(z),
\label{psi} \\
\psi^{1,2}(z) & \equiv & \frac{1}{\sqrt{k+N+2}} Q^{(N+1)^\ast}(z),  
\qquad
\cdots,
\qquad 
\psi^{N,2}(z) \equiv \frac{1}{\sqrt{k+N+2}} Q^{(2N)^\ast}(z),
\nonu
\eea 
one  can construct other representations 
described in \cite{GG1305}.
They are symmetric combination or antisymmetric combination of the state
with (\ref{psi}) 
as follows:
\bea
 |(0;[2,0, \cdots, 0])>= \sum_{(ij)}  \psi^{i,\alpha}_{-\frac{1}{2}} 
 \psi^{j,\beta}_{-\frac{1}{2}} |0>, \quad
  |(0;[0,1, 0, \cdots, 0])>= \sum_{[ij]}  \psi^{i,\alpha}_{-\frac{1}{2}} 
 \psi^{j,\beta}_{-\frac{1}{2}} |0>.
\label{highereigen}
\eea
Let us calculate the conformal dimensions for these states
(\ref{highereigen}).
 Then one can find the conformal dimensions of above states by 
using the following results
 \bea
 T(z)  \sum_{(ij)}  \psi^{i,\alpha}  \psi^{j,\beta}(w) |_{\frac{1}{(z-w)^2}} & = &
\left[ \frac{k}{(N+k+2)}  \right] 
\sum_{(ij)}  \psi^{i,\alpha}  \psi^{j,\beta}(w), \nonu \\
  T(z)  \sum_{[ij]}  \psi^{i,\alpha}  \psi^{j,\beta}(w) |_{\frac{1}{(z-w)^2}}
& = & 
\left[ \frac{(k+2)}{(N+k+2)} \right] 
\sum_{[ij]}  \psi^{i,\alpha}  \psi^{j,\beta}(w).
\label{condim}
 \eea
That is the conformal dimension for the former 
is given by 
$h(0;[2,0, \cdots, 0])= \frac{k}{(N+k+2)}$ (\ref{condim}) 
and the one for the latter 
is given by $h(0;[0,1, 0, \cdots, 0])=\frac{(k+2)}{(N+k+2)} $ 
(\ref{condim}) which is the same as $(C.8)$ and $(C.9)$
of \cite{GG1305}
respectively.
 
Similarly, one can find the 
conformal dimension of $p$-fold antisymmetric
product
 \bea
|(0;[0^{p-1}, 1, 0, \cdots, 0])>
=\sum_{[i_1 i_2 \cdots i_p]}  \psi^{i_1,\alpha_1}_{-\frac{1}{2}} 
\psi^{i_2,\alpha_2}_{-\frac{1}{2}}  \cdots \psi^{i_p,\alpha_p}_{-\frac{1}{2}} |0>.
\label{state}
\eea
From the explicit result in the second order pole  $
 \{ T(z) \sum_{[i_1 i_2 \cdots i_p]}  \psi^{i_1,\alpha_1}
\psi^{i_2,\alpha_2} \cdots \psi^{i_p,\alpha_p} (w) \}_{-2}$, 
one has 
\bea
\left[ \frac{p(p+2+2k)}{4(k+N+2)} \right]
\sum_{[i_1 i_2 \cdots i_p]}  \psi^{i_1,\alpha_1}
\psi^{i_2,\alpha_2} \cdots \psi^{i_p,\alpha_p} (w),
\label{interinter}
\eea
and one can read off 
$h(0;[0^{p-1}, 1, 0, \cdots, 0])=  \frac{p(p+2+2k)}{4(k+N+2)}$
with (\ref{state}) and (\ref{interinter})
which is the same as the last equation of Appendix $C$ in \cite{GG1305}.
We have checked up to $6$-fold products in $G=SU(11)$
case (that is $N=9$)
\footnote{ As described in \cite{GG1305}, 
the `short' representations are given by the following conditions
\bea
\left[ G_{2 a} \right]_{-\frac{1}{2}} |(f;0)>_{+} 
=\left[ G_{1 a} \right]_{-\frac{1}{2}} |(f;0)>_{-}  =0,
\nonu \\
\left[ G_{ a 1} \right]_{-\frac{1}{2}} |(0;\bar{f})>_{+} 
= \left[ G_{ a 2} \right]_{-\frac{1}{2}} |(0; \bar{f})>_{-}  =0,
\nonu
\eea
where $a=1,2$. These correspond to $(2.39)$ of \cite{GG1305}. }.

There exists other higher representation.
Let us check the 
conformal dimension of 
$|(f;\bar{f})> \equiv \frac{1}{\sqrt{N+k+2}} Q^{\bar{A}}_{-\frac{1}{2}} |(f;0)>$
as in \cite{GG1305}.
For example, in the coset of $G=SU(3+2)$,
one can calculate 
the following eigenvalue equation
\bea
T_{0} |(f;\bar{f})> &=& \frac{1}{\sqrt{5+k}}
\left( \left[ T_{0}, Q^{\bar{A}}_{-\frac{1}{2}} \right]
+Q^{\bar{A}}_{-\frac{1}{2}} T_{0}  \right) |(f;0)>
\nonu \\
&=& \left[ h(0;\bar{f})+h(f;0) \right]  |(f;\bar{f})>
\nonu \\
&+& \frac{1}{\sqrt{5+k}} Q^{\bar{A}}_{-\frac{1}{2}}
\left( \frac{5i \sqrt{6}+3 \sqrt{10}}{24(5+k)} V^{12}_0
+\frac{-5i \sqrt{6}+3 \sqrt{10}}{24(5+k)} V^{12^*}_0 \right) |(f;0)>
\nonu \\
&=& \left[ h(0;\bar{f})+h(f;0) +
\left(
\begin{array}{ccc|cc}
 \frac{1}{3(5+k)} & 0 & 0 & 0 & 0 \\
 0 & \frac{1}{3(5+k)} & 0 & 0 & 0 \\
 0 & 0 & \frac{1}{3(5+k)} & 0 & 0 \\
\hline 
0 & 0 & 0 & -\frac{1}{2(5+k)} & 0 \\
 0 & 0 & 0 & 0 & -\frac{1}{2(5+k)} \\
\end{array}
\right)\right]  |(f;\bar{f})>
\nonu \\
&=& \left[ \frac{2k+3}{4(5+k)}+\frac{9}{4(5+k)} 
-\frac{1}{2(5+k)} \right]  |(f;\bar{f})>
=\frac{1}{2}   |(f;\bar{f})>,
\label{expp}
\eea
where $ h(0;\bar{f})=  h(0;f)$.
Note that 
when the operator 
$ \left[ T_{0}, Q^{\bar{A}}_{-\frac{1}{2}} \right]$ acts on the state $|(f;0)>$, 
the lower order pole (as well as the highest order pole) 
can contribute to the eigenvalue equation.
In the coset theory of $G=SU(N+2)$, the above matrix in (\ref{expp}) can be 
generalized to 
\bea
\frac{1}{(N+k+2)} \, \mbox{diag}(\frac{1}{N}, \cdots, \frac{1}{N}, 
-\frac{1}{2}, -\frac{1}{2}).
\label{genexpexp}
\eea
Then the conformal dimension for generic $N$ can be described as
\bea
T_{0} |(f;\bar{f})> &=&  
\left[ h(0;\bar{f})+h(f;0) -\frac{1}{2(N+k+2)}\right]  |(f;\bar{f})>
\label{half} \\
&=&  \left[ \frac{(2k+3)}{4(N+k+2)} +\frac{(2N+3)}{4(N+k+2)}
-\frac{1}{2(N+k+2)}\right]  |(f;\bar{f})>=\frac{1}{2}   |(f;\bar{f})>,
\nonu
\eea
where the result of (\ref{genexpexp}) is used and 
this  (\ref{half}) was observed in \cite{GG1305}.

Therefore, in this section, the ratios of the three point functions can be 
summarized by (\ref{threethree}). They  
can be obtained from the previous relations 
(\ref{spinoneexpexp}), 
(\ref{express}) and  (\ref{nonw3large}).

\section{ The large $\mathcal N = 4$ linear superconformal algebra and its 
extension in the coset theory}

For the $16$ currents of large ${\cal N}=4$ linear superconformal algebra,
the work of \cite{Saulina} leads to the complete expressions 
in terms of the coset fields.
In this section, some of the recapitulation of \cite{Saulina} in our 
notations are given 
and one would like to construct the $16$ higher spin currents.

\subsection{ The $16$ currents 
of $\mathcal N =4$ linear superconformal algebra using 
the Kac-Moody currents}

The $16$ currents and the $16$ higher spin currents 
are constructed 
in the following coset theory
\bea
\frac{G}{H} = 
\frac{SU(N+2)}{SU(N)}.
\label{coset1}
\eea
The following three indices are defined in the corresponding 
group $G$, subgroup $H$ and the coset $\frac{G}{H}$ respectively  
\bea 
G \quad \mbox{indices} &:& a,b,c,\cdots,
\nonu \\
H \quad \mbox{indices} &:& a',b',c',\cdots,
\nonu \\
\frac{G}{H} \quad \mbox{indices} &:& \tilde{a},\tilde{b},\tilde{c},
\cdots.
\label{abnot}
\eea
The number of coset indices is given by 
the difference between $(N+2)^2-1$ and $(N^2-1)$ and therefore 
 the dimension of the coset is given by  $(4N+4)$.
For given $(N+2) \times (N+2)$ matrix the $(4N+4)$ coset indices 
can be associated with the following locations with asterisk
\bea
\left(\begin{array}{rrrrr|rr}
 &&&&&{\ast} & {\ast}\\
 &&&&&{\ast} & {\ast}\\
 &&&&&\vdots & \vdots \\
 &&&&&{\ast} & {\ast}\\
 &&&&&{\ast} & {\ast}\\ \hline
 {\ast} & {\ast} & \cdots & {\ast} & {\ast}& {\ast} & {\ast} \\ 
 {\ast} & {\ast} & \cdots & {\ast} & {\ast} & {\ast} & {\ast} \\ 
\end{array}\right)_{(N+2) \times (N+2)}.
\label{linearmatrix}
\eea
Compared to the previous case in (\ref{4nmatrix}),
there are extra $2 \times 2 $ matrix corresponding
$SU(2) \times U(1)$. 
One  can further 
divide the linear coset indices (\ref{abnot}) as
$\tilde{a}=(\bar{a}, \hat{a})$ 
where the index $\hat{a}$ associates with the above 
$2 \times 2$ matrix and runs over $4$ values.
Of course, the remaining $\bar{a}$ index runs over $4N$ values
as in an extension of large ${\cal N}=4$ nonlinear 
superconformal algebra in the section $2$.   

Let us consider 
four spin-$\frac{3}{2}$ currents of large $\mathcal N =4$ linear 
superconformal algebra.
It is known that the spin-$\frac{3}{2}$ current corresponding to 
the $\mathcal N =1$ supersymmetry generator
consists of two parts. One of them contains the 
spin-$1$ current as well as the spin-$\frac{1}{2}$ current
and the other contains the cubic term in the spin-$\frac{1}{2}$ current. 
See also \cite{AF}.
By generalizing the two coefficient tensors to possess
the three additional supersymmetry indices, 
one can write down 
\bea
{\bf G}^{\mu}(z) = A(k,N) \left[ \, h^{\mu}_{\tilde{a} \tilde{b}} \,
Q^{\tilde{a}} \, V^{\tilde{b}}+B(k,N) S^{\mu}_{\tilde{a} \tilde{b} \tilde{c}} 
Q^{\tilde{a}} Q^{\tilde{b}} Q^{\tilde{c}} \right](z)
, \qquad (\mu=0,1,2,3),
\label{linGansatz}
\eea
where the relative coefficients 
$A(k,N)\equiv \frac{i}{(k+N+2)}$ and 
$B(k,N) \equiv -\frac{1}{6(k+N+2)}$ 
are taken from the ${\cal N}=1$ 
supersymmetry generator and moreover  
the two tensors are given by 
$h^0_{\tilde{a} \tilde{b}} \equiv g_{\tilde{a} \tilde{b}}$ and  
$S^{0}_{\tilde{a} \tilde{b} \tilde{c}} \equiv f_{\tilde{a} \tilde{b} \tilde{c}}$
for $\mu=0$ index.
The new objects 
$h^i_{\tilde{a} \tilde{b}}$ and $S^{i}_{\tilde{a} \tilde{b} \tilde{c}} $ 
for other three indices $i =1,2,3$ 
are undetermined numerical constants
\footnote{One uses the boldface notation for the $16$ currents plus
 $16$ higher spin currents
in the linear version.
For the previous $11$ currents and $16$ higher spin 
currents in the nonlinear version, 
the boldface notation is not used.}.

Then one obtains the explicit OPE between 
the spin-$\frac{3}{2}$ current (\ref{linGansatz}) and itself as follows:
\bea
{\bf G}^{\mu}(z) \, {\bf G}^{\nu}(w) &=&
\frac{1}{(z-w)^3} \, A^2 \, \left[ - k(k+N+2) \, h^{\mu}_{\tilde{a} \tilde{b}}
 \, h^{\nu \tilde{a} \tilde{b}} 
 + 6 B^2  (k+N+2)^3 S^{\mu}_{\tilde{a} \tilde{b} \tilde{c}}  
 S^{\nu \tilde{a} \tilde{b} \tilde{c}}  \right]
\nonu \\
&+& \frac{1}{(z-w)^2} \, A^2 \, \left[  (k+N+2) 
\, h^{\mu}_{\tilde{a} \tilde{b}} \, h^{\nu \tilde{a}}_{\,\,\,\,\,\, \tilde{d}}
\, f^{\tilde{b} \tilde{d}}_{\,\,\,\,\,\, e} \, V^e
+k \, h^{\mu}_{\tilde{a} \tilde{b}} \,
h^{\nu \tilde{b}}_{\tilde{c}} \, Q^{\tilde{a}} \, Q^{\tilde{c}} \right.
\nonu \\
&-& \left. 18 B^2 (k+N+2)^2 S^{\mu}_{\tilde{a} \tilde{b} \tilde{c}}
S^{\nu \tilde{a} \tilde{b}}_{\,\,\,\,\,\,\,\,\, \tilde{d}}  
\, Q^{\tilde{c}} Q^{\tilde{d}}\right] (w)
\nonu \\
&+& \frac{1}{(z-w)} \, A^2 \left[ - (k+N+2) \, h^{\mu}_{\tilde{a} \tilde{b}} 
\, h^{\nu \tilde{a}}_{\,\,\,\,\,\, \tilde{d}} \, V^{\tilde{b}} \, V^{\tilde{d}} 
+ k \, h^{\mu}_{\tilde{a} \tilde{b}} \, h^{\nu \tilde{b}}_{\tilde{c}} \, \pa 
\, Q^{\tilde{a}} \, Q^{\tilde{c}}  \right.
\nonu \\
&-&  h^{\mu}_{\tilde{a} \tilde{b}} \, h^{\nu}_{\tilde{c} \tilde{d}}
 f^{\tilde{b} \tilde{d}}_{\,\,\,\,\,\, e} \, Q^{\tilde{a}} \, Q^{\tilde{c}} \,
V^{e} 
-6B(k+N+2)  S^{(\mu}_{\tilde{a} \tilde{b} \tilde{c}} h^{\nu) \tilde{a}}_{\,\,\,\,\,\, \tilde{d}}
\, Q^{\tilde{b}} \, Q^{\tilde{c}} \,V^{\tilde{d}} 
\nonu \\
&-& 
9 B^2 (k+N+2)  S^{\mu}_{\tilde{a} \tilde{b} \tilde{c}}  
 S^{\nu \tilde{a} }_{\,\,\,\,\,\, \tilde{d} \tilde{e}} \, Q^{\tilde{b}} Q^{\tilde{c}}
 Q^{\tilde{d}} Q^{\tilde{e}}
\nonu \\
 &-&\left. 18 B^2 (k+N+2)^2  S^{\mu}_{\tilde{a} \tilde{b} \tilde{c}}  
 S^{\nu \tilde{a} \tilde{b} }_{\,\,\,\,\,\,\,\, \tilde{d} } \, \pa Q^{\tilde{c}} Q^{\tilde{d}}
 \right] (w) + \cdots.
\label{linGGope}
\eea
Compared to the similar calculation for the OPE between the 
spin-$\frac{3}{2}$ current and itself in the nonlinear version,  
the result in (\ref{linGGope}) contains 
the $S^{\mu}$ tensor dependent terms.

By using the defining $\mathcal N =4$ linear superconformal 
algebra (\ref{N4linearalg}) and 
above OPE (\ref{linGGope}), one identifies the 16 currents
in terms of spin-$1$ current and spin-$\frac{1}{2}$ current.
From the particular expression when $\mu=\nu=0$, one can 
read off the spin-$2$ current explicitly.
That is, the first order pole is given by 
 ${\bf G}^0 (z)\, {\bf G}^0 (w) |_{\frac{1}{(z-w)}} = 2 {\bf T}(w) $, 
and the corresponding expression is obtained from  the 
OPE (\ref{linGGope}).
Then one obtains
\bea
{\bf T}(z) &=& 
\frac{1}{2(k+N+2)^2} \left[ (k+N+2) \, V_{\tilde{a}} \, V^{\tilde{a}} 
+k \, Q_{\tilde{a}} \, \pa \, Q^{\tilde{a}} 
+f_{\tilde{a} \tilde{b} c'} \, Q^{\tilde{a}} \, Q^{\tilde{b}} \, V^{c'}  \right.
\nonu \\
&+& \left. 
\frac{1}{2} f_{\tilde{a} \tilde{b} \tilde{c}} f^{\tilde{a} \tilde{b}}_{\,\,\,\,\,\, \tilde{d}} 
\pa Q^{\tilde{c}}  Q^{\tilde{d}} 
+\frac{1}{4(k+N+2)} f_{\tilde{a} \tilde{b} \tilde{e}} f_{\tilde{c} \tilde{d}}^{\,\,\,\,\,\, \tilde{e}} 
Q^{\tilde{a}}  Q^{\tilde{b}}  Q^{\tilde{c}}  Q^{\tilde{d}} 
\right] (z).
\label{tlinear}
\eea
Compared to the corresponding spin-$2$ current in (\ref{11currents}),
the dummy variables are summed over the extra $4$ indices corresponding to 
the lower $2\times 2$ matrix in (\ref{linearmatrix}). 
 
From the $V^{\tilde{b}} \, V^{\tilde{d}}(w)$ term of 
the following relation 
${\bf G^{(\mu} }(z) \, {\bf G^{\nu)} }(w) |_{\frac{1}{(z-w)}} = 
2 \delta^{\mu \nu } {\bf T}(w) $, 
one has the identity
\bea
h^{\mu}_{\tilde{a} \tilde{b} } \, h^{\nu \tilde{a} }_{\,\,\,\,\,\, \tilde{d} }
+h^{\nu}_{\tilde{a} \tilde{b} } \, h^{\mu \tilde{a} }_{\,\,\,\,\,\, \tilde{d} }
&=& 
2 \, \delta^{\mu \nu} \, g_{\tilde{b} \tilde{d}}, \qquad (\mu,\nu=0,1,2,3),
\label{hcondition}
\eea
which corresponds to $(2.19)$ of \cite{Saulina}.
The left hand side of (\ref{hcondition}) comes from the above
OPE (\ref{linGGope}) and the right hand side 
comes from the (\ref{tlinear}).  
Thus $h^i_{\tilde{a} \tilde{b}}$ are almost complex structures.

From the $Q^{\tilde{a}} \, Q^{\tilde{b}} \, V^{\tilde{e}}$ term of 
${\bf G^{(\mu} } (z) \, {\bf G^{\nu)} } (w) |_{\frac{1}{(z-w)}} = 
2 \delta^{\mu \nu } {\bf T}(w)$
\footnote{ In the nonlinear version of \cite{AK1411}, 
there was no second identity of (\ref{tensorid}).
Because there exists a quadratic term for  the spin-$1$ current
 in the first order pole $\frac{1}{(z-w)}$ of 
Appendix (\ref{N4scalgebra}), 
the different identity was required.},
the following identities are satisfied
\bea
h^{(\mu}_{\tilde{a} \tilde{c}} h^{\nu)}_{\tilde{b} \tilde{d}}
f^{\tilde{c} \tilde{d}}_{\,\,\,\,\,\, \tilde{e}} 
&=&
h^{\tilde{d} \,\,\, (\mu }_{\,\,\,\, \tilde{e} }
S^{\nu)}_{\tilde{d} \tilde{a} \tilde{b}},
\nonu \\
h^{i}_{\tilde{a} \tilde{c}} f^{\tilde{c} }_{\,\,\,\, \tilde{b} e'}  &=&
h^{i}_{\tilde{b} \tilde{c}} f^{\tilde{c} }_{\,\,\,\, \tilde{a} e'},  
\label{tensorid}
\eea
which correspond to 
$(2.20)$ and $(2.21)$ of \cite{Saulina} respectively.
The second equation of (\ref{tensorid}) 
can be seen from the equation $(3.11)$ of \cite{GK}.
By using the three identities given in (\ref{hcondition}) and 
(\ref{tensorid}), the complete expression for the 
$S^{i}_{\tilde{a} \tilde{b} \tilde{c} }$ tensor is as follows: 
\bea
S^{i}_{\tilde{a} \tilde{b} \tilde{c} } = 
h^i_{\tilde{a} \tilde{d}} h^i_{\tilde{b} \tilde{e}} h^i_{\tilde{c} \tilde{f}}
f^{\tilde{d} \tilde{e} \tilde{f}}, \qquad (i=1,2,3),
\label{Sdef}
\eea
corresponding to $(2.27)$ of \cite{Saulina}.

Thus, the spin-$\frac{3}{2}$ currents are given by, together with 
the three almost complex structures and the structure constants
where the result in (\ref{Sdef}) is substituted,
\bea
{\bf G}^{0} (z) &=& \frac{i}{(k+N+2)} \left[ \, 
Q_{\tilde{a}} \, V^{\tilde{a}}- \frac{1}{6(k+N+2)} f_{\tilde{a} \tilde{b} \tilde{c}} 
Q^{\tilde{a}} Q^{\tilde{b}} Q^{\tilde{c}} \right](z)
,
\nonu \\
{\bf G}^{j}(z) &=& \frac{i}{(k+N+2)} \left[ \, h^{j}_{\tilde{a} \tilde{b}} \,
Q^{\tilde{a}} \, V^{\tilde{b}}- \frac{1}{6(k+N+2)} 
h^j_{\tilde{a} \tilde{d}} h^j_{\tilde{b} \tilde{e}} h^j_{\tilde{c} \tilde{f}}
f^{\tilde{d} \tilde{e} \tilde{f}} 
Q^{\tilde{a}} Q^{\tilde{b}} Q^{\tilde{c}} \right](z).
\label{linearspin3half}
\eea
Compared to the corresponding spin-$\frac{3}{2}$ currents in 
the nonlinear version, the above expressions (\ref{linearspin3half})
have cubic term in the spin-$\frac{1}{2}$ current and 
the index $\tilde{a}$ contains $4$ indices corresponding to 
the $2 \times 2$ matrix as before. 

Let us determine other currents.
The six spin-$1$ currents ${\bf A}^{\pm i}(z)$ can be obtained 
from the second order pole $\frac{1}{(z-w)^2}$ terms of the OPE 
(\ref{linGGope}) and
 the corresponding OPE in (\ref{N4linearalg})
 \footnote{We have used  the following tensor identities,
 \bea
 S^{0 \tilde{a} \tilde{b} }_{\,\,\,\,\,\,\,\, \tilde{c}} 
 S^{1}_{\tilde{a} \tilde{b} \tilde{d}} 
 + S^{2 \tilde{a} \tilde{b} }_{\,\,\,\,\,\,\,\, \tilde{c}} 
 S^{3}_{\tilde{a} \tilde{b} \tilde{d}}
 =4 h^1_{\tilde{c} \tilde{d}},
 \quad
  S^{0 \tilde{a} \tilde{b} }_{\,\,\,\,\,\,\,\, \tilde{c}} 
 S^{2}_{\tilde{a} \tilde{b} \tilde{d}} 
 + S^{3 \tilde{a} \tilde{b} }_{\,\,\,\,\,\,\,\, \tilde{c}} 
 S^{1}_{\tilde{a} \tilde{b} \tilde{d}}
 =4 h^2_{\tilde{c} \tilde{d}},
  \quad
    S^{0 \tilde{a} \tilde{b} }_{\,\,\,\,\,\,\,\, \tilde{c}} 
 S^{3}_{\tilde{a} \tilde{b} \tilde{d}} 
 + S^{1 \tilde{a} \tilde{b} }_{\,\,\,\,\,\,\,\, \tilde{c}} 
 S^{2}_{\tilde{a} \tilde{b} \tilde{d}}
 =4 h^3_{\tilde{c} \tilde{d}}.
 \nonu 
 \eea
 },
\bea
{\bf A}^{+i}(z)&=&-\frac{1}{4(N+1)} \left[ h^i_{\tilde{a} \tilde{b} } f^{\tilde{a} \tilde{b}}_{\,\,\,\,\,\, c} V^c
+\frac{1}{(k+N+2)} \left( h^i_{\tilde{c} \tilde{d} } 
+\frac{1}{2} S^i_{\tilde{a} \tilde{b} \tilde{c} }f^{\tilde{a} \tilde{b}}_{\,\,\,\,\,\, d}\right) 
Q^{\tilde{c}} Q^{\tilde{d}}\right](z),
\nonu \\
{\bf A}^{-i}(z)&=&-\frac{1}{4(k+N+2)} h^i_{\tilde{a} \tilde{b} } Q^{\tilde{a}} Q^{\tilde{b}}(z),
\label{aplusminus}
\eea
corresponding to $(2.36)$ and $(2.37)$ of \cite{Saulina}.
From the second order poles in the 
OPEs ${\bf A}^{\pm i}(z) \, {\bf G}^{\mu}(w)$ with (\ref{linearspin3half}) and 
(\ref{aplusminus}), the four fermionic 
spin-$\frac{1}{2}$ currents 
${\bf \Gamma}^\mu(z)$ corresponding to $(2.43)$ 
of \cite{Saulina} can be fixed as 
follows:
\bea
{\bf \Gamma}^0 (z) =-\frac{i }{4(N+1)}
h^j_{\tilde{a} \tilde{b} }
f^{\tilde{a} \tilde{b}}_{\,\,\,\,\,\, \tilde{c}} h^{j \tilde{c}}_{ \,\,\,\, \tilde{d} } Q^{\tilde{d}}(z),
\qquad
{\bf \Gamma}^j (z) =-\frac{i }{4(N+1)} h^j_{\tilde{a} \tilde{b} }
f^{\tilde{a} \tilde{b}}_{\,\,\,\,\,\, \tilde{c}} Q^{\tilde{c}} (z),
\label{Gamma}
\eea
where $j=1,2,3$ and there is no sum over $j$ in the first equation of
(\ref{Gamma}).
From the OPE 
${\bf \Gamma}^\mu(z) \, {\bf G}^{\nu}(w)$ when the index $\mu$
is the same as the index $\nu$, the spin-$1$ current 
${\bf U}(z)$ corresponding to $(2.44)$ 
of \cite{Saulina} can be determined by
\bea
{\bf U} (z) =-\frac{1}{4(N+1)} h^j_{\tilde{a} \tilde{b} }
f^{\tilde{a} \tilde{b}}_{\,\,\,\,\,\, \tilde{c}} h^{j \tilde{c}}_{ \,\,\,\, \tilde{d} } \left[
 V^{\tilde{d}}
-\frac{1}{2(k+N+2)}  
f^{\tilde{d} }_{\,\,\,\, \tilde{e} \tilde{f}} Q^{\tilde{e}} Q^{\tilde{f}}
\right](z),
\label{ulinear}
\eea
where there is no sum over an index $j$.
One can easily see that the first (the second) 
term of (\ref{ulinear}) comes from the 
OPE between ${\bf \Gamma}^{\mu}(z)$ in (\ref{Gamma}) 
and the spin-$1$ term (the cubic term in the spin-$\frac{1}{2}$ current) 
in ${\bf G}^{\mu}(w)$ in (\ref{linearspin3half})
\footnote{
As in the footnote \ref{foot}, one can calculate 
the corresponding ${\bf U}$ charges for the coset fields 
as follows:
\bea
i {\bf U}(z)\, 
\left( 
\begin{array}{c}
Q^{\bar{A}} \nonu \\
 Q^{\bar{A}^*}
\end{array}
\right)(w) & = & 
\mp \frac{1}{(z-w)} \left[  \frac{1}{2} \sqrt{ \frac{N+2}{N}} \right] 
\left( 
\begin{array}{c}
Q^{\bar{A}} \nonu \\
 Q^{\bar{A}^*} 
\end{array}
\right)(w) +\cdots, 
\nonu \\
i {\bf U}(z) 
\left(
\begin{array}{c}
V^{\bar{A}} \nonu \\
 V^{\bar{A}^*} 
\end{array}
\right)(w) & = &  \mp \frac{1}{(z-w)} \left[  \frac{1}{2} 
\sqrt{ \frac{N+2}{N}} \right] 
\left(
\begin{array}{c}
V^{\bar{A}} \nonu \\
 V^{\bar{A}^*} 
\end{array}
\right)(w) +\cdots.
\nonu 
\eea
These are proportional to the $U$ charges in the footnote \ref{foot}.}.

The defining OPE equations for the $16$ currents are given in Appendix $C$
for convenience. Appendix $D$ contains the explicit 
form for the complex structures.

\subsection{ The higher spin currents in the coset (\ref{coset1})}

Let us consider 
the higher spin currents in an extension of  large 
$\mathcal N =4$ linear superconformal algebra.
Again, it is crucial to find the lowest higher spin-$1$ current 
and it is straightforward to obtain the remaining 
higher spin currents.
Let us denote the $16$ higher spin currents as follows:
\bea
\left(1, \frac{3}{2}, \frac{3}{2}, 2 \right)
& : & ({\bf T^{(1)}}, {\bf T_{+}^{(\frac{3}{2})}}, {\bf T_{-}^{(\frac{3}{2})}}, {\bf T^{(2)}}), 
\nonu \\
 \left( \frac{3}{2}, 2, 2, \frac{5}{2} \right) & : & 
({\bf U^{(\frac{3}{2})}}, {\bf U_{+}^{(2)}}, {\bf U_{-}^{(2)}}, {\bf U^{(\frac{5}{2})}} ), \nonu \\
\left(\frac{3}{2}, 2, 2, \frac{5}{2} \right) & : & 
({\bf V^{(\frac{3}{2})}}, {\bf V^{(2)}_{+}}, {\bf V^{(2)}_{-}}, {\bf V^{(\frac{5}{2})}}),  \nonu \\
\left(2, \frac{5}{2}, \frac{5}{2}, 3 \right) & : &
 ({\bf W^{(2)}}, {\bf W_{+}^{(\frac{5}{2})}}, {\bf W_{-}^{(\frac{5}{2})}}, {\bf W^{(3)}}).
\label{lowestlinearhigher}
\eea
Here the  
 $16$ higher spin currents are primary under the 
stress energy tensor given in (\ref{tlinear}). 
In the convention of \cite{BCG1404}, 
the higher spin-$3$ current is a quasiprimary current 
under the stress energy tensor.

\subsubsection{  The higher spin-$1$ current}
 
The natural ansatz for the higher spin-$1$ current is 
\bea
{\bf T^{(1)}}(z) &=&
 A_a \, V^a (z) 
+  B_{\tilde{a} \tilde{b}} \, Q^{\tilde{a}} \, Q^{\tilde{b}} (z). 
\label{higherspinonelinear}
\eea
By requiring that 
the OPEs between the above higher spin-$1$ current and  six spin-$1$
currents are regular,
the higher spin-$1$ current is a primary under the stress energy 
tensor ${\bf T}(w)$
and the OPEs between the higher spin-$1$ current 
and  $4$ free fermions  ${\bf \Gamma}^{\mu}(w)$ 
(and ${\bf U}(w)$) are regular \cite{BCG1404},
all the unknown coefficients in (\ref{higherspinonelinear}) 
are determined completely.
Then, it turns out that the higher spin-$1$ current 
(\ref{higherspinonelinear}) in the linear version 
is the same as the 
higher spin-$1$ current in the nonlinear version
\footnote{
Of course, we required same normalization with the higher spin-$1$ 
current $T^{(1)}$ in the nonlinear version
and take the same sign as follows
\bea
{\bf T^{(1)} } (z)\, {\bf T^{(1)} } (w)= \frac{1}{(z-w)^2} \left[ \frac{2 N k }{N+k+2}\right]+\cdots.
\nonu
\eea }
\bea
{\bf T^{(1)}}(z)=T^{(1)}(z).
\label{higherspinone}
\eea
Now the lowest higher spin-$1$ current is completely 
fixed and it is straightforward to calculate the remaining higher spin 
currents as mentioned before.

\subsubsection{  The higher spin-$\frac{3}{2}$ currents in 
(\ref{lowestlinearhigher})}

Let us define the 
four higher spin-$\frac{3}{2}$ currents  ${\bf G'^{\mu} } (w)$
from the first order pole of the following OPE
\bea
{\bf G^{\mu} }(z) \, {\bf T^{(1)}} (w) 
&=&
\frac{1}{(z-w)} \, {\bf G'^{\mu}} (w) + \cdots.
\label{linGprime}
\eea
It turns out that the first order pole in (\ref{linGprime}) provides
\bea
{\bf G'^{\mu} } (z) = G'^{\mu}  (z),
\label{linnonlinear}
\eea
which is given by (\ref{gprimemu}).
This implies that the OPE between the extra terms between ${\bf G^{\mu} }(z)$
and $G^{\mu} (z)$ in Appendix (\ref{gsformula}) and higher spin-$1$ current
is regular.

By calculating the OPEs between the spin-$\frac{3}{2}$ currents 
in (\ref{linearspin3half}) and the higher spin-$1$ current
in (\ref{higherspinone}) and by subtracting the spin-$\frac{3}{2}$
currents in the left hand side with correct coefficients, 
the following higher spin-$\frac{3}{2}$ currents 
can be obtained 
\bea
{\bf T_{+}^{(\frac{3}{2})} }(z) & = & 
\frac{1}{2} \left( {\bf G}'_{21}  - {\bf G}_{21} \right) (z)
=\frac{1}{2} \left(  G'_{21}  - {\bf G}_{21} \right) (z),
\nonu \\
{\bf T_{-}^{(\frac{3}{2})} } (z) &=&
\frac{1}{2} \left( {\bf G}'_{12}  + {\bf G}_{12} \right) (z)
=\frac{1}{2} \left( G'_{12}  + {\bf G}_{12} \right) (z),
\nonu \\
{\bf U^{(\frac{3}{2})} } (z) & = &
\frac{1}{2} \left( {\bf G}'_{11}  - {\bf G}_{11} \right) (z)
=\frac{1}{2} \left( G'_{11}  - {\bf G}_{11} \right) (z),
\nonu \\
{\bf V^{(\frac{3}{2})} } (z) &=&
\frac{1}{2} \left( {\bf G}'_{22}  + {\bf G}_{22} \right) (z)
=\frac{1}{2} \left(  G'_{22}  + {\bf G}_{22} \right) (z).
\label{3halflinear}
\eea
In each last line, the conditions in (\ref{linnonlinear}) are used.
Furthermore, the relations in the footnote \ref{doubleindex} 
between the double index and a single index 
hold in this case.
Compared to the corresponding higher spin-$\frac{3}{2}$ currents
in the nonlinear version \cite{AK1411},
the four corresponding OPEs generating (\ref{3halflinear}) 
are the same as the ones in the nonlinear version. See also the first four 
equations in Appendix $E$. In other words, those four equations remain 
unchanged if one uses the currents without a boldface.

\subsubsection{  The remaining higher spin currents}

So far, the higher spin-$1$ and four higher spin-$\frac{3}{2}$ currents 
are obtained from (\ref{higherspinone}) and (\ref{3halflinear}).

How does one obtain the other higher spin currents?
Let us consider the six higher spin-$2$ currents in 
(\ref{lowestlinearhigher}). 
For example,
the  following higher spin-$2$ current can be described as follows:
\bea
{\bf U}^{(2)}_{-} (w)&=&  \{ {\bf G}_{12} \,  {\bf U}^{(\frac{3}{2})} \}_{-1} (w)
-\frac{1}{2 } \pa \{ {\bf G}_{12} \,  {\bf U}^{(\frac{3}{2})} \}_{-2} (w)
\nonu \\
&=& \frac{1}{2} \{ {\bf G}_{12} \, G'_{11}  \}_{-1} (w)
- \frac{1}{2} \{ {\bf G}_{12} \, {\bf G}_{11}  \}_{-1} (w)
-\frac{1}{2 } \pa \{ {\bf G}_{12} \,  {\bf U}^{(\frac{3}{2})} \}_{-2} (w)
\nonu \\
&=& \frac{1}{2} \{ {\bf G}_{12} \, G'_{11}  \}_{-1} (w).
\label{aboveexpression}
\eea
The first line of (\ref{aboveexpression})
comes from the seventh OPE in Appendix $E$. In the second line, the equation 
(\ref{3halflinear}) is used. It turns out that the second and third terms
are canceled each other. Then the higher spin-$2$ current can be obtained 
from the first order pole of the OPE between 
 $ {\bf G}_{12}(z)$ and  $ G'_{11}(w)$. 

Similarly, 
the other remaining higher spin-$2$ currents can be simplified as 
follows:
\bea
{\bf V}^{(2)}_{+} (w) & = & \frac{1}{2} \{ {\bf G}_{21} \, G'_{22}  \}_{-1} (w),
\nonu \\
{\bf V}^{(2)}_{-} (w) & = & 
\frac{1}{2} \{ {\bf G}_{12} \, G'_{22}  \}_{-1} (w),
\nonu \\
{\bf U}^{(2)}_{+} (w) & = & 
\frac{1}{2} \{ {\bf G}_{21} \, G'_{11}  \}_{-1} (w),
\nonu \\
{\bf T}^{(2)} (w) & = & \frac{1}{2} \{ {\bf G}_{21} \, G'_{12}  \}_{-1} (w) - 
\frac{1}{2} \pa {\bf T}^{(1)} (w),
\nonu \\
{\bf W}^{(2)} (w) & = & 
\frac{1}{2} \{ {\bf G}_{11} \, G'_{22}  \}_{-1} (w) - 
\frac{1}{2} \pa {\bf T}^{(1)} (w).
\label{expressionrecent}
\eea
In (\ref{expressionrecent}), because the derivative of higher spin-$1$
current can be obtained easily, 
the nontrivial parts for the higher spin currents 
are given by the OPE 
${\bf G}_{\alpha \beta}(z) \, G'_{\rho \sigma}(w)$.

Let us  write down ${\bf G}_{\alpha \beta}(z)$ and $G'_{\rho \sigma}(z)$
explicitly with simplified notations 
\bea
{\bf G}_{\alpha \beta}(z) &=& \frac{i}{(k+N+2)} \left[ \, 
h^{\alpha \beta}_{\tilde{a} \tilde{b}} \,
Q^{\tilde{a}} \, V^{\tilde{b}}- \frac{1}{6(k+N+2)} 
S^{\alpha \beta}_{\tilde{a} \tilde{b} \tilde{c}}
Q^{\tilde{a}} Q^{\tilde{b}} Q^{\tilde{c}} \right](z),
\nonu \\
G'_{\rho \sigma}(z) &=& 
 \frac{i}{(k+N+2)}  
d^{\rho \sigma}_{\bar{a} \bar{b}} \,
Q^{\bar{a}} \, V^{\bar{b}} (z).
\label{twoexpression}
\eea
where $\alpha, \beta, \rho, \sigma=1,2$ and 
in (\ref{twoexpression}) the following notations are introduced
\bea
h^{11}_{\tilde{a} \tilde{b}} & \equiv & 
\frac{1}{\sqrt{2}} \left( h^1 - i h^2 \right)_{\tilde{a} \tilde{b}}, 
\quad
h^{12}_{\tilde{a} \tilde{b}} \equiv
-\frac{1}{\sqrt{2}} \left( h^3 - i h^0 \right)_{\tilde{a} \tilde{b}},
\nonu \\
h^{22}_{\tilde{a} \tilde{b}} & \equiv &
\frac{1}{\sqrt{2}} \left( h^1 + i h^2 \right)_{\tilde{a} \tilde{b}},
\quad
h^{21}_{\tilde{a} \tilde{b}} \equiv
-\frac{1}{\sqrt{2}} \left( h^3 + i h^0 \right)_{\tilde{a} \tilde{b}},
\nonu \\
S^{11}_{\tilde{a} \tilde{b} \tilde{c}} & \equiv &
\frac{1}{\sqrt{2}} \left( S^1 - i S^2 \right)_{\tilde{a} \tilde{b} \tilde{c} }, 
\qquad
S^{12}_{\tilde{a} \tilde{b} \tilde{c}} \equiv
-\frac{1}{\sqrt{2}} \left( S^3 - i S^0 \right)_{\tilde{a} \tilde{b} \tilde{c}},
\nonu \\
S^{22}_{\tilde{a} \tilde{b} \tilde{c}} & \equiv &
\frac{1}{\sqrt{2}} \left( S^1 + i S^2 \right)_{\tilde{a} \tilde{b} \tilde{c}},
\qquad
S^{21}_{\tilde{a} \tilde{b} \tilde{c}} \equiv
-\frac{1}{\sqrt{2}} \left( S^3 + i S^0 \right)_{\tilde{a} \tilde{b} \tilde{c}},
\nonu \\
d^{11}_{\bar{a} \bar{b}} & \equiv &
\frac{1}{\sqrt{2}} \left( d^1 - i d^2 \right)_{\bar{a} \bar{b}}, 
\qquad
d^{12}_{\bar{a} \bar{b}} \equiv
-\frac{1}{\sqrt{2}} \left( d^3 - i d^0 \right)_{\bar{a} \bar{b}},
\nonu \\
d^{22}_{\bar{a} \bar{b}} & \equiv &
\frac{1}{\sqrt{2}} \left( d^1 + i d^2 \right)_{\bar{a} \bar{b}},
\qquad
d^{21}_{\bar{a} \bar{b}} \equiv
-\frac{1}{\sqrt{2}} \left( d^3 + i d^0 \right)_{\bar{a} \bar{b}}.
\label{coeffcoeffcoeff}
\eea
From the explicit forms in (\ref{twoexpression}), one can 
calculate the OPEs between them and arrives at
the following first order pole as follows with (\ref{coeffcoeffcoeff}):
\bea
 {\bf G}_{\alpha \beta}(z) \,  G'_{\rho \sigma}(w) |_{\frac{1}{(z-w)}} &=&
 \frac{1}{(k+N+2)^2} \left[ (k+N+2)  h^{\alpha \beta}_{\bar{a} \bar{b}}
 d^{\rho \sigma \bar{a}}_{ \,\,\,\,\,\,\,\,\, \bar{d}}
 V^{\bar{b}} V^{\bar{d}} -k  h^{\alpha \beta}_{\bar{a} \bar{b}}
 d^{\rho \sigma \bar{b}}_{ \,\,\,\, \bar{c}}
 \pa Q^{\bar{a}} Q^{\bar{c}} \right.
 \nonu \\
 &+&\left.  h^{\alpha \beta}_{\tilde{a} \tilde{b}} d^{\rho \sigma}_{\bar{c} \bar{d}} 
 f^{\tilde{b} \bar{d} }_{\,\,\,\,\,\, e} Q^{\tilde{a}} Q^{\bar{c}} V^e
 -\frac{1}{2} S^{\alpha \beta}_{\bar{a} \tilde{b} \tilde{c}}
  d^{\rho \sigma \bar{a}}_{ \,\,\,\,\,\,\,\,\, \bar{e}} Q^{\tilde{b}} Q^{\tilde{c}} 
  V^{\bar{e}} \right](w).
\label{opefirstorder}
\eea
Then by taking the appropriate six cases in (\ref{opefirstorder}),
one obtains the final six higher spin-$2$ currents together with 
the derivative term of higher spin-$1$ current. 

For the higher spin-$\frac{5}{2}$ currents, 
one should calculate the OPE between the spin-$\frac{3}{2}$ currents 
and the higher spin-$2$ currents obtained in previous paragraph.
Then it is straightforward to calculate the OPEs between the 
spin-$\frac{3}{2}$ currents and (\ref{opefirstorder}).
Once the higher spin-$\frac{5}{2}$ currents are obtained, then 
one should repeat the above procedure. That is, 
in order to obtain the final higher spin-$3$ current, 
the OPEs between the spin-$\frac{3}{2}$ currents and the first order poles 
where the composite fields have the conformal dimension $\frac{5}{2}$ are 
needed. 
Although the complete expressions are not written down in this paper, 
one can follow the procedures in the nonlinear version \cite{AK1411}.
See also Appendix $F$
which contains the precise relations between the 
higher spin currents in the nonlinear and linear versions 
for general $N$ and $k$.

\section{  
 Three-point functions in an extension of 
large ${\cal N}=4$ linear superconformal algebra
}

As in section $3$, one calculates the three point functions 
for the higher spin currents in the extension of 
large ${\cal N}=4$ linear superconformal algebra. 

\subsection{  Eigenvalue equations for spin-$2$ current 
 acting on the 
states $|(f;0)>$ and $|(0;f)>$
in the large 
${\cal N}=4$ linear superconformal algebra }

Let us  define the ${\bf U}$-charge as in  \cite{npb1989}.
\bea
i {\bf U}_0 |(f;0)> & = & {\bf u}(f;0) |(f;0)>, \nonu \\
i {\bf U}_0 |(0;f)>  & = & {\bf u}(0;f) |(0;f)>.
\label{eigenu}
\eea
From the explicit expression in (\ref{ulinear}),
one can obtain the eigenvalues  $ {\bf u}(f;0)  $ and $ {\bf u}(0;f)  $
as follows: 
\bea
{\bf u}(f;0) & = & -\frac{1}{2} \sqrt{ \frac{N}{N+2}}, 
\nonu \\
{\bf u}(0;f) & = & \frac{1}{2} \sqrt{ \frac{N+2}{N}}.
\label{eigenvalueu}
\eea
One has the explicit expressions
\footnote{
\label{footagain}
For $N=3$, one considers the following matrix 
acting on the states
\bea
i {\bf U}_0 |(f;\star)>=\left(
\begin{array}{ccc|cc}
 \frac{1}{\sqrt{15}} & 0 & 0 & 0 & 0 \\
 0 & \frac{1}{\sqrt{15}} & 0 & 0 & 0 \\
 0 & 0 & \frac{1}{\sqrt{15}} & 0 & 0 \\
\hline
 0 & 0 & 0 & -\frac{1}{2}  \sqrt{\frac{3}{5}} & 0 \\
 0 & 0 & 0 & 0 & -\frac{1}{2}  \sqrt{\frac{3}{5}} \\
\end{array}
\right) |(f;\star)>.
\nonu
\eea
Then one can vary $N$ for $N=5,7,9$ as before and one sees the 
general $N$ behavior as in (\ref{eigenvalueu}).
One can show that the spin-$1$ current ${\bf U}(z)$
is equivalent  to the previous spin-$1$ current $U(z)$ in the 
footnote \ref{foot}.
In other words, ${\bf U}(z) = -\frac{i}{2 \sqrt{N(N+2)}}  U(z)$.
From the explicit form for the spin-$1$ current,
the relevant term can be described as  
$
i {\bf U}(z) \sim  
\frac{1}{2(5+k)} \sqrt{\frac{5}{3}} \sum_{a=1}^{6} Q^a Q^{a^*}(z)$.
Then one can calculate  
the following OPE
$
i {\bf U}(z) \, Q^{\bar{A}^*} (w)= 
\frac{1}{(z-w)} \left[ \frac{1}{2} \sqrt{\frac{5}{3}}  \right] Q^{\bar{A}^*}(w)
+\cdots$.
Therefore, the eigenvalue is given by $ \frac{1}{2} \sqrt{\frac{5}{3}}$ 
for $N=3$.}.

From the spin-$2$ current in (\ref{gsformula}),
one has the following relation
\bea
{\bf T}_0 |(f;0)>
&\sim& \left[ T-\frac{1}{(k+N+2)} 
 {\bf UU} \right]_0 |(f;0)>
 \nonu \\
&=& \left[ h(f;0)+\frac{1}{(k+N+2)} 
 {\bf u}^2 (f;0) \right]  |(f;0)>
 \nonu \\
 &=&
 \left[ \frac{(N+1)(N+3)}{2(N+2)(k+N+2)} \right] |(f;0)>.
\label{tf0linear}
\eea
In the first line, the spin-$\frac{1}{2}$ current dependent 
terms are ignored. See also (\ref{Gamma}) where ${\bf \Gamma^{\mu}}$
term contains the spin-$\frac{1}{2}$ current. 
In the second and third lines, the relations (\ref{t0f0}), (\ref{eigenu}) 
and (\ref{eigenvalueu}) 
are used. 
The eigenvalue in the equation 
(\ref{tf0linear})  
is the same value as  $h'(f;0)$ appeared in \cite{GG1305}.
One can see this by writing $\frac{(N+1)(N+3)}{2(N+2)}$ as 
$\left[\frac{(N+2)}{2} -\frac{1}{2(N+2)}\right]$ 
which is nothing but the quadratic 
Casimir of fundamental representation in $SU(N+2)$. 
Therefore, the $\hat{u}$ part and $u$ part of \cite{GG1305}
in the conformal dimension are canceled each other completely.

Because the OPE between 
$  {\bf \Gamma^{\mu}} (z)$ and $ Q^{\bar{A^\ast}}(w)$ is regular, 
$ \pa {\bf \Gamma^{\mu} \Gamma_{\mu}}(z)$ term  does not contribute 
to the eigenvalue equation.
Then  one obtains  
the zero mode eigenvalue  equation of ${\bf T}(z) $ for the state $|(0;f)>$
as follows
\footnote{
One determines the zero mode eigenvalue for the `light' state as follows: 
\bea
{\bf T}_0 |(f;f)>
&\sim& \left[ T-\frac{1}{(k+N+2)} 
 {\bf UU} \right]_0 |(f;f)>
= \left[ h(f;f)+\frac{1}{(k+N+2)} 
 {\bf u}^2 (f;f) \right]  |(f;f)>
 \nonu \\
 &=&
 \left[ \frac{(N+1)^2}{N(N+2)(k+N+2)} \right] |(f;f)>
 \rightarrow  \frac{\lambda}{N} |(f;f)> \rightarrow 0,
\nonu
\eea
where we used $ {\bf u} (f;f) = \frac{1}{\sqrt{N(N+2)}}$.
See also the footnote \ref{footagain} for the $3 \times 3$ 
diagonal elements, $\frac{1}{\sqrt{15}}$, when $N=3$.
In the last line, the large $N$ limit (\ref{limit}) is taken. }:
\bea
{\bf T}_0 |(0;f)>
&\sim& \left[ T-\frac{1}{(k+N+2)} 
 {\bf UU} \right]_0 |(0;f)>
 \nonu \\
&=& \left[ h(0;f)+\frac{1}{(k+N+2)} 
 {\bf u}^2 (0;f) \right]  |(0;f)>
 \nonu \\
 &=&
 \left[ \frac{(Nk+2N+1)}{2N(k+N+2)} \right] |(0;f)>.
\label{t0flinear}
\eea
In the third line, the relation (\ref{hzerof}) and (\ref{eigenvalueu}) 
are used. 
The  eigenvalue in the equation (\ref{t0flinear}) 
is  exactly the same as  $h'(0;f)$ described in \cite{GG1305}.
One can also understand this by writing
the eigenvalue as
$\frac{1}{2}- \left[ \frac{(\frac{N}{2}-\frac{1}{2N})}{(k+N+2)} \right]$
where the second term is a quadratic Casimir in the fundamental representation 
in $SU(N)$ and the first term $\frac{1}{2}$ is the conformal 
dimension from an excitation number. 
The $\hat{u}$ part and $u$ part of \cite{GG1305}
in the conformal dimension are canceled each other completely.

The large $N$ limit (\ref{limit}) for (\ref{tf0linear}) 
and (\ref{t0flinear}) leads to
\bea
{\bf T}_0 |(f;0)> & = & \frac{1}{2} \lambda  |(f;0)>, \nonu \\
{\bf T}_0 |(0;f)> & = & \frac{1}{2} (1-\lambda)  |(0;f)>.
\label{Expexp}
\eea
which are  exactly the same as the ones in the subsection $3.1
$ in the nonlinear version. In other words, the extra 
${\bf U U}$ term in (\ref{tf0linear}) and (\ref{t0flinear})
does not contribute to the eigenvalue equation in this large 
$N$ limit.

\subsection{  Eigenvalue equations for higher spin-$1$ current
 acting on the 
states $|(f;0)>$ and $|(0;f)>$}

In this case, the previous relations (\ref{t1f0}) and (\ref{t10f}) hold
because of (\ref{higherspinone}). 

\subsection{  Eigenvalue equations for higher spin currents 
of spins $2$ and $3$
 acting on the 
states $|(f;0)>$ and $|(0;f)>$}


Let us calculate the eigenvalue equation for the higher spin-$2$ current.
It turns out that one obtains the following result
\bea
{\bf T}^{(2)}_0 |(f;0)>_{\pm} & = & \pm \left[ \frac{N}{2(N+k+2)}\right] 
|(f;0)>_{\pm},
\nonu \\
{\bf T}^{(2)}_0 |(0;f)>_{\pm} & = & \pm \left[ \frac{k}{2(N+k+2)} \right] 
|(0;f)>_{\pm}.
\label{t2nonlinear}
\eea
Compared to the previous expression (\ref{t2eigen}) in the nonlinear version,
the above eigenvalues (\ref{t2nonlinear}) appear in the factors in 
(\ref{t2eigen}). The extra terms in the higher 
spin-$2$ current in the linear version
contribute to the eigenvalue equation also and can be added 
to the right hand side of
(\ref{t2eigen}). Then the above simple result occurs.  
Furthermore, the previous relation (\ref{t2f0}) holds in this case.
More explicitly, the expressions 
$\frac{(N -2k)}{(2N k +N +k)}$ and 
$-\frac{(N-2k)}{(2 N k +N +k)}$ (with an overall factor)
in the eigenvalues of first two terms in (\ref{t2eigen})
arise from the eigenvalue equations for the zero mode 
of the extra terms in the fifth equation of 
Appendix (\ref{twoexpression1}).
Similarly,  the expressions 
$\frac{(k -2N)}{(2N k +N +k)}$ and $-\frac{(k-2N)}{(2 N k +N +k)}$
in the eigenvalues of the 
last two terms in (\ref{t2eigen}) can be analyzed.

For the other higher spin-$2$ current,
the similar calculation gives the following result
\bea
{\bf W}^{(2)}_0 |(f;0)>_{\pm} & = & \mp \left[\frac{N}{2(N+k+2)}\right] 
|(f;0)>_{\pm},
\nonu \\
{\bf W}^{(2)}_0 |(0;f)>_{\pm} & = & \pm \left[ \frac{k}{2(N+k+2)} \right] 
|(0;f)>_{\pm}.
\label{w2nonlinear}
\eea
The eigenvalue equations (\ref{w2nonlinear}) look similar to the 
previous ones in (\ref{t2nonlinear}) up to the signs.
Furthermore, compared to the previous ones in the nonlinear version,
one sees the eigenvalues in the factors of (\ref{w2eigen}).
In this case also,  
the extra terms in the higher spin-$2$ current in the linear version
contribute to the eigenvalue equation also and can be added 
to the right hand side of
(\ref{w2eigen}). Then the above very simple result can be obtained.
Simple linear combinations between these higher spin-$2$  currents 
will give rise to simple eigenvalue
equations which will appear in next subsection.
One sees the previous relation (\ref{w2f0}) 
in this case.
The expressions 
$-\frac{(N-2k)}{(2N k + N+k)} $ and  $\frac{(N-2k)}{(2N k + N+k)}$ 
in the eigenvalues of the 
first two equations in (\ref{w2eigen}) can be analyzed from the 
last equation of Appendix (\ref{twoexpression1}). 
Also the expressions 
$\frac{(k-2N)}{(2N k + N +k)}$ and $-\frac{(-2N+k)}{(2N k +
N +k)}$ appearing in the eigenvalues of the 
last two equations in (\ref{w2eigen}) 
can be described similarly.

Furthermore, if one replaces the fundamental representation
$f$ with the antifundamental representation $\bar{f}$ in (\ref{t2nonlinear})
and (\ref{w2nonlinear}),
then the right hand sides remain unchanged.

By taking  the large $N$ 't Hooft limit (\ref{limit}), one obtains 
\bea
{\bf T}^{(2)}_0 |(f;0)>_{\pm} & = & \pm \frac{1}{2} \lambda |(f;0)>_{\pm},
\nonu \\
{\bf T}^{(2)}_0 |(0;f)>_{\pm} & = & \pm \frac{1}{2} (1-\lambda) |(0;f)>_{\pm},
\nonu \\
{\bf W}^{(2)}_0 |(f;0)>_{\pm} & = & \mp \frac{1}{2} \lambda |(f;0)>_{\pm},
\nonu \\
{\bf W}^{(2)}_0 |(0;f)>_{\pm} & = & \pm \frac{1}{2} (1-\lambda) |(0;f)>_{\pm}.
\label{reducedexp}
\eea
Surprisingly, these eigenvalue equations (\ref{reducedexp}) are exactly 
the same as the previous ones 
in the nonlinear version (\ref{express}).

For the final higher spin-$3$ current
\footnote{
\label{partial}
Explicitly 
one obtains the first three diagonal matrix elements 
$-\frac{52 (k-3) (5 k+9)}{15 (k+5)^2 (13 k+17)}$, 
and the remaining two diagonal matrix elements
$\frac{2 \left(65 k^2+742 k+549\right)}{5 (k+5)^2 (13 k+17)}$ for
$N=3$.}, 
the following eigenvalue equations
hold
\bea
{\bf W}^{(3)}_0 |(f;0)> &=& {\bf w^{(3)}}(f;0) |(f;0)>,
\nonu \\
{\bf W}^{(3)}_0 |(0;f)> &=& {\bf w^{(3)}}(0;f)
|(0;f)>,
\nonu \\
{\bf w^{(3)}}(f;0) & \equiv & 
 \frac{ 2N }{3(N+2)(N+k+2)^2(4N+5+(3N+4)k)}
\times \left[ (5N^3+26N^2+50N+30) \right. 
\nonu \\
& + & \left. (6N^3+36N^2+71N+43)k + (3N^2+10N+8)k^2 \right],
\label{w3linear}
\\
{\bf w^{(3)}}(0;f) & \equiv & 
 -\frac{2k \left[ (4N^3+23N^2+18N) 
+(3N^3+24N^2+16N-3)k + (6N^2+5N)k^2 \right]}{3N (N+k+2)^2(4N+5+(3N+4)k)}.
\nonu
\eea
Based on the partial result in the footnote \ref{partial},
one can further calculate those quantities for $N=5,7,9$.
It is not difficult to obtain the  denominators in $44$ and $55$ elements 
for generic $N$.
It is nontrivial to see the generic $N$ behavior for the numerators.  
The numerator is a quadratic in $k$ and one can introduce  
the three independent polynomials in $N$ with the highest power $3$
(with four unknown coefficients) 
in the quadratic in $k$, in the linear in $k$ and in the constant terms
respectively.
Now one can use the four constraint equations from the above 
$N=3,5,7,9$ cases. It turns out 
the above unknown coefficients are completely determined uniquely.      
For the eigenvalue ${\bf w^{(3)}}(0;f)$, one can analyze similarly.
The numerator behaves nontrivially. So one introduces 
three polynomials of order $3$ with undetermined coefficients.  
They can be fixed by solving the relevant equations 
for $N=3,5,7,9$.

Furthermore, if one replaces the fundamental representation
$f$ with the antifundamental representation $\bar{f}$ in (\ref{w3linear}),
then the right hand sides have minus signs.

It is obvious that 
there is no $N \leftrightarrow k$ symmetry which is 
a 
different aspect compared to the $W^{(3)}_0 $ case in the nonlinear version.
However, under the large $N$ 't Hooft limit (\ref{limit}), 
the above expressions (\ref{w3linear}) become simple form as follows:
\bea
{\bf W}^{(3)}_0 |(f;0)> & = & \frac{2}{3} \lambda (1+\lambda)|(f;0)>,
\nonu \\
{\bf W}^{(3)}_0 |(0;f)> & = & -\frac{2}{3} (1-\lambda) (2-\lambda)|(0;f)>.
\label{reducedexpexp}
\eea
This is exactly the same as  the ones in (\ref{nonw3large}) 
for the $ W^{(3)}_0$ eigenvalue equations in the nonlinear version
\footnote{
One also calculates the eigenvalue equation for the `light' state
as follows:
\bea
{\bf W}^{(3)}_0 |(f;f)>=
\left[ \frac{4(N-k)(4N^3+19N^2+22N+6+(3N^3+10N^2+8N)k )}
{3N(N+2)(N+k+2)^2 (4N+5+(3N+4)k)} \right]
|(f;f)> \rightarrow \frac{4 \lambda (2 \lambda -1)}{3N} |(f;f)>
\rightarrow 0.
\nonu
\eea
As in the footnote \ref{partial}, the first three elements in 
$N=3$ case are relevant to this expression.
By following the prescription in (\ref{w3linear}), the explicit 
$N$ dependence can be fixed completely for $N=3,5,7,9$.
Note that there are two cubic polynomials in $N$ in the numerator.
}.

\subsection{ 
 Eigenvalue equations for higher spin currents 
of spins $2$ and $3$ 
 acting on the 
states $|(f;0)>$ and $|(0;f)>$
in the  basis of \cite{BCG1404} }

The eigenvalue equations in the nonlinear version \cite{BCG1404} 
are the same as the ones in the  linear version. 
From the explicit relations in \cite{Ahn1504},
\bea
V_1^{(1), \pm 1}(z) & = & 
2i \left( {\bf U_{\mp}^{(2)}} - 
{\bf V_{\pm}^{(2)}} \right)(z), 
\nonu \\
V_1^{(1), \pm 2}(z) & = & 
-2 \left( {\bf U_{\mp}^{(2)}} + 
{\bf V_{\pm}^{(2)}} \right)(z), 
\nonu \\
V_1^{(1), \pm 3}(z) & = & 
\pm 2i \left( {\bf T^{(2)}} \mp
{\bf W^{(2)}} \right)(z),
\label{rel}
\eea
one can rewrite the eigenvalue equations 
as follows:
\bea
\left[V_1^{(1), + 3} \right]_0 |(f;0)>_{\pm}  & = &  
\pm \left[ \frac{2i N}{(N+k+2)} \right] |(f;0)>_{\pm},
\nonu \\
\left[V_1^{(1), + 3} \right]_0 |(0;f)> & = &  
0,
\nonu \\
\left[V_1^{(1), - 3} \right]_0 |(f;0)>  & = &  
0,
\nonu \\
\left[V_1^{(1), - 3} \right]_0 |(0;f)>_{\pm} & = &  
\mp \left[ \frac{2i k}{(N+k+2)} \right]
 |(0;f)>_{\pm}.
\label{relrel}
\eea
The corresponding large $N$ 't Hooft limit (\ref{limit})
in (\ref{relrel})
provides the following 
result,
\bea
\left[V_1^{(1), + 3} \right]_0 |(f;0)>_{\pm}  & = & 
\pm 2i \lambda |(f;0)>_{\pm}, \nonu \\
\left[V_1^{(1), + 3} \right]_0 |(0;f)>  & = & 0,
\nonu \\
\left[V_1^{(1), - 3} \right]_0 |(f;0)> & = & 0, 
\nonu \\ 
\left[V_1^{(1), - 3} \right]_0 |(0;f)>_{\pm} & = & \mp
2i (1- \lambda)
 |(0;f)>_{\pm}.
\label{inthis}
\eea
In this case (\ref{inthis}), 
the eigenvalue equation contains zero values  
and one cannot take one of the ratios between them.

By introducing the following quantities
\bea
V_{1}^{(1) +\pm } (z)\equiv \left[ V_{1}^{(1)+1} \pm i V_{1}^{(1)+2} \right](z),
\nonu \\
V_{1}^{(1) - \pm } (z) \equiv \left[ V_{1}^{(1)-1} \pm i V_{1}^{(1)-2} \right](z),
\label{othercombi}
\eea 
one can calculate 
\footnote{
More precisely, one has 
\bea
\left[ V_{1}^{(1) -+}  \right]_0  
 \frac{1}{\sqrt{k+N+2}} Q^{(a+N)^*}_{-\frac{1}{2}} |0>
&=&
\left[ \frac{4k i }{(N+k+2)}\right] 
\frac{1}{\sqrt{k+N+2}} Q^{a^*}_{-\frac{1}{2}} |0>, 
\nonu \\
\left[ V_{1}^{(1) --}  \right]_0  \frac{1}{\sqrt{k+N+2}} Q^{a^*}_{-\frac{1}{2}} |0> 
&=&
\left[ \frac{4k i }{(N+k+2)} \right]  
\frac{1}{\sqrt{k+N+2}} Q^{(a+N)^*}_{-\frac{1}{2}} |0>,
\nonu
\eea
where $a=1, 2, \cdots, N$.}
\bea
\left[ V_{1}^{(1) + \pm}  \right]_0  |(f;0)>_{\mp}&=&
- \left[ \frac{4N i }{(N+k+2)} \right] 
|(f;0)>_{\pm}   \rightarrow  -4i \lambda |(f;0)>_{\pm}, 
\nonu \\
\left[ V_{1}^{(1) -\pm}  \right]_0  |(0;f)>_{\mp}&=&
\left[ \frac{4k i }{(N+k+2)} \right] 
|(0;f)>_{\pm}  \rightarrow  4i (1-\lambda) |(0;f)>_{\pm}.
\label{simeigen}
\eea
The quadratic combinations will be obtained later
\footnote{
One can also present the three point functions together with 
(\ref{othercombi}) and (\ref{simeigen})
as follows:
\bea
<\overline{{\cal O}}_{+,\pm } 
{\cal O}_{+,\mp }  V_{1}^{(1) + \pm}  > &=&  -4i \lambda 
<\overline{{\cal O}}_{+,\pm } 
{\cal O}_{+,\pm } >,
\nonu \\
<\overline{{\cal O}}_{-,\pm } 
{\cal O}_{-,\mp }  V_{1}^{(1) - \pm}  > &=&   4i (1-\lambda)
<\overline{{\cal O}}_{-,\pm } 
{\cal O}_{-,\pm } >.
\nonu
\eea}.
 
Furthermore, there exists the following relation \cite{Ahn1504}
which was found for $N=3$, together with (\ref{tlinear}) and 
(\ref{higherspinone})
\bea
V_{2}^{(1)}(z) & = & 
4 \left[ {\bf W^{(3)}} + 
 \frac{4 (k-N)}{(4N+5+(3N+4)k)} \left(  {\bf T} \, {\bf T^{(1)}} -
\frac{1}{2} \pa^2 {\bf T^{(1)}}
\right)  \right](z)
\nonu \\
& = & 
4 \left[ {\bf W^{(3)}} + 
 \frac{4 (k-N)}{(4N+5+(3N+4)k)}   {\bf T^{(1)}} \, {\bf T}    \right](z).  
\label{spin3change}
\eea
This holds for any value of $N$ which is obtained 
by varying the $N$ values.
Note that the $N,k$ dependence in the second term is very simple.
In the last line of (\ref{spin3change}), the derivative term  
by changing the commutator \cite{BS} in 
$
\left[  {\bf T^{(1)}} , \, {\bf T}   \right] (z)
 =  -\frac{1}{2} \pa^2 {\bf T^{(1)}} (z)$
is used in order
to simplify the zero mode eigenvalue equation. 
For $N=k$, the above spin-$3$ current
$V_{2}^{(1)}(z)$ becomes a primary current under the stress energy tensor 
(\ref{tlinear}).  

It turns out that one obtains very simple eigenvalue equations
\bea
\left[ V_{2}^{(1)} \right]_0 |(f;0)> 
  & = & 
   \left[ \frac{8N(2N+k+3)}{3(N+k+2)^2} \right] |(f;0)>,
   \nonu \\
\left[ V_{2}^{(1)} \right]_0 |(0;f)> 
  & = & 
-\left[\frac{8k(2k+N+3)}{3(N+k+2)^2}\right] |(0;f)>.
    \label{resres}
\eea
Note that one can also check the correctness of (\ref{w3linear})
by looking at (\ref{resres}) and (\ref{spin3change}).
The reason is as follows: 
Because the zero mode of ${\bf T^{(1)}} {\bf T}$ acting on the 
two states are known for general $N$ and $k$. 
The coefficient in the second term of (\ref{spin3change})
depends on $N,k$ explicitly. 
By combining these two contributions, one obtains (\ref{w3linear})
exactly. 
 
From this (\ref{resres}), one has the following relation
\bea
\left( \left[ V_{2}^{(1)} \right]_0 |(f;0)> \right)_{N \leftrightarrow k,
0 \leftrightarrow f } 
=- \left[ V_{2}^{(1)} \right]_0 |(0;f)>. 
\label{above}
\eea
Therefore, in this particular basis, 
there exists a $N \leftrightarrow k$ symmetry (\ref{above})
up to signs.

The large $N$ 't Hooft limit (\ref{limit}) in (\ref{resres})
leads to the following result 
\bea
\left[ V_{2}^{(1)} \right]_0 |(f;0)> 
   & = &  \frac{8}{3} \lambda (1+\lambda)|(f;0)>,
  \nonu \\
  \left[ V_{2}^{(1)} \right]_0 |(0;f)> 
   & = &   -\frac{8}{3} (1-\lambda) (2-\lambda)|(0;f)>.
\label{largeexp}
\eea
Note the four times 
eigenvalues of ${\bf W}^{(3)}_0$ are the same as those of 
$V_{2}^{(1)}$ in the large $N$ 't Hooft limit (\ref{largeexp}).
 The second term in the bracket of (\ref{spin3change}) does not contribute 
to the eigenvalue equation  in the large $N$ 't Hooft limit
\footnote{
One also sees the following result
\bea
\left[ V_{2}^{(1)} \right]_0 |(f;f)> 
 = \left[ \frac{16(N-k)}{3(N+k+2)^2} \right] |(f;f)> 
\rightarrow \frac{16 \lambda (2 \lambda-1)}{3N} |(f;f)> \rightarrow 0. 
\nonu
\eea
}.

One can also calculate the eigenvalue equations for the 
sum of the square of spin-$2$ currents
acting on the two states with (\ref{rel})
and it turns out that
\bea
\left[ \sum_{i=1}^3 V_{1}^{(1) +i} V_{1}^{(1) +i}  \right]_0 |(f;0)>
& = & - \left[ \frac{12N (5N+4k+4)}{(N+k+2)^2} \right]|(f;0)>,
\nonu \\ 
\left[ \sum_{i=1}^3 V_{1}^{(1) +i} V_{1}^{(1) +i}  \right]_0 |(0;f)>
& = & - \left[ \frac{48 k}{(N+k+2)^2} \right] |(0;f)>,
\nonu \\
\left[ \sum_{i=1}^3 V_{1}^{(1) -i} V_{1}^{(1) -i}  \right]_0 |(f;0)>
& = & - \left[ \frac{48 N }{(N+k+2)^2} \right]|(f;0)>,
\nonu 
\\
\left[ \sum_{i=1}^3 V_{1}^{(1) -i} V_{1}^{(1) -i}  \right]_0 |(0;f)>
& = & - \left[ \frac{12k (5k+4N+4)}{(N+k+2)^2} \right] |(0;f)>.
\label{last}
\eea
One can 
interpret 
the eigenvalues, 
$-\frac{12N \times 2}{(N+k+2)^2}$, $-\frac{24k}{(N+k+2)^2}$,
$-\frac{24N}{(N+k+2)^2}$ and $\frac{-12k \times 2 }{(N+k+2)^2}$
appearing in the right hand side of (\ref{last})
come from the extra terms between the 
left hand side of (\ref{spin2rel}) and 
the left hand side of (\ref{last}).
There is a $N \leftrightarrow k $ symmetry between the first and the last
(and also the second and the third) in (\ref{last}).
One also has the large $N$ 't Hooft limits 
as follows:
$-12 \lambda (4+ \lambda)$, $-\frac{48 \lambda (1- \lambda)}{N}$, 
$-\frac{48 \lambda^2 }{N}$ and $ -12 (1-\lambda) (5- \lambda)$ 
respectively.
In the footnote \ref{footfoot}, the similar calculations
were done for the spin-$1$ currents.
Here the conformal dimensions of $V_1^{(1) \pm i}(z)$ are 
given by two and the sum over 
$SU(2)$ indices in the quadratic of the higher spin-$2$ currents is taken. 
The zero mode eigenvalue equations depend on $N$ and $k$. 
It would be interesting to study the representation theory 
concerning on the higher spin currents further by generalizing 
the previous works in \cite{npb1989}.
Under the large $N$ 't Hooft limit, 
the nonzero eigenvalue corresponding to 
the quadratic higher spin-$2$ currents $V_1^{(1),+i}$
appears in the state $|(f;0)>$ while 
the nonzero eigenvalue corresponding to 
the quadratic higher spin-$2$ currents $V_1^{(1),-i}$
appears in the state $|(0;f)>$.
 The corresponding three point functions can be described 
without any difficulty  
\footnote{
For the light state, one obtains
$\left[ \sum_{i=1}^3 V_{1}^{(1) +i} V_{1}^{(1) +i}  \right]_0 |(f;f)>
 =  - \left[ \frac{48 (2k+1)}{(N+k+2)^2} \right]|(f;f)>
\ra -\frac{96 \lambda(1-\lambda) }{N} |(f;f)> 
$ and
$\left[ \sum_{i=1}^3 V_{1}^{(1) -i} V_{1}^{(1) -i}  \right]_0 |(f;f)>
 =  - \left[ \frac{48 (2N+1)}{(N+k+2)^2} \right]|(f;f)>
\ra -\frac{96 \lambda^2 }{N} |(f;f)> 
$
and the corresponding relations in the nonlinear version 
are given by
$
\left[ \sum_{i=1}^3 \tilde{V}_{1}^{(1) +i} \tilde{V}_{1}^{(1) +i}  \right]_0 |(f;f)>
 =  - \left[ \frac{96 k}{(N+k+2)^2} \right]|(f;f)>
\ra -\frac{96 \lambda(1-\lambda) }{N} |(f;f)> 
$ and 
$\left[ \sum_{i=1}^3 \tilde{V}_{1}^{(1) -i} \tilde{V}_{1}^{(1) -i}  \right]_0 |(f;f)>
 =  - \left[ \frac{96 N}{(N+k+2)^2} \right] |(f;f)>
\ra -\frac{96 \lambda^2 }{N} |(f;f)> 
$. All of these go to zero under the large $N$ 't Hooft limit.}.

From the eigenvalues 
\bea
 {\bf u} (0;[2,0, \cdots, 0])=  {\bf u} (0;[0,1, 0, \cdots, 0])=  \sqrt{ \frac{N+2}{N}},
\label{recentexp}
\eea
which can be obtained from (\ref{highereigen})
and the spin-$1$ current ${\bf U}(z)$,
one can find the conformal dimensions in the linear version
together with (\ref{recentexp})
\bea
h' (0;[2,0, \cdots, 0])&=&h  (0;[2,0, \cdots, 0]) +\frac{1}{(k+N+2)}  {\bf u}^2 (0;[2,0, \cdots, 0])
\nonu \\
& = &  \frac{(Nk+N+2)}{N(k+N+2)},
\nonu \\
h' (0;[0,1, 0, \cdots, 0])&=&h  (0;[0,1, 0, \cdots, 0]) 
+\frac{1}{(k+N+2)}  {\bf u}^2 (0;[0,1, 0, \cdots, 0])
\nonu \\
& = & \frac{(Nk+3N+2)}{N(k+N+2)}.
\label{hprimeexpexp}
\eea
Note that in (\ref{hprimeexpexp}), 
the differences between the conformal dimensions 
in the nonlinear and linear versions can be seen 
from $(C.8)$ and $(C.9)$ of \cite{GG1305}.
Similarly, from the state (\ref{state}), one obtains
the following eigenvalue
\bea
{\bf u} (0;[0^{p-1}, 1, 0, \cdots, 0]) = \frac{p}{2}  \sqrt{ \frac{N+2}{N}},
\label{Eigeneigeneigen}
\eea
and the following conformal dimension together with 
(\ref{Eigeneigeneigen}) can be obtained
\bea
h'(0;[0^{p-1}, 1, 0, \cdots, 0])&=&h  (0;[0^{p-1}, 1, 0, \cdots, 0])
 +\frac{1}{(k+N+2)}  {\bf u}^2 (0;[0^{p-1}, 1, 0, \cdots, 0])
\nonu \\
& = &
 \frac{p(Np+Nk+N+p)}{2N(k+N+2)}.
 \label{explastlast} 
\eea
Also in this case (\ref{explastlast}),
the difference between the conformal dimensions
can be seen from the last equation of \cite{GG1305}
\footnote{From the relation
\bea
\left[ -{\bf U U} \right]_0 |(f;\bar{f})>
&=& \left[ {\bf u}^2 (f;0) + {\bf u}^2 (0; \bar{f}) 
+ \mbox{diag} \left( -\frac{1}{N}, \cdots, -\frac{1}{N}, \frac{1}{2}, \frac{1}{2}\right)
\right]  |(f;\bar{f})>
\nonu  \\
&=& \left[ \frac{(N+1)^2}{N(N+2)} \right]  |(f;\bar{f})>,
\nonu
\eea
where the eigenvalue 
${\bf u} (0; \bar{f}) = -\frac{1}{2} \sqrt{\frac{N+2}{N}}$
is used, 
one obtains the following conformal dimension
$
h'(f;\bar{f})= h(f;\bar{f}) + \frac{1}{(k+N+2)} \frac{(N+1)^2}{N(N+2)} 
=\frac{1}{2} + \frac{(N+1)^2}{N(N+2)(N+k+2)}$.  
The previous result (\ref{half})
is substituted.}.

Therefore, in this section, the three point functions can be 
summarized by (\ref{Expexp}), 
(\ref{reducedexp}) and  (\ref{reducedexpexp}).  
As in the nonlinear version one obtains the following ratios for the 
three point functions  
\bea
\frac{<\overline{{\cal O}}_{+ } 
{\cal O}_{+ } {\bf T^{(1)}}>}{< \overline{{\cal O}}_{- } 
{\cal O}_{- } {\bf T^{(1)}}>}
 &= & 
 \left[\frac{\lambda}{1-\lambda} \right],
\nonu \\
\frac{<\overline{{\cal O}}_{+ } 
{\cal O}_{+ } {\bf T^{(2)}}>}{< \overline{{\cal O}}_{- } 
{\cal O}_{- } {\bf T^{(2)}}>}
 &= & 
 \left[\frac{\lambda}{1-\lambda} \right],
\nonu \\
 \frac{< \overline{{\cal O}}_{+ } 
{\cal O}_{+ } {\bf {W}^{(2)}}>}{< \overline{{\cal O}}_{- } 
{\cal O}_{- }
{\bf {W}^{(2)}}>}
& = &  - \left[ \frac{\lambda}{1-\lambda} \right],
\nonu \\
\frac{<\overline{{\cal O}}_{+ } 
{\cal O}_{+ } {\bf W^{(3)}}> }{< \overline{{\cal O}}_{- } 
{\cal O}_{- }  {\bf W^{(3)}}> }
&=&- \left[ \frac{\lambda(1+\lambda)}{(1-\lambda)(2-\lambda)}
\right].
\nonu
\eea
These are exactly the same as the ones in (\ref{threethree}).
Under the large $N$ 't Hooft limit, 
the ratios of the three point functions in the nonlinear and linear 
versions are equivalent to each other. 

\section{Conclusions and outlook }

In this paper, the three point functions in 
the large ${\cal N}=4$ holography in the large $N$ 't Hooft limit 
are obtained. In the three point functions, 
$\mbox{scalar}-\overline{\mbox{scalar}}-\mbox{current}$, 
the two scalars are characterized by the coset primaries
corresponding to the two states $|(0;f)>$ and $|(f;0)>$ 
in the minimal representations of the coset. 
The currents are given by the spin-$1$ currents, 
spin-$2$ current of the large ${\cal N}=4$
(non)linear superconformal algebra, the higher spin-$1$ current, 
the higher spin-$2$ currents and 
the higher spin-$3$ current. 

$\bullet$ Three point functions in the bulk

As in \cite{AK1308}, it is an open problem to obtain 
the asymptotic symmetry algebra of the higher spin theory 
on the $AdS_3$ space. Once this is found explicitly, then 
one can compare the results of this paper with the corresponding three point 
functions which can be obtained indirectly in the bulk.

$\bullet$  The general $s$ dependence of three point function 

From the three point functions with spins $2, 3$, one can 
expect that the ratio of three point function for given spin $s$
looks like as $\frac{\la (1+\la ) \cdots (s-2 +\la)}{(1-\la)(2-\la) \cdots 
(s-1-\la)}$ up to signs.
It is an open problem to see whether this is general behavior or not.  
For the higher spin-$4$ current, one can apply the present method to 
the orthogonal coset theory where one has 
the higher spin-$4$ current in the lowest ${\cal N}=4$ higher spin current.
It is a good exercise to see whether one sees 
the above three point function with $s=4$ and determines whether 
the behavior looks like the one in \cite{Ahn1111}. 

$\bullet$ An extension of small ${\cal N}=4$ linear superconformal algebra

As described in the introduction, 
the small ${\cal N}=4$ linear superconformal algebra 
can be obtained by taking the large level limit in the large 
${\cal N}=4$ linear superconformal algebra.
Then one can obtain an extension of the small ${\cal N}=4$ linear
superconformal algebra from the extension of large 
${\cal N}=4$ linear superconformal algebra.  
Therefore, in this construction, the complete OPEs 
between the $16$ currents of 
large ${\cal N}=4$ linear superconformal algebra 
and the $16$ lowest higher spin currents for general $N$ and $k$
should be obtained.

$\bullet$ Oscillator formalism for the higher spin currents

According to the original Vasiliev's oscillator formalism, some part of 
calculations in the higher spin currents of large 
${\cal N}=4$ nonlinear superconformal algebra  
are obtained in \cite{GG1305}.
It is an open problem to see whether one can see the oscillator formalism
in an extension of the large ${\cal N}=4$ linear superconformal algebra.   

$\bullet$ The operator product expansion of the $16$ higher spin currents
in ${\cal N}=4$ superspace 

So far, the complete OPEs between the $16$ currents and the $16$ higher spin
currents for general $N$ and $k$ are not known although its nonlinear 
version appears in \cite{BCG1404} where the coset field realizations 
are not checked. 
The first step is to write down the complete OPEs in an extension of
large ${\cal N}=4$ linear superconformal algebra in the ${\cal N}=4$
superspace because it is more plausible to consider the linear version 
rather than the nonlinear version. 

$\bullet$ Three point functions in the coset theory 
which contains an orthogonal Wolf space

One can also consider the different large ${\cal N}=4$ holography
based on the orthogonal Wolf space \cite{AP1410}. 
The nonlinear version contains the Wolf space 
$\frac{SO(N+4)}{SO(N) \times SU(2) \times SU(2)}$
while the linear version contains 
the coset $\frac{SO(N+4)}{SO(N) \times SU(2)} \times U(1)$.
In this case, the minimal representations 
contain the two states $|(0;v)>$ and $|(v;0)>$
where the former is the vector representation in the $SO(N)$
among the singlets in the $SO(N+4)$ and the latter is the vector 
representation in the $SO(N+4)$ and at the same time is the singlet under
the $SO(N)$. The relevant previous works on this direction are 
given in \cite{Ahn1106,GV1106,Ahn1202,AP1310}.

$\bullet$ The next $16$ higher spin currents 

So far, the higher spin currents in the context of the three point function
are the member of the $16$ lowest higher spin currents.
One can consider the next $16$ higher spin currents where 
the bosonic currents contain the higher spin currents with spins $2, 3, 4$.
One would like to see the behaviors of the three point functions 
and satisfy whether they behave as above. 
One expects that as the spin increases, the $N$ dependence for several $N$ 
in the fractional 
coefficient functions in the level $k$ becomes complicated. 
In order to extract the general $N$ behavior, 
one needs more information about the OPEs for the several $N$. 
Furthermore, the basis in \cite{BCG1404} is more useful because 
they already presented the defining OPEs between the $16$ currents 
in the large ${\cal N}=4$ linear superconformal algebra and the 
next $16$ higher spin currents. 

$\bullet$ Three point functions involving the fermionic (higher spin)
currents

In \cite{MZ1211}, the three point functions which contain 
the fermionic (higher spin) currents have been described.
See also \cite{CHR1211}.
In this paper, we have only considered the bosonic (higher spin) currents
in the three-point functions.
It would be interesting to discover the three point functions 
with fermionic (higher spin) currents explicitly.

\vspace{.7cm}

\centerline{\bf Acknowledgments}

We would like to thank 
R. Gopakumar and C. Peng for discussions. 
This work was supported by the Mid-career Researcher Program through
the National Research Foundation of Korea (NRF) grant 
funded by the Korean government (MEST) 
(No. 2012-045385/2013-056327/2014-051185).
CA acknowledges warm hospitality from 
the School of  Liberal Arts (and Institute of Convergence Fundamental
Studies), Seoul National University of Science and Technology.

\newpage

\appendix

\renewcommand{\thesection}{\large \bf \mbox{Appendix~}\Alph{section}}
\renewcommand{\theequation}{\Alph{section}\mbox{.}\arabic{equation}}

\section{ The generators of $SU(N+2)$ in complex basis }

The $SU(N+2)$ generators can be expressed in the complex basis.
There exist $(N+1)$ Cartan generators in $SU(N+2)$ denoted by 
$H_1, H_2, \cdots, H_{N+1}$ which are defined by \cite{Georgi}
\bea
[H_m]_{ij}=\frac{1}{\sqrt{2m(m+1)}} 
\left( \sum_{k=1}^m \delta_{ik} \delta_{jk} - m \delta_{i,m+1} \delta_{j,m+1}
\right).
\label{matrixH}
\eea 
Then one can define the $\frac{(N+1)}{2}$ diagonal generators with 
(\ref{matrixH}) as follows:
\bea
T_{p+1} & = & i H_1+ H_2,
\nonu \\
T_{p+2} & = & i H_3+ H_4,
\nonu \\
\vdots
\nonu \\
T_{p+\frac{(N+1)}{2}} & = & i H_N + H_{N+1},
\label{Tss}
\eea
where one introduces $p \equiv \frac{(N+2)^2-1}{2} -\frac{(N+1)}{2}$. 
The last generator in (\ref{Tss}) has a subscript $\frac{(N+2)^2-1}{2}$
which is the last element of adjoint index in complex basis.
The remaining $\frac{(N+1)}{2}$ 
half of diagonal matrix can be obtained by taking the complex conjugation
from (\ref{Tss}).

Among off-diagonal matrices, the $(2N+1)$ matrices have nonzero elements  
as follows:
\bea
T_1 & = & \left(\begin{array}{rrrrr|rr}
 &&&&&0&0  \\ 
 &&&&&0&0  \\
 &&0&&&\vdots&\vdots   \\
 &&&&&0&0  \\
 &&&&&0&0  \\ \hline
 1&0&\cdots&0&0&0&0  \\ 
 0&0&\cdots&0&0&0&0 \\ 
\end{array}\right),
\qquad
T_2 = \left(\begin{array}{rrrrr|rr}
 &&&&&0&0  \\ 
 &&&&&0&0  \\
 &&0&&&\vdots&\vdots   \\
 &&&&&0&0  \\
 &&&&&0&0  \\ \hline
 0&1&\cdots&0&0&0&0  \\ 
 0&0&\cdots&0&0&0&0 \\ 
\end{array}\right), \cdots
\nonu \\
T_N & = &  \left(\begin{array}{rrrrr|rr}
 &&&&&0&0  \\ 
 &&&&&0&0  \\
 &&0&&&\vdots&\vdots   \\
 &&&&&0&0  \\
 &&&&&0&0  \\ \hline
 0&0&\cdots&0&1&0&0  \\ 
 0&0&\cdots&0&0&0&0 \\ 
\end{array}\right), 
\qquad
T_{N+1} = \left(\begin{array}{rrrrr|rr}
 &&&&&0&0  \\ 
 &&&&&0&0  \\
 &&0&&&\vdots&\vdots   \\
 &&&&&0&0  \\
 &&&&&0&0  \\ \hline
 0&0&\cdots&0&0&0&0  \\ 
 1&0&\cdots&0&0&0&0 \\ 
\end{array}\right),
\nonu \\
T_{N+2} & = & \left(\begin{array}{rrrrr|rr}
 &&&&&0&0  \\ 
 &&&&&0&0  \\
 &&0&&&\vdots&\vdots   \\
 &&&&&0&0  \\
 &&&&&0&0  \\ \hline
 0&0&\cdots&0&0&0&0  \\ 
 0&1&\cdots&0&0&0&0 \\ 
\end{array}\right), \cdots,
\qquad
T_{2N} = \left(\begin{array}{rrrrr|rr}
 &&&&&0&0  \\ 
 &&&&&0&0  \\
 &&0&&&\vdots&\vdots   \\
 &&&&&0&0  \\
 &&&&&0&0  \\ \hline
 0&0&\cdots&0&0&0&0  \\ 
 0&0&\cdots&0&1&0&0 \\ 
\end{array}\right),
\nonu \\
T_{2N+1} & = & \left(\begin{array}{rrrrr|rr}
 &&&&&0&0  \\ 
 &&&&&0&0  \\
 &&0&&&\vdots&\vdots   \\
 &&&&&0&0  \\
 &&&&&0&0  \\ \hline
 0&0&\cdots&0&0&0&0  \\ 
 0&0&\cdots&0&0&1&0 \\ 
\end{array}\right).
\label{gengen}
\eea
The remaining $(2N+1)$ off diagonal matrices 
can be obtained by taking the transpose for 
these matrices (\ref{gengen}). 
Then 
the $4N$ coset generators in the nonlinear version are given by
(\ref{gengen})
\bea
T_1, \, T_2, \, \cdots, \,  T_{2N}, \, T^{\dagger}_{1}(=T_{1^\ast}), 
\, T^{\dagger}_{2}(=T_{2^\ast}), \, \cdots \, , \, T^{\dagger}_{2N}(=T_{2N^\ast}).
\label{one}
\eea
The $(4N+4)$ coset generators in the linear version are given by
(\ref{Tss}) and (\ref{gengen})
\bea
T_1,T_2,\cdots,  T_{2N}, T^{\dagger}_{1}, T^{\dagger}_{2}, \cdots, 
T^{\dagger}_{2N};
T_{2N+1}, T^{\dagger}_{2N+1}, T_{q}, T^{\dagger}_{q},
\,\,\, q \equiv p+\frac{(N+1)}{2} = \frac{(N+2)^2-1}{2}.
\label{two}
\eea
That is, 
compared to (\ref{one}) and (\ref{two}),
the extra four generators are given by 
$T_{2N+1}$ and $T_q$ (and their conjugated ones).
Then the remaining $\frac{N(N-1)}{2}$ off-diagonal 
generators live in the lower half triangle matrix of $N \times N$
matrix. Similarly, the same number of off-diagonal generators 
live in the upper triangle matrix (by conjugation).
Note that the sum of the numbers $N(N-1)$ and $2(2N+1)$ is equal to
the difference of $(N+2)^2-1$ and $(N+1)$ as expected. 




The metric is  
\bea
g_{ab} = \mbox{Tr} (T_a T_b)=
\left(
\begin{array}{cc}
0 & 1 \\
1 & 0 \\
\end{array}
\right), 
 a,b=1,2, \cdots, \frac{(N+2)^2-1}{2}, 1^{\ast}, 2^{\ast}, \cdots, 
(\frac{(N+2)^2-1}{2})^{\ast}.
\label{metric}
\eea
This is consistent with the description of subsection $2.1$.
 The nonvanishing metric components in (\ref{metric}) are
\bea
 g_{ A A^\ast }=g_{  A^\ast A }=1,
\label{metricappendix}
\eea 
where $A=1,2,\cdots,\frac{(N+2)^2-1}{2} $.
For convenience, let us write down 
the four Cartan generators 
$T_{11}$ and $T_{12}$ ($T_{11^{\ast}}$ and $T_{12^{\ast}}$) 
in $SU(5)$. One obtains
$T_{11} =\mbox{diag} (\frac{1}{2 \sqrt{3}}+\frac{i}{2}, \frac{1}{2 \sqrt{3}}-\frac{i}{2}, -\frac{1}{\sqrt{3}}, 0, 0)$ and
$T_{12} =\mbox{diag} (\frac{1}{2 \sqrt{10}}+\frac{i}{2 \sqrt{6}},
\frac{1}{2 \sqrt{10}}+\frac{i}{2 \sqrt{6}},
\frac{1}{2 \sqrt{10}}+\frac{i}{2 \sqrt{6}},
\frac{1}{2 \sqrt{10}}-\frac{1}{2} i \sqrt{\frac{3}{2}},
-\sqrt{\frac{2}{5}})$ and their conjugated ones ($T_{11^{\ast}}$ and 
$T_{12^{\ast}}$). 
The remaining $7$ generators (and $7$ conjugated ones) 
can be obtained from (\ref{gengen}) and the $3$ generators
(and $3$ conjugated ones) live in the $3 \times 3$ matrix.   

\section{ The large 
$\mathcal N = 4$ nonlinear superconformal algebra}

The large 
$\mathcal N = 4$ nonlinear superconformal algebra
can be summarized by $11$ currents as follows: 
one spin-$2$ current $T(z)$, four spin-$\frac{3}{2}$ currents $G^{\mu}(z)$, 
six 
spin-$1$ currents $A^{\pm i}(z)$.
The explicit OPEs are given by \cite{GS}
\bea 
T(z) \, T(w) &=& \frac{1}{(z-w)^4} \, \frac{\hat{c}}{2} +
\frac{1}{(z-w)^2} \, 2 T(w) +\frac{1}{(z-w)} \, \pa T(w) +\cdots,
\nonu \\
T(z) \, \phi(w) & = & \frac{1}{(z-w)^2}
\, h_{\phi} \, \phi(w) + \frac{1}{(z-w)} \, \pa \phi(w) + \cdots,
\nonu \\
G^{\mu}(z) \, G^{\nu}(w) &=& \frac{1}{(z-w)^3} \,
\frac{2}{3} \delta^{\mu \nu} c_{\mbox{Wolf}}
- \frac{1}{(z-w)^2} \, \frac{8}{(k+N+2)} ( N \, \alpha^{+i}_{\mu \nu} \, A^{+}_i      
+k \, \alpha^{-i}_{\mu \nu} \, A^{-}_i  )(w)
\nonu \\
& + & \frac{1}{(z-w)} \, \left[ 2 \delta^{\mu \nu} T
-  \frac{4}{(k+N+2)} \pa ( N \, \alpha^{+i}_{\mu \nu} \, A^{+}_i      
+k \, \alpha^{-i}_{\mu \nu} \, A^{-}_i  ) \right.
\label{N4scalgebra} \\
&-& \left.  \frac{8}{(k+N+2)}  ( \alpha^{+i} A^{+}_i      
- \alpha^{-i} A^{-}_i  )_{\rho (\mu}
( \alpha^{+j} A^{+}_j      
- \alpha^{-j} A^{-}_j  )_{\nu)}^{\,\,\,\,\, \rho} \right](w) +\cdots,
\nonu \\
A^{\pm i}(z) \, G^{\mu} (w) &=& 
\frac{1}{(z-w)} \, \alpha^{\pm i}_{\mu \nu} \, G^{\nu} (w)
+\cdots,
\nonu \\
A^{\pm i}(z) \, A^{\pm j}(w) &=& 
-\frac{1}{(z-w)^2} \, \frac{1}{2} \, \delta^{ij} \, \hat{k}^{\pm} 
+ \frac{1}{(z-w)} \, \epsilon^{ijk}  A^{\pm k} (w) + \cdots,
\nonu
\eea
where two central charges  and
two levels of $SU(2)_{ \hat{k}^{+} } \times SU(2)_{\hat{k}^{-} }$
are given by
\bea
\hat{c}=\frac{3(k+N+2k N)}{(k+N+2)},
\quad c_{\mbox{Wolf}}   =  
\frac{6 k N}{(2+k+N)},
\quad
\hat{k}^{+}=k,  
\quad
\hat{k}^{-}=N.
\label{character}
\eea
The conformal dimensions for the currents  are 
$h_{A^{\pm i}}=1$, $h_{G^{\mu}}=\frac{3}{2}$.
The quantity $\alpha^{\pm i} $ is defined by
\bea
\alpha^{\pm i}_{\mu \nu} &=&
\frac{1}{2} \left( \pm \delta_{i \mu} \delta_{ \nu 0} \mp \delta_{i \nu} \delta_{ \mu 0}
+\ep_{i \mu \nu} \right).
\nonu
\eea
Note that the nonlinear terms in the OPEs between the spin-$\frac{3}{2}$
currents occur.
The two levels are given by $k$ and $N$ in (\ref{character})
which are the quantities in the Wolf space coset.
The equation (\ref{11currents}) is an explicit coset realization of 
the large ${\cal N}=4$ nonlinear superconformal algebra.

\section{ The large 
$\mathcal N = 4$ linear superconformal algebra}

For convenience, the full expressions for the large ${\cal N}=4$
linear superconformal algebra can be summarized by \cite{npb1988}
\bea
{\bf T}(z) \, {\bf T}(w) & = & \frac{1}{(z-w)^4} \, \frac{c}{2} +\frac{1}{(z-w)^2}
\, 2 {\bf T}(w) + \frac{1}{(z-w)} \, \pa {\bf T}(w) + \cdots,
\nonu \\
{\bf T} (z) \, \phi (w) & = & \frac{1}{(z-w)^2}
\, h_{\phi} \, \phi(w) + \frac{1}{(z-w)} \, \pa \phi(w) + \cdots,
\nonu \\
{\bf G}^{\mu}(z) \, {\bf G}^{\nu}(w) &=& \frac{1}{(z-w)^3} \,
\frac{2}{3} \delta^{\mu \nu} c
- \frac{1}{(z-w)^2} \, \frac{8}{(k^{+}+k^{-})} ( k^{-} \, \alpha^{+i}_{\mu \nu} \, {\bf A}^{+}_i      
+k^{+} \, \alpha^{-i}_{\mu \nu} \, {\bf A}^{-}_i  )(w)
\nonu \\
& + & \frac{1}{(z-w)} \, \left[ 2 \delta^{\mu \nu} {\bf T}
-  \frac{4}{(k^{+}+k^{-})} 
\pa ( k^{-} \, \alpha^{+i}_{\mu \nu} \, {\bf A}^{+}_i      
+k^{+} \, \alpha^{-i}_{\mu \nu} \, {\bf A}^{-}_i  ) \right](w) +\cdots,
\nonu \\
{\bf A}^{\pm i}(z) \, {\bf G}^{\mu }(w) & = & \mp \frac{1}{(z-w)^2} \,  \frac{2k^{\pm}}{(k^{+}+k^{-})}  
\,  \alpha_{\mu \nu}^{\pm i } \, 
{\bf \Gamma}^{\nu }(w) + \frac{1}{(z-w)} \,  \alpha_{\mu \nu }^{\pm i} \, 
{\bf G}^{\nu}(w) + \cdots,
\nonu \\
{\bf A}^{\pm i}(z) \, {\bf A}^{\pm j}(w) & = & - \frac{1}{(z-w)^2} \, \frac{1}{2} k^{\pm} 
\delta^{ij}  +\frac{1}{(z-w)} \, \ep^{ijk} {\bf A}^{\pm k }(w) +\cdots,
\nonu \\
{\bf \Gamma}^{\mu }(z) \, {\bf G}^{\nu}(w) & = & \frac{1}{(z-w)} \, 
 \left[ 2 ( \alpha_{\mu \nu}^{+i} {\bf A}_i^{+}
- \alpha_{\mu \nu}^{-i} {\bf A}_i^{-}) + \delta^{\mu \nu} \, {\bf U}  \right](w) 
+\cdots,
\nonu \\
{\bf A}^{\pm i}(z) \, {\bf \Gamma}^{\mu}(w) & = & \frac{1}{(z-w)} \,  
\alpha_{\mu \nu}^{\pm i} \, 
{\bf \Gamma}^{\nu}(w)+ \cdots,
\nonu \\
{\bf U}(z) \, {\bf G}^{\mu}(w) & = & \frac{1}{(z-w)^2} \, {\bf \Gamma}^{\mu}(w)
 + \cdots,
\nonu \\
{\bf \Gamma}^{\mu}(z) \, {\bf \Gamma}^{\nu}(w) & = & -\frac{1}{(z-w)} \,
 \frac{(k^{+}+k^{-})}{2} \, \delta^{\mu \nu} + \cdots, 
\nonu \\
{\bf U}(z) \, {\bf U}(w)  & = & -\frac{1}{(z-w)^2} \, \frac{(k^{+}+k^{-})}{2}
+ \cdots,
\label{N4linearalg}
\eea
where the indices run over $\mu,\nu=0,1,2,3$ and 
$i,j,k=1,2,3$.
The conformal dimensions are given by 
$h_{{\bf \Gamma}^{\mu}}=\frac{1}{2}$, 
$h_{{\bf A}^{\pm i}}=1$,
$h_{{\bf U}}=1$ and 
$h_{{\bf G}^{\mu}}=\frac{3}{2}$.
The levels are given by 
$k^{+}=k+1$ and $k^{-}=N+1$.
The central charge is given by 
 $c=\hat{c}+3=\frac{6(k+1)(N+1)}{(N+k+2)}$.
The equations (\ref{tlinear}), (\ref{linearspin3half}), (\ref{aplusminus}),
(\ref{Gamma}) and (\ref{ulinear}) are explicit coset realization 
 of the large ${\cal N}=4$ linear superconformal algebra.
Compared to the nonlinear version in Appendix $B$, 
there exist four spin-$\frac{1}{2}$ currents ${\bf \Gamma^{\mu}}(z)$ 
and spin-$1$ current ${\bf U}(z)$.
The explicit relations between the 
$11$ currents in the nonlinear and linear versions will 
appear in Appendix $F$. 

One can introduce the $U(1)$ charge for the ${\cal N}=2$ superconformal 
algebra  by reading off the 
second order pole of ${\bf G}^{\mu}(z) \, {\bf G}^{\nu}(w)$
as follows:
\bea
U_{{\cal N}=2}(z) \equiv -2 i \left( \gamma {\bf A}^{+3}
+(1-\gamma) {\bf A}^{-3} \right) (z), \qquad 
\gamma \equiv \frac{(N+1)}{(k+N+2)}.
\label{chargeu}
\eea
Then one obtains the following OPEs with (\ref{chargeu})
where one can read off the corresponding $U(1)$ charges
\bea
U_{{\cal N}=2}(z) \, Q^{\bar{A}}(w) & = & 
\frac{1}{(z-w)}  (1-\gamma)  Q^{\bar{A}}(w)+\cdots, 
(\bar{A}=1,2, \cdots, N),
\nonu \\
U_{{\cal N}=2}(z) \, Q^{\bar{A}}(w) & = & 
-\frac{1}{(z-w)}  (1-\gamma)  Q^{\bar{A}}(w)+\cdots, 
(\bar{A}=N+1, N+2, \cdots, 2N),
\nonu \\
U_{{\cal N}=2}(z) \, Q^{\bar{A}^*}(w) & = & 
-\frac{1}{(z-w)} (1-\gamma)  Q^{\bar{A}^*}(w)+\cdots, 
(\bar{A}^*=1^*,2^*, \cdots, N^*),
\nonu \\
U_{{\cal N}=2}(z) \, Q^{\bar{A}^*}(w) & = & 
\frac{1}{(z-w)}  (1-\gamma)  Q^{\bar{A}^*}(w)+\cdots, 
(\bar{A}^*=(N+1)^*, (N+2)^*, \cdots, 2N^*),
\nonu \\
U_{{\cal N}=2}(z) \, V^{\bar{A}}(w) & = & 
\frac{1}{(z-w)} \gamma  V^{\bar{A}}(w)+\cdots, 
(\bar{A}=1,2, \cdots, N),
\nonu \\
U_{{\cal N}=2}(z) \, V^{\bar{A}}(w) & = & 
-\frac{1}{(z-w)}  \gamma V^{\bar{A}}(w)+\cdots, 
(\bar{A}=N+1, N+2, \cdots, 2N),
\nonu \\
U_{{\cal N}=2}(z) \, V^{\bar{A}^*}(w) & = & 
-\frac{1}{(z-w)} \gamma V^{\bar{A}^*}(w)+\cdots, 
(\bar{A}^*=1^*,2^*, \cdots, N^*),
\label{chargecharge}
 \\
U_{{\cal N}=2}(z) \, V^{\bar{A}^*}(w) & = & 
\frac{1}{(z-w)}   \gamma  V^{\bar{A}^*}(w)+\cdots, 
(\bar{A}^*=(N+1)^*, (N+2)^*, \cdots, 2N^*).
\nonu
\eea
From the explicit charges in (\ref{chargecharge}),
one can construct any multiple product of spin-$1$ currents and 
spin-$\frac{1}{2}$ currents for given $U_{{\cal N}=2}$ charge.

\section{ The complex structures in the extension of large ${\cal N}=4$
linear superconformal algebra  }

One can represent three almost complex structures
$h^{1}_{\tilde{a}  \tilde{b}}$,  $h^{2}_{\tilde{a}  \tilde{b}}$ and 
$h^{3}_{\tilde{a}  \tilde{b}} (\equiv h^{1}_{\tilde{a}  \tilde{c}}
h^{2 \tilde{c}}_{ \,\,\,\,\,\, \tilde{b}} )$ in terms of 
the $(4N+4) \times (4N+4) $ matrix. 
We represent $h^{i}_{\tilde{a}  \tilde{b}}$ using the 
following four block matrices.
\bea
h^{i}_{\tilde{a}  \tilde{b}}=\left(
\begin{array}{c|c}
h^{i}_{\bar{a}  \bar{b}} & h^{i}_{\bar{a}  \hat{b}}=0  \\  
\hline
h^{i}_{\hat{a}  \bar{b}}=0 & h^{i}_{\hat{a}  \hat{b}} \\
\end{array}
\right),
\label{hexpexp}
\eea
where the first block is given by $4N \times 4N$ matrix, the 
second is $4N \times 4$ matrix,
the third is $4 \times 4N$ matrix, and the last block is 
given by $4 \times 4$ matrix.
Note that the coset indices are decomposed into 
$\tilde{a} = (\bar{a}, \hat{a})$ where $\bar{a} =1, 2, \cdots, 2N, 1^{\ast},
2^{\ast}, \cdots, 2N^{\ast}$
and $\hat{a}$ runs over the remaining four indices.


The first $4N \times 4N $ block matrices in (\ref{hexpexp}) 
are exactly the same as
the complex structures
in Wolf space coset as follows: 
\bea
h^1_{\bar{a} \bar{b}} & = & 
\left(
\begin{array}{cccc}
0 & 0  & 0 & -i \\
0 & 0 & -i & 0 \\
0 & i & 0 & 0 \\
i & 0 & 0 & 0 \\
\end{array}
\right), \qquad
h^2_{\bar{a} \bar{b}} = 
\left(
\begin{array}{cccc}
0 & 0  & 0 & 1 \\
0 & 0 & -1 & 0 \\
0 & 1 & 0 & 0 \\
-1 & 0 & 0 & 0 \\
\end{array}
\right), \nonu \\
h^{3}_{\bar{a} \bar{b}}
& \equiv & h^{1}_{\bar{a} \bar{c}}  \, h^{2 \bar{c} }_{\,\,\,\,\,\, \bar{b}}
=  
\left(
\begin{array}{cccc}
0 & 0  & i & 0 \\
0 & 0 & 0 & -i \\
-i & 0 & 0 & 0 \\
0 & i & 0 & 0 \\
\end{array}
\right),
\label{exphexp}
\eea
where each entry in the above matrices (\ref{exphexp}) 
is $N \times N$ matrix.
Recall that the coset index $\bar{a}$ in the nonlinear version 
runs over the $4N$ indices 
(\ref{one}): 
the first $N$ indices are given by $1, 2, \cdots, N$, 
the second $N$ indices are given by $(N+1), (N+2), \cdots, 2N$, 
third $N$ indices are given by $1^\ast, 2^\ast, \cdots, N^\ast$, 
and the last $N$ indices are given by 
$(N+1)^\ast, (N+2)^\ast, \cdots, 2N^\ast$.

Let us represent the last $4 \times 4$ block matrices $h^{i}_{\hat{a}  \hat{b}}$
in (\ref{hexpexp})
as follows:
\bea
h^1_{\hat{a} \hat{b}} & = & 
\left(
\begin{array}{cccc}
0 & 0  & 0 & -\frac{1}{a} \\
0 & 0 & a & 0 \\
0 & -a & 0 & 0 \\
\frac{1}{a} & 0 & 0 & 0 \\
\end{array}
\right), \qquad
h^2_{\hat{a} \hat{b}} = 
\left(
\begin{array}{cccc}
0 & 0  & 0 & \frac{i}{a} \\
0 & 0 & i a & 0 \\
0 & -ia & 0 & 0 \\
-\frac{i}{a} & 0 & 0 & 0 \\
\end{array}
\right), \nonu \\
h^{3}_{\hat{a} \hat{b}}
& \equiv & h^{1}_{\hat{a} \hat{c}}  \, h^{2 \hat{c} }_{\,\,\,\,\,\, \hat{b}}
= 
\left(
\begin{array}{cccc}
0 & 0  & -i & 0 \\
0 & 0 & 0 & i \\
i & 0 & 0 & 0 \\
0 & -i & 0 & 0 \\
\end{array}
\right),
\label{lastexp}
\eea
where the element $a$ in (\ref{lastexp}) which depends on $N$ is 
given by
\bea
a \equiv -\frac{1}{2(N+1)} \left( 
\sqrt{2N(N+1)} + i  \sqrt{2(N+1)(N+2)}
\right).
\label{adef}
\eea
The $\hat{a}$ indices are given by  $(2N+1)$, $q$, $(2N+1)^\ast$ and  
$q^\ast$
where the index $q$ was defined in (\ref{two}).
The $N$ dependences in (\ref{adef}) are rather 
unusual but  they can be fixed by 
the several $N$ cases.

\section{ The defining OPEs for the higher spin currents 
in the extension of large ${\cal N}=4$
linear superconformal algebra  }

The relevant OPEs for the higher spin currents for  general $N$ 
(starting from \cite{Ahn1504})
are given by
\bea
\left(
\begin{array}{c}
{\bf G}_{12} \nonu \\
{\bf G}_{21}  \end{array}
\right)(z) \, {\bf T^{(1)}}(w)
& = & \frac{1}{(z-w)} \left[ \left(
\begin{array}{c}
-{\bf  G}_{12} \nonu \\
{\bf  G}_{21}  \end{array}
\right) + 2 
{\bf T_{\mp}^{(\frac{3}{2})}}
\right](w) +\cdots,
\nonu \\
\left(
\begin{array}{c}
{\bf  G}_{11} \nonu \\
{\bf  G}_{22}  \end{array}
\right)(z) \, {\bf T^{(1)}}(w)
& = & \frac{1}{(z-w)} \left[ \left(
\begin{array}{c}
{\bf  G}_{11} \nonu \\
-{\bf  G}_{22}  \end{array}
\right) + 2 
\left(
\begin{array}{c}
{\bf U^{(\frac{3}{2})}} \nonu \\
{\bf V^{(\frac{3}{2})}}  \end{array}
\right)
\right](w) +\cdots,
\nonu \\
 \left(
\begin{array}{c}
{\bf  G}_{12} \nonu \\
{\bf G}_{21}
 \end{array}
\right)(z) \, {\bf T_{\pm}^{(\frac{3}{2})}}(w)
& = & 
\mp \frac{1}{(z-w)^3} \frac{2(k+1)(N+1)}{(N+k+2)} +
\frac{1}{(z-w)^2} \left[ \frac{2i(N+1)}{(N+k+2)} {\bf A}^{+3} \right.
\nonu \\
&-& \left. \frac{2i(k+1)}{(N+k+2)} {\bf A}^{-3} + {\bf T^{(1)}}
\right](w) 
\nonu \\
& + & 
\frac{1}{(z-w)} \left[  
\frac{1}{2} \pa \, \mbox{(pole-2)}
\mp {\bf T^{(2)}}  \mp {\bf T} \right](w) +
\cdots, 
\nonu \\
\left( 
\begin{array}{c}
{\bf G}_{12} \nonu \\
{\bf G}_{21}
 \end{array}
\right)(z)  \left(
\begin{array}{c}
{\bf U^{(\frac{3}{2})}} \nonu \\
{\bf V^{(\frac{3}{2})}} 
\end{array}
\right)(w)
& = & 
-\frac{1}{(z-w)^2}   \frac{2i(N+1)}{(N+k+2)} {\bf A}^{+ \mp} (w)
\nonu \\
&+& \frac{1}{(z-w)} \left[ 
\frac{1}{2} \pa \, \mbox{(pole-2)}
+  \left(
\begin{array}{c}
{\bf U_{-}^{(2)}} \nonu \\
{\bf V_{+}^{(2)}}
 \end{array}
\right)
\right](w) +\cdots,
\nonu \\
\left( 
\begin{array}{c}
{\bf G}_{12} \nonu \\
{\bf G}_{21}
 \end{array}
\right)(z)  \left(
\begin{array}{c}
{\bf V^{(\frac{3}{2})}} \nonu \\
{\bf U^{(\frac{3}{2})}} 
\end{array}
\right)(w)
& = & 
-\frac{1}{(z-w)^2}   \frac{2i(k+1)}{(N+k+2)} {\bf A}^{- \pm} (w)
\nonu \\
&+& \frac{1}{(z-w)} \left[ 
\frac{1}{2} \pa \, \mbox{(pole-2)}
+  \left(
\begin{array}{c}
{\bf V_{-}^{(2)}} \nonu \\
{\bf U_{+}^{(2)}}
 \end{array}
\right)
\right](w) +\cdots,
\nonu \\
\left( 
\begin{array}{c}
{\bf G}_{11} \nonu \\
{\bf G}_{22}
 \end{array}
\right)(z)  \left(
\begin{array}{c}
{\bf V^{(\frac{3}{2})}} \nonu \\
{\bf U^{(\frac{3}{2})}} 
\end{array}
\right)(w)
& = & 
\pm \frac{1}{(z-w)^3} \frac{2(N+1)(k+1)}{(N+k+2)}
+\frac{1}{(z-w)^2} \left[ - \frac{2i(N+1)}{(N+k+2)} {\bf  A}^{+3}  \right.
\nonu \\ 
&-& \left.
\frac{2i(k+1)}{(N+k+2)} {\bf A}^{-3} + {\bf T^{(1)}} \right](w)
\nonu \\
&+& \frac{1}{(z-w)} \left[  
\frac{1}{2} \pa \, \mbox{(pole-2)}
\pm  {\bf W^{(2)}} \pm  {\bf T} \right](w) +\cdots,
\nonu \\
\left(
 \begin{array}{c}
{\bf G}_{12} \nonu \\
{\bf G}_{21}
 \end{array}
\right)(z) \, 
\left(
\begin{array}{c}
{\bf V_{+}^{(2)}} \nonu \\
{\bf U_{-}^{(2)}}
 \end{array}
\right)(w) & = &
\frac{1}{(z-w)^2} \frac{(N+2k+3)}{(N+k+2)} \left[ 
\left(
 \begin{array}{c}
-{\bf G}_{22} \nonu \\
{\bf G}_{11}
 \end{array}
\right) + 2   \left(
\begin{array}{c}
{\bf V^{(\frac{3}{2})}} \nonu \\
{\bf U^{(\frac{3}{2})}} 
\end{array}
\right)
\right](w) 
\nonu \\
& + & \frac{1}{(z-w)} \left[  
\frac{1}{3} \pa \, \mbox{(pole-2)}
 \mp     \left(
\begin{array}{c}
{\bf V^{(\frac{5}{2})}} \nonu \\
{\bf U^{(\frac{5}{2})}} 
\end{array}
\right) \right](w) + \cdots, 
\nonu \\
\left(
\begin{array}{c}
{\bf G}_{12} \nonu \\
{\bf G}_{21}
 \end{array}
\right)(z) \, {\bf W^{(2)}}(w) & = & 
\frac{1}{(z-w)^2} \frac{(k-N)}{2(N+k+2)} \left[  \left(
\begin{array}{c}
{\bf G}_{12} \nonu \\
{\bf G}_{21}
 \end{array}
\right) \mp 2 {\bf T_{\mp}^{(\frac{3}{2})}} \right](w)
\nonu \\
&+ &  \frac{1}{(z-w)} \left[ 
\frac{1}{3} \pa \, \mbox{(pole-2)}  
+  {\bf W_{\mp}^{(\frac{5}{2})}} \right](w) +\cdots, 
\nonu \\
 \left(
\begin{array}{c}
{\bf G}_{12} \nonu \\
{\bf G}_{21}
 \end{array}
\right)(z) \, {\bf W_{\pm}^{(\frac{5}{2})}}(w) & = & 
\mp
\frac{1}{(z-w)^3} \frac{8(k-N)}{3(N+k+2)} {\bf T^{(1)}}(w) 
\nonu \\
& + &  
\frac{1}{(z-w)^2} \left[ \frac{4(k-N)}{3(N+k+2)} {\bf T^{(2)}} +
4 {\bf W^{(2)}} \right](w) \nonu \\
& + & \frac{1}{(z-w)} \left[ \frac{1}{4} \pa \, \mbox{(pole-2)} 
\mp {\bf W^{(3)}} 
\right. 
\label{opeopeexp} \\
& \mp & \left. \frac{4(k-N)}{((4N+5)+(3N+4)k)} \left(
{\bf T} \, {\bf T^{(1)}} -\frac{1}{2} \pa^2 {\bf T^{(1)}} 
\right) \right](w)  +\cdots,
\nonu
\eea
where  the notations in \cite{Ahn1504} have the following relations
\bea
A_1\equiv -{\bf A}^{+1}, \qquad  A_2\equiv {\bf A}^{+2}, \qquad 
 A_3\equiv -{\bf A}^{+3}, 
\qquad
B_i\equiv {\bf A}^{-i}, \qquad i =1,2,3.
\nonu
\eea
Note that the structure constants appearing in 
(\ref{opeopeexp}) are rather simple form 
and one can easily figure out the general $N$ dependence for given 
several $N$ results.

\section{ The precise relations between the higher spin currents 
in the linear and nonlinear versions}

The exact relations between the $11$ currents in the nonlinear and 
linear versions are given by   \cite{GS}
\bea
T(z)&=&{\bf T}(z)+\frac{1}{(k+N+2)} 
\left( \bf UU +\pa \Gamma^{\mu} \Gamma_{\mu}\right)(z),
\nonu \\
G^{\mu}(z)&=&{\bf G^{\mu}}(z)\nonu \\
& + & \frac{2}{(k+N+2)} \left( {\bf U \Gamma^{\mu}}
-\frac{1}{3(k+N+2)} \epsilon_{\mu \nu \rho \sigma} 
{\bf \Gamma^{\nu} \Gamma^{\rho} \Gamma^{\sigma}}
-2 \alpha_{\mu \nu}^{+i} {\bf \Gamma^{\nu} } A^{+i}
+2 \alpha_{\mu \nu}^{-i} {\bf \Gamma^{\nu} } A^{-i}
\right)(z),
\nonu \\
A^{\pm i}(z) &=&
{\bf A}^{\pm i}(z)-\frac{1}{(k+N+2)} \alpha_{\mu \nu}^{\pm i}
{\bf \Gamma^{\mu} \Gamma^{\nu}} (z).
\label{gsformula}
\eea

For the higher
spin-$\frac{3}{2}$ currents,
because of ${\bf G}'^{\mu}(z) = G'^{\mu}(z)$ in (\ref{linnonlinear}), 
the differences between the higher spin-$\frac{3}{2}$ currents 
in the nonlinear and linear versions  (\ref{3halflinear}) 
come from 
the differences between the spin-$\frac{3}{2}$ current $G^{\mu}$ 
and the spin-$\frac{3}{2}$ current 
${\bf G^{\mu}}$ in (\ref{gsformula}).
See the equation (\ref{3halflinear}) (and 
$(3.21)$ and $(3.22)$ of \cite{AK1411}).

For the higher spin-$2$ currents, one has
\bea
U^{(2)}_{+}(z)&=& {\bf U^{(2)}_{+} }(z)+\frac{1}{(N+k+2)} \left[ 
{\bf \Gamma}_{11} {\bf G}'_{21}- {\bf \Gamma}_{21} {\bf G}'_{11} 
+ 2 A^{+3} A^{--} \right](z),
\nonu \\
U^{(2)}_{-} (z) &=& {\bf U^{(2)}_{-} }(z)+\frac{1}{(N+k+2)} \left[ 
{\bf \Gamma}_{12} {\bf G}'_{11}- {\bf \Gamma}_{11} {\bf G}'_{12} 
+2 A^{+-} A^{-3} \right](z),
\nonu \\
V^{(2)}_{+} (z) &=& {\bf V^{(2)}_{+} }(z)+\frac{1}{(N+k+2)} \left[ 
{\bf \Gamma}_{21} {\bf G}'_{22}- {\bf \Gamma}_{22} {\bf G}'_{21}
-2 A^{++} A^{-3} \right] (z),
\nonu \\
V^{(2)}_{-}(z) &=& {\bf V^{(2)}_{-} }(z)+\frac{1}{(N+k+2)} \left[ 
{\bf \Gamma}_{22} {\bf G}'_{12}- {\bf \Gamma}_{12} {\bf G}'_{22}
-2 A^{+3} A^{-+} \right](z),
\nonu \\
 T^{(2)}(z) &=& {\bf T^{(2)} }(z)+\frac{1}{(N+k+2)} \left[
{\bf \Gamma}_{22} {\bf G}'_{11}- {\bf \Gamma}_{11} {\bf G}'_{22}
+A^{+i} A^{+i} + A^{-i} A^{-i} +2 A^{+3} A^{-3}\right](z)
\nonu \\
&+& \frac{(N+k)}{(2Nk+N+k)} T(z),
\nonu \\
W^{(2)} (z)&=& {\bf W^{(2)} }(z)+\frac{1}{(N+k+2)} \left[
{\bf \Gamma}_{12} {\bf G}'_{21}- {\bf \Gamma}_{21} {\bf G}'_{12}
+A^{+i} A^{+i} + A^{-i} A^{-i} -2 A^{+3} A^{-3}\right](z)
\nonu \\
& + & T(z).
\label{twoexpression1}
\eea
In (\ref{twoexpression1}), the $N$ dependence can be determined 
without any difficulty.

For the higher spin-$\frac{5}{2}$ currents, one obtains
\bea
U^{(\frac{5}{2})} (z)&=& {\bf U^{(\frac{5}{2})}} (z) + \frac{1}{(N+k+2)}
\left[ -2 \pa {\bf  \Gamma_{11}} {\bf T^{(1)}} +  {\bf \Gamma_{11}}  \pa {\bf T^{(1)}}
+2 {\bf  \Gamma_{11} W^{(2)}} - 2 {\bf \Gamma_{12} U^{(2)}_{+} } 
-2 {\bf  \Gamma_{21} U^{(2)}_{-} } \right.
\nonu \\
&+& i {\bf A}^{+-} {\bf G}'_{21}+i {\bf A}^{+3} {\bf G}'_{11}
-i {\bf A}^{--} {\bf G}'_{12} -i  {\bf A}^{-3} {\bf G}'_{11}  + {\bf U G}'_{11}
-i A^{+-} G'_{21} 
+ 2i  A^{--} G'_{12}
\nonu \\
&+& \left. 2i A^{-3} G'_{11} -i A^{+-} G_{21} - 2i  A^{-3} G_{11}
+\frac{2}{3} \pa G'_{11} 
-\frac{2}{3} \pa G_{11}\right](z),
\nonu \\
V^{(\frac{5}{2})} (z)&=& {\bf V^{(\frac{5}{2})}} (z) + \frac{1}{(N+k+2)}
\left[ 2 \pa {\bf  \Gamma_{22}} {\bf T^{(1)}} -  {\bf \Gamma_{22}}  \pa {\bf T^{(1)}}
+2 {\bf  \Gamma_{12} V^{(2)}_{+}} + 2 {\bf \Gamma_{21} V^{(2)}_{-} } 
+2 {\bf  \Gamma_{22} W^{(2)} } \right.
\nonu \\
&+& i {\bf A}^{++} {\bf G}'_{12}+i {\bf A}^{+3} {\bf G}'_{22}
-i {\bf A}^{-+} {\bf G}'_{21} -i  {\bf A}^{-3} {\bf G}'_{22}  - {\bf U G}'_{22}
 -2i A^{++} G'_{12}  
-  2i A^{+3} G'_{22}
\nonu \\
&+& \left. i A^{-+} G'_{21} -2i A^{+3} G_{22} -i A^{-+} G_{21}
+\frac{2}{3} \pa G'_{22} 
+\frac{2}{3} \pa G_{22}  \right](z),
\label{5halfexpression} \\
W^{(\frac{5}{2})}_{+} (z)&=& {\bf W^{(\frac{5}{2})}_{+}} (z) + \frac{1}{(N+k+2)}
\left[ -2 \pa {\bf  \Gamma_{21}} {\bf T^{(1)}}  + {\bf \Gamma_{21}}  \pa {\bf T^{(1)}}
+2 {\bf  \Gamma_{21} T^{(2)}} + 2 {\bf \Gamma_{11} V^{(2)}_{+} } 
+2 {\bf  \Gamma_{22} U^{(2)}_{+} } \right.
\nonu \\
&+& i {\bf A}^{++} {\bf G}'_{11}-i {\bf A}^{+3} {\bf G}'_{21}
+i {\bf A}^{--} {\bf G}'_{22} -i  {\bf A}^{-3} {\bf G}'_{21} + {\bf U G}'_{21}
-i A^{++} G'_{11} 
 + 2i A^{+3} G'_{21}
\nonu \\
&-& \left.  i A^{--} G'_{22} +2i A^{-3} G'_{21} -i A^{++} G_{11} +2i A^{+3} G_{21} 
+i A^{--} G_{22}  
-  2i A^{-3} G_{21} -\frac{4}{3} \pa G_{21}
  \right](z),
\nonu \\
W^{(\frac{5}{2})}_{-} (z)&=& {\bf W^{(\frac{5}{2})}_{-}} (z) + \frac{1}{(N+k+2)}
\left[ 2 \pa {\bf  \Gamma_{12}} {\bf T^{(1)}}  - {\bf \Gamma_{12}}  \pa {\bf T^{(1)}}
+2 {\bf  \Gamma_{12} T^{(2)}} - 2 {\bf \Gamma_{11} V^{(2)}_{-} } 
-2 {\bf  \Gamma_{22} U^{(2)}_{-} } \right.
\nonu \\
&+& i {\bf A}^{+-} {\bf G}'_{22}-i {\bf A}^{+3} {\bf G}'_{12}
+i {\bf A}^{-+} {\bf G}'_{11} -i  {\bf A}^{-3} {\bf G}'_{12}  - {\bf U G}'_{12}
-i A^{+-} G'_{22} +2i A^{+3} G'_{12}
\nonu \\
&-& \left.  i A^{-+} G'_{11} +2i  A^{-3} G'_{12} +i A^{+-} G_{22} -2i A^{+3} G_{12} 
-i A^{-+} G_{11}  
+  2i A^{-3} G_{12}
-\frac{4}{3} \pa G_{12}
 \right](z).
\nonu
\eea
In this case, the expression (\ref{5halfexpression})
reveals the $N$ dependence clearly.

Finally, for the higher spin-$3$ current,
the following relation holds
\bea
W^{(3)}(z)&=&
\left[ {\bf W^{(3)}} + a_1 {\bf T^{(1)} T } 
\right. \nonu \\
& + & \left.
 \frac{1}{(N+k+2)} \left( \sum_{\mu=0}^{3} ( {\bf \Gamma^{\mu} \pa G'^{\mu} } 
-3  {\bf \pa \Gamma^{\mu}  G'^{\mu} } )
 +2( {\bf  \pa U  T^{(1)} } - {\bf U \pa T^{(1)} } )\right) \right](z) 
\nonu \\
&+& a_2 T^{(1)} T (z)
-
\frac{1}{(N+k+2)} \left[
2 G_{12}  G_{21} -2i A^{++} U^{(2)}_{-} -2i A^{--} V^{(2)}_{-}
-4i A^{+3} T^{(2)} \right.
\nonu \\
&+& a_4 A^{-3} T
-4i  A^{-3} T^{(2)} +i  A^{+3} \pa T^{(1)} 
+i A^{-3} \pa T^{(1)} - i T^{(1)} \pa A^{+3}
-  i T^{(1)} \pa A^{-3} - \pa W^{(2)}  
\nonu \\
 &-& \left. \pa T \right](z)
- \frac{1}{(N+k+2)^2} \left[
 8i A^{+3} A^{+-} A^{++} 
+8i A^{-3} A^{+-} A^{++} 
+8i A^{+3} A^{+3} A^{+3} \right.
\nonu \\
&+&16 i A^{+3} A^{+3} A^{-3} 
+  8i A^{+3} A^{-3} A^{-3}
+a_5  \pa A^{+-} A^{++} +a_6  A^{+-} \pa A^{++}
\nonu \\
& -& (k+1)  A^{-+} \pa A^{--}
+  6 A^{+3} \pa A^{+3} 
+a_7 A^{+3} \pa A^{-3} +a_8 A^{-3} \pa A^{+3}
-2 A^{-3} \pa A^{-3} \nonu \\
& + & \left. a_9 \pa^2 A^{+3}
- \frac{8i k }{3} \pa^2 A^{-3}  
 + a_3 A^{+3} T 
+(k-1) A^{--} \pa A^{-+}
\right](z),
\label{w3generalexpexp}
\eea
where the coefficients in (\ref{w3generalexpexp}) are given by
\bea
a_1 & = & \frac{4(k-N)}{(4N+5+(3N+4)k)}, 
\nonu \\
a_2 & = & \frac{8(N-k)}{(5N+4+(6N+5)k)}, 
\nonu \\
a_3 & = & 
\frac{8 i (N+k+2)(4N^2+4N+(6N^2+13N+4)k+(8N+5)k^2)}{ (2Nk+N+k) (5N+4+(6N+5)k)},
\nonu \\
a_4 & = & \frac{8 i (-2N^2-3N + k) k }{ (2Nk+N+k) (5N+4+(6N+5)k)},
\nonu \\
a_5 & = & (N-1),
\nonu \\
a_6 & = & -(N+1),
\nonu \\
a_7 & = & 2(k-N-1),
\nonu \\
a_8 & = & - 2(k-N-3),
\nonu \\
a_9 & = & \frac{i (5N+3)}{3}.
\label{coeffcoeff}
\eea
In this case, the general $N$ behavior can be obtained 
by taking some power of $N$ in the fractional expressions of
the level $k$  from the several $N$ results. 
It turns out that the quadratic in $N$ occurs. See, for example, 
the numerator 
of the coefficient $a_3$ in (\ref{coeffcoeff}). 


\end{document}